\documentclass[11pt]{article}
\pdfoutput = 1

\usepackage[utf8]{inputenc}
\usepackage[dvipsnames]{xcolor}
\usepackage{amsmath}
\usepackage{verbatim}
\usepackage{amssymb}
\usepackage{physics}
\usepackage{amsfonts}
\usepackage{cite}
\usepackage{array}
\usepackage{setspace}
\usepackage{float}
\usepackage{url}
\usepackage{tikz}
\usepackage{graphicx}
\usepackage{fancybox}
\usepackage{tensor}
\usepackage{tikz}
\usepackage{cite}
\usepackage{caption}
\usepackage{subcaption}
\usepackage{slashed}

\allowdisplaybreaks

\usepackage[margin = 2.2cm]{geometry}
\usepackage[ragged]{footmisc}
\usepackage{hyperref}
\hypersetup{
    colorlinks,
    citecolor=blue,
    filecolor=black,
    linkcolor=blue,
    urlcolor=blue,
    linktocpage=true
}

\def\Th{\Theta}
\def\th{\Theta}

\def\nn{\nonumber}

\def\G{\Gamma}
\def\D{\Delta}

\def\J{\mathbf{j}}
\def\Tr{\operatorname{Tr}}

\newcommand{\be}{\begin{equation}}
\newcommand{\ee}{\end{equation}}
\newcommand{\bea}{\begin{align}}
\newcommand{\eea}{\end{align}}
\newcommand{\bi}{\begin{itemize}}
\newcommand{\ei}{\end{itemize}}

\newcommand{\lr}[1]{\left( #1 \right)}

\def\t{\text}

\def\rb{\rangle}
\def\lb{\langle}

\newcommand{\Ebrk}{E_\text{brk.}}

\def\cO{\mathcal{O}}

\def\a{\alpha}
\def\g{\gamma}
\def\nn\nonumber
\def\ve{\varepsilon}
\def\nn{\nonumber}
\def\rb{\rangle}
\def\lb{\langle}
\def\ol{\overline}

\def\t{\text}
\def\ol{\overline}

\def\bps{\ol{\psi}}

\def\bpsra{\ol{\psi}_{r,1}}
\def\bpsrb{\ol{\psi}_{r,2}}

\def\psla{{\psi}_{l}^1}
\def\pslb{{\psi}_{l}^2}
\def\psra{{\psi}_{r}^1}
\def\psrb{{\psi}_{r}^2}

\numberwithin{equation}{section}

\begin{document}

\thispagestyle{empty}

\begin{center}
\vspace*{.4cm}

     {\LARGE \bf 
The evaporation of black holes in supergravity}
    
    \vspace{0.4in}
    {\bf Guanda Lin$^1$, Luca V.~Iliesiu$^1$, Mykhaylo Usatyuk$^2$}

    \vspace{0.4in}
    {${}^1$ Center for Theoretical Physics and Department of Physics, Berkeley, CA, 94720, USA}
    
    {${}^2$ Kavli Institute for Theoretical Physics, Santa Barbara, CA 93106, USA}
    \vspace{0.1in}
    
    {geoff\_guanda\_lin@berkeley.edu, liliesiu@berkeley.edu, musatyuk@kitp.ucsb.edu \ \ \ }
\end{center}

\vspace{0.4in}
\begin{abstract}
In supergravity, charged rotating black holes are generically driven towards becoming extremal and supersymmetric through the emission of Hawking radiation. Eventually, as the black hole approaches the BPS bound and is close to becoming supersymmetric, quantum gravity corrections become critical to describing the emission of Hawking radiation, making the QFT in curved spacetime approximation inaccurate. In this paper, we compute how such quantum gravity corrections affect the spectrum of Hawking radiation for black holes in $\mathcal N=2$ supergravity in flatspace. We show that due to such corrections, the spectrum of emitted Hawking radiation for both spin-0 and spin-$1/2$ particles deviates drastically at low temperatures from the naively expected black-body spectrum. Rather remarkably, the spectrum exhibits a discrete emission line from direct transitions from near-BPS to BPS states, providing the first controlled example where the discreteness of the black hole energies is visible in the emitted Hawking radiation. Similar quantum gravity effects drastically modify the absorption cross-section: BPS black holes are transparent to certain frequencies, while near-BPS black holes appear much larger than the semi-classical prediction.
\end{abstract}

\pagebreak
\setcounter{page}{1}
\tableofcontents

\section{Introduction}

\begin{figure}[t!]
    \centering
    \includegraphics[width=1\linewidth]{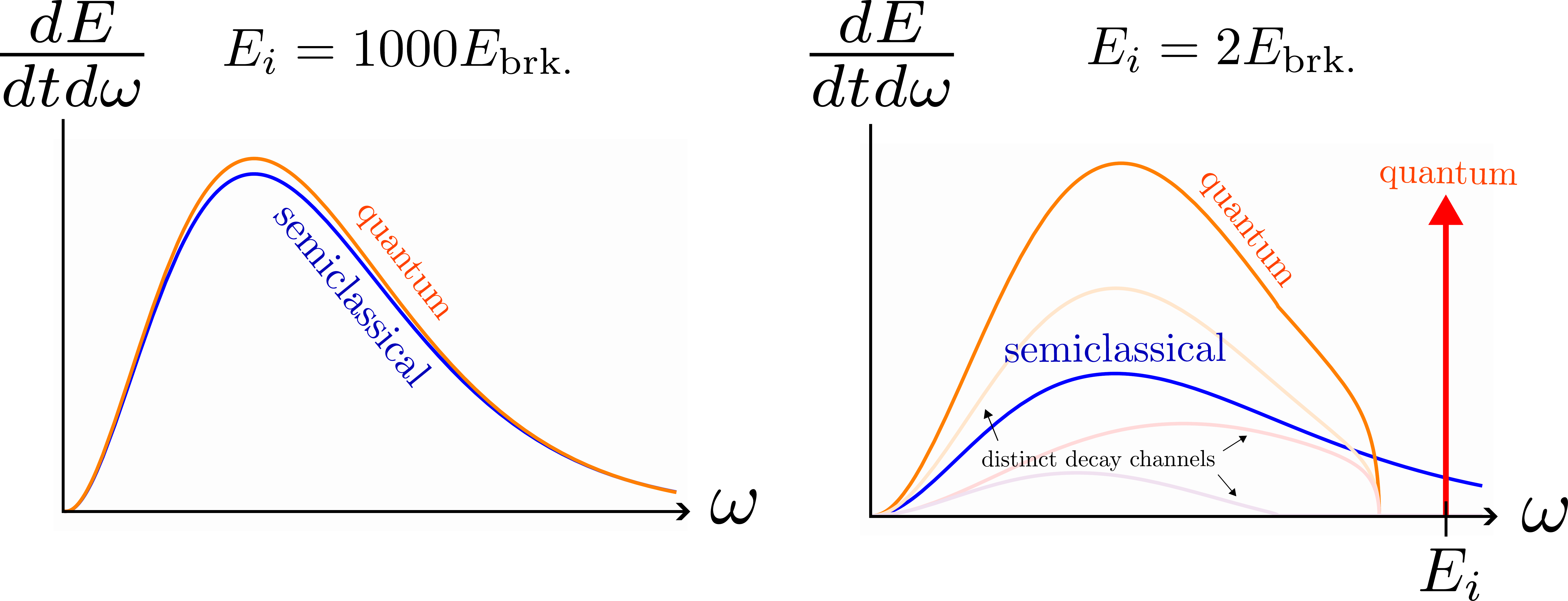}
    \caption{Comparison of the \textcolor{blue}{semiclassical prediction} vs. \textcolor{orange}{quantum corrected} Hawking radiation into a massless scalar field. The energy flux is plotted for a black hole that is initially in an energy eigenstate $|E_i, \Psi\rb$ above extremality with zero angular momentum $j=0$. \textbf{Left:} At large energies $E_i \gg \Ebrk$ the quantum flux approaches the semiclassical answer. \textbf{Right:} At low energies $E_i \sim \Ebrk$ there are very large deviations from the semiclassical answer. The lightly colored curves are the fluxes from \textcolor{magenta}{distinct decay channels}, and sum to the \textcolor{orange}{quantum flux}. The black hole can emit all of its energy into a single particle and transition to a \textcolor{red}{BPS state}, which gives a Dirac delta function represented by the red arrow. Due to the gaps in the spectrum, transitions into \textcolor{magenta}{near-BPS states} abruptly cut off when the states cease to exist. Furthermore, due to large quantum fluctuations, the scalar particle can oftentimes more easily escape the black hole potential barrier, giving a larger quantum flux than the semiclassical prediction. Distinct decay channels are not plotted in the first figure for visual clarity. } 
    \label{fig:dedt_micro}
\end{figure}

\begin{figure}
    \centering
    \hspace*{-1cm}
    \includegraphics[width=1.1\linewidth]{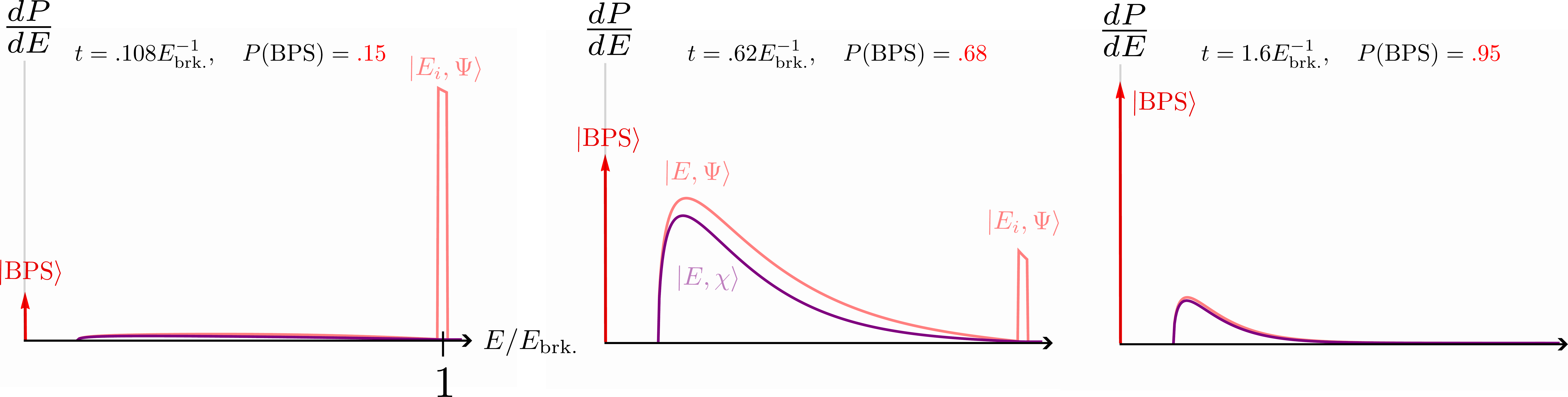}
    \caption{We plot the evolution of the probability density $\frac{d P}{d E}$ for a black hole in an \textcolor{magenta}{initial energy eigenstate} with $J=0$ as it evolve in time due to Hawking radiation. Starting in an initial state, the black hole evolves to states with smaller energy, within different supermultiplets ($\textcolor{magenta}{|E,\Psi\rb}, \textcolor{purple}{|E,\chi\rb},$ within the same supermultiplet or $ \textcolor{red}{|\t{BPS}\rb}$). At late times, the black hole evolves towards \textcolor{red}{BPS states} with $M=Q$, and we include the probability to end up in the BPS state as a function of time. An additional video of the evolution over time can also be found at \cite{videolink}.}
    \label{fig:statesevolution}
\end{figure}

\begin{figure}[t!]
    \centering
    \includegraphics[width=1\linewidth]{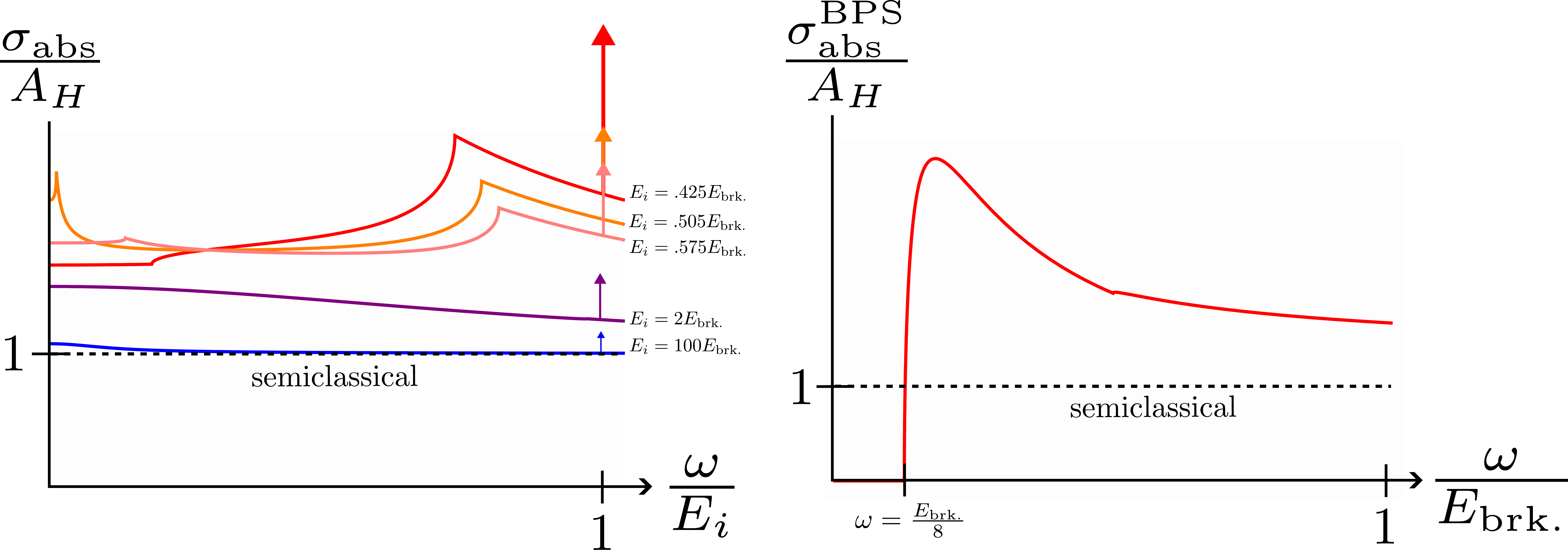}
    \caption{Comparison of the semiclassical prediction (black dashed) vs. \textcolor{orange}{quantum corrected} (all colors) absorption cross section for a massless scalar. \textbf{Left:} The black hole is in an initial energy eigenstate with $J=0$ and energy $E_i$, and the absorption cross section is shown for a variety of initial energies. At \textcolor{blue}{large energies} $E_i$ above extremality, the cross section approaches the semiclassical answer. At very \textcolor{red}{small energies} the quantum cross section deviates significantly, with resonance peaks in the plot indicating that there are either: new black hole states at $E_i + \omega$ that the mode $\omega$ can be absorbed by, or that stimulated emission can no longer occur at frequency $\omega$ since the final BH states do not exist at $E_i-\omega$. There are always delta function peaks at $E_i=\omega$ from stimulated emission into BPS states, though they decrease in magnitude as the initial black hole deviates further from the BPS bound. We have schematically plotted these delta functions in the plot. \textbf{Right:} Absorption cross section for an initial BPS state with $E_i=0$. The black hole is transparent to an incoming wave with $\omega < \frac{\Ebrk}{8}$ since there are no final states at $E=\omega$ to transition into.}
    \label{fig:absorption}
\end{figure}

Consider a charged rotating black hole in a theory of supergravity in flatspace.\footnote{Here, we shall assume that we are in $\mathcal N\geq 2$ supergravity in $d=4$ spacetime dimensions. We shall also assume that the black hole carries non-zero charge $Q$ under the gauge field in the supermultiplet of the graviton and gravitino with $Q e^{-1}\gg 1$.} When such a black hole preserves none of the supersymmetries of the theory, it will emit Hawking radiation. Because of supersymmetry, the particles that the black hole emits, as well as the black holes themselves, satisfy a BPS bound that imposes that their masses are lower bounded by their charges.\footnote{For example, in pure $\mathcal N=2$ supergravity the BPS bound is $m>q$ where $m$ is the mass of the particle or black hole while $q$ is the charge under the $U(1)$ gauge field in the graviton supermultiplet.} Because of this bound, black holes in supergravity are driven not only towards extremality but also towards becoming supersymmetric. As this limit is approached, the semiclassical picture of Hawking radiation stops making sense: a single emitted quantum carries away an energy greater than the black hole has available above the BPS bound, naively making the resulting object violate this bound. Since violating this bound is forbidden in supergravity, the semi-classical picture of near-BPS black holes must consequently fail, and quantum gravity effects should become critical in this regime \cite{Preskill:1991tb,Maldacena:1998uz,Page:2000dk}.\footnote{For charged black holes in pure gravity or supergravity, the energy scale at which we expect both the backreaction and the quantum gravity effects to become important is $\Ebrk = \frac{M_\text{Pl}}{Q^3}$. } The aim of this paper is to quantify precisely how this semi-classical picture fails by computing the new rates of Hawking radiation once quantum gravity corrections in supergravity are taken into account. The spectrum of Hawking radiation that we discover for near-BPS black holes indeed has drastic deviations from the approximately thermal spectrum seen in the semi-classical analysis; rather, the spectrum exhibits a discrete emission line, similar to those seen in atomic physics. The main results of the paper are summarized in figures \ref{fig:dedt_micro} and \ref{fig:statesevolution} for the emission of Hawking radiation. Similarly, in figure \ref{fig:absorption}, we summarize the results for the absorption of radiation by the black hole in a plane wave scattering experiment.\footnote{In figures \ref{fig:dedt_micro}, \ref{fig:statesevolution}, and \ref{fig:absorption}, we will take the initial state to be $\ket{E, \Psi}$ within the supermultiplet labeled by $\mathbf J= \frac{1}{2}$ with $J=0$. We will discuss the supermultiplet structure of black holes in $\mathcal N =2$ supergravity in detail in section \ref{sec:4}. }

While quantifying the exact way in which the semi-classical picture of Hawking radiation fails for near-BPS black holes has so far been a mystery, there are numerous clues that have helped guide our analysis. In recent years, our understanding of the thermodynamics of near-extremal black holes has been drastically improved. The near-extremal geometry has a set of metric fluctuations that have an arbitrarily small classical action and, consequently, can have an arbitrarily large quantum mechanical variance \cite{Iliesiu:2020qvm,  Ghosh:2019rcj, Iliesiu:2022onk, Kapec:2023ruw, Rakic:2023vhv, Kolanowski:2024zrq, Maulik:2024dwq}.  Luckily, these modes, known as the Schwarzian modes because their action takes the form of a Schwarzian derivative, can be integrated out exactly, and their effect on the entropy of near-extremal black holes has been precisely quantified. For black holes in conventional non-supersymmetric gravitational theories, such quantum corrections drastically affect their density of states at low energies, making it vanish as extremality is approached \cite{Stanford:2017thb, Iliesiu:2020qvm, Iliesiu:2022onk}. Such quantum corrections not only affect the thermodynamic properties of such black holes but also affect their dynamics. In contrast to the semi-classical prediction, the spectrum of Hawking radiation becomes highly non-thermal, ensuring that, in contrast to the semi-classical prediction, such black holes always remain subextremal \cite{Brown:2024ajk, Biggs:2025nzs,  Mertens:2019bvy, Blommaert:2020yeo}. Since the emitted particles have very small energy, emitting even a single Hawking quanta takes an enormous amount of time in the near-extremal regime \cite{Brown:2024ajk}. Because of that, it is difficult for a hypothetical observer to probe the large quantum gravity fluctuations in the black hole geometry by measuring the spectrum of the emitted Hawking radiation. However, because these large fluctuations also affect the absorption cross-section of such black holes, the hypothetical observer can instead conceivably probe large quantum gravity effects through a scattering experiment \cite{Emparan:2025sao}.  For black holes in supergravity, not only fluctuations of the metric but also their superpartners, specific fluctuations of the gravitino, have arbitrarily small classical action \cite{Heydeman:2020hhw, Iliesiu:2021are, Boruch:2022tno, Iliesiu:2022kny, Heydeman:2024ohc, Ezroura:2024xba,  Heydeman:2024ezi}. Expanding the action in terms of these fluctuations, one no longer finds the Schwarzian derivative, but instead, for black holes in $\mathcal N\geq 2$ supergravity in flatspace, one finds its supersymmetric extension, the $\mathcal N=4$ super-Schwarzian \cite{Heydeman:2020hhw}.  Taking both bosonic and fermionic quantum gravity fluctuations into account results in a completely different entropy for near-BPS black holes: in contrast to the non-supersymmetric case, the density of states now has a large degeneracy of BPS states, followed by a gap that scales with the black hole charges and then by a smooth density of states \cite{Stanford:2017thb, Heydeman:2020hhw}.\footnote{For example for near-BPS black holes in pure $\mathcal N=2$ supergravity for a black hole with charge $Q$, the gap is given by $E_\text{gap} = \frac{\Ebrk}8 = \frac{M_\text{Pl}}{8Q^3}$. This gap also agrees with prior conjectures made from the weak coupling stringy construction of such a black hole \cite{Callan:1996dv,Maldacena:1996ds,Maldacena:1997ih}. } The existence of this gap indeed suggests that the Hawking radiation emitted by near-BPS black holes has an associated discrete emission line coming from direct transitions from near-BPS to BPS states. However, the exact computation of the Hawking radiation flux for such black holes that explains how these emission lines arise has so far not been performed. To calculate such a flux, we carefully compute the two-point function in the $\mathcal N=4$ super-Schwarzian theory, including the effects of
backreaction, as well as greybody factors and metric fluctuations.\footnote{On a technical level, this extends the prior results found for correlation functions in the $\mathcal N=1$ and  $\mathcal N=2$ super-Schwarzian theories \cite{Mertens:2017mtv, Fan:2021wsb, Lin:2022zxd, Lin:2022rzw}. } These technical results allow us to recover the full history of black hole evaporation in supergravity, drastically correcting the QFT in curved spacetime calculation as the black hole approaches the BPS bound.\vspace{-0.2cm}

\subsection*{Plan for paper}

The rest of this paper is organized as follows. In section \ref{sec:N=4-super-Schw-and-EFT-for-SUSY-BHs} we review how the $\mathcal N=4$ super-Schwarzian emerges as the EFT that describes the large quantum gravity fluctuations seen for near-BPS black holes in $\mathcal N\geq 2$ supergravity. We also explain how the fields in the supergravity theory organize themselves in the representation of the near-horizon isometry of such black holes. In section \ref{eq:correlators-in-super-Schw}, we compute the two-point function in the $\mathcal N=4$ super-Schwarzian, reformulating the theory in terms of an $\mathcal N=4$ super-Liouville quantum mechanics whose Hamiltonian can be exactly diagonalized. With these correlation functions, in section \ref{sec:4} we then compute the spontaneous emission rate for Hawking radiation when including quantum gravity effects by using Fermi's golden rule. To exemplify our results, we focus on massless spin-$0$ (see figure \ref{fig:dedt_micro}) and $1/2$ particles that belong to supermultiplets in supergravity, a hypermultiplet. Assuming there are no vector multiplets, this turns out to be the dominant emission channel, which allows us to describe the evaporation history of the black hole (see figure  \ref{fig:statesevolution}). Finally, we compute the absorption cross-section of the black hole and find that, just like the emitted flux, it also receives drastic corrections (see figure \ref{fig:absorption}).

\section{The $\mathcal{N}=4$ super-Schwarzian as an EFT for near-BPS black holes}
\label{sec:N=4-super-Schw-and-EFT-for-SUSY-BHs}

In this section, we will review the necessary ingredients for obtaining the flux of Hawking radiation from scalar and fermionic fields emitted by nearly-supersymmetric black holes in $\mathcal N=2$ supergravity. We start by reviewing the classical features of such black holes in section \ref{subsec:N=2SUGRA}. Then, in section \ref{subsec:super-Schw}, we proceed to analyze the super-Schwarzian EFT that is responsible for describing the large quantum fluctuations of the metric around the black hole geometry. Finally, in section \ref{subsec:symm-NHR}, we describe how matter supermultiplets in supergravity decompose into representations of the near-horizon isometry and analyze the properties of the Hawking radiation emitted into such supermultiplets when solely using the QFT in curved spacetime approximation.   

\subsection{$\mathcal{N}=2$ Supergravity and supersymmetric black holes}
\label{subsec:N=2SUGRA}

We review the necessary aspects of ungauged pure $\mathcal{N}=2$ supergravity in $d=4$ following the conventions of \cite{Freedman:2012zz}. The field content is the metric $g_{M N}$, two spin-$\frac{3}{2}$ (four-component) Majorana gravitini $\Psi_{M}^I$ labelled by $I=1,2$, and the $U(1)$ gauge field $A_M$ with field strength $F_{M N}$. The indices $M,N$ are spacetime indices while spinor indices are suppressed. The action is 
\begin{equation}
\begin{aligned} \label{eqn:I_Sugra}
    S = \frac{1}{8 \pi G_N}\int d^4 x \sqrt{-g} \left(\frac{1}{2} R-\ol{\Psi}_{I M} \Gamma^{M N P} D_N \Psi_P^I \right. &-\frac{1}{4} F_{M N} F^{M N}  +  
     \frac{\varepsilon^{I J}}{2 \sqrt{2}} \ol{\Psi}_I^M\left(F_{M N}+i \star F_{M N} \Gamma_5\right) \Psi_J^N  \\ & \left. + \frac{\varepsilon^{I^{\prime}J^{\prime}}\varepsilon^{IJ}}{8} \ol{\Psi}^M_{I^{\prime}} \ol{\Psi}^N_{J^{\prime}} R_{MNPQ} \Psi^P_I \Psi^Q_J \right) \,.
\end{aligned}
\end{equation}
In the tetrad formalism the metric is given by $g_{M N} = e_{M}^{~a} e_{N}^{~b} \eta_{a b}$ with $\eta$ the flat space metric and $a,b$ tangent space indices.\footnote{Some other conventions are that $\Gamma_5$ is the standard Chirality matrix in flat space. Objects with $a,b$ indices have been transformed $F_{a b}=e_a^{~M} e_b^{~N} F_{M N}$.} Spacetime gamma matrices are defined in the standard way $\Gamma^M=e^{M}_{~a} \Gamma^a$ with flat space gamma matrices $\Gamma^a$. The covariant derivative acts on the gravitini according to $D_N \Psi_P = (\nabla_N+\frac{1}{4}\omega_N^{a b} \Gamma_{a b}) \Psi_P $ with $\omega$ the spin connection, and $\epsilon^{I J}$ is the antisymmetric symbol. In the second line, we have written the four-fermion interaction. For simplicity, we will from now on work in units where $G_N=1$.

The local supersymmetry transformations that preserve the action are
\begin{equation}
\begin{aligned} \label{eqn:susy_transf_N=2}
& \delta_\epsilon e_M^a=\frac{1}{2} \bar{\epsilon}^I \Gamma^a \Psi_{M I}+\text { h.c.}\,,\\
& \delta_\epsilon A_M=\frac{1}{\sqrt{2}} \varepsilon^{I J} \bar{\epsilon}_I \Psi_{M J}+\text { h.c.}\,, \\
& \delta_\epsilon \Psi_M^I=\left(\partial_M+\frac{1}{4} \omega_M^{a b} \Gamma_{a b}\right) \epsilon^I-\frac{1}{4 \sqrt{2}} \Gamma^{a b} F_{a b} \Gamma_M \varepsilon^{I J} \epsilon_J\,.
\end{aligned}
\end{equation}
The Majorana spinor $\epsilon_I(x)$ is an arbitrary function of the spacetime coordinates, and we have eight real functions worth of independent transformations. There is also a $SU(2)_R$ global symmetry under which the gravitini rotate $\Psi^I \to U^I_{~J} \Psi^J$.\footnote{This is not the $SU(2)$ symmetry appearing later and will not be important for us.}

\subsection*{Supersymmetric black holes}
At the classical level, the solutions of pure $\mathcal{N}=2$ supergravity we will be interested in are standard reissner-Nordstr\"om (RN) black holes with metric, gauge field, and gravitini
\begin{gather}
d s^2=-f(r) d t^2+ \frac{d r^2}{f(r)}+r^2 (d\theta^2 + \sin^2\theta d\phi^2)\,, \qquad f(r)=\frac{(r-r_+)(r-r_-)}{r^2}\,, \\
A = \frac{Q}{r} d t, \quad F=\frac{Q}{r^2} dr \wedge d t,\qquad \Psi^I_M = 0\,.
\end{gather}
where $r_\pm = G_N M \pm \sqrt{(G_N M)^2 -G_N Q^2}$ are the inner and outer horizons of the BH. The solution presented above is purely electric, but there are also magnetic and dyonic solutions for which the analysis performed in this paper is applicable; however, for brevity, we will solely focus on the electric solutions.  At the classical level, the temperature and entropy are given by
\be
\label{eq:thermodynamic-relations}
T=\frac{r_+-r_-}{4\pi r_+^2}, \qquad S=\frac{\pi r_+^2}{G_N}= S_0+\frac{4 \pi^2 }{\Ebrk}T + \ldots\,, \qquad S_0 = \pi Q^2\,.
\ee
In the second line, we have expanded the entropy around $T=0$ and listed the entropy $S_0$ at extremality. At zero temperature, we have $r_+=r_-$, which results in $M=Q/\sqrt{G_N}$, giving the zero temperature entropy. We have also defined an emergent energy scale
\be \label{eqn:Ebrk}
\Ebrk = \frac{M_{\t{pl}}}{Q^3}\,,
\ee
which will be important in the quantum analysis. It is at this energy scale above extremality that we expect quantum gravity effects to become important \cite{Preskill:1991tb,Maldacena:1998uz,Page:2000dk}, and therefore the thermodynamic relations \eqref{eq:thermodynamic-relations} are only to be trusted if the energy above extremality is much larger than $\Ebrk$. 

\paragraph{Killing spinors and symmetry group.} We now summarize the amount of supersymmetry preserved by these solutions. We would like to find supersymmetry transformations \eqref{eqn:susy_transf_N=2} with spinor profiles $\epsilon_I$ such that the background is left invariant. Since the gravitini vanish on-shell, the only condition to enforce is $\delta_\epsilon \Psi_M^I=0$ given by
\be
\left(\partial_M+\frac{1}{4} \omega_M^{a b} \Gamma_{a b}\right) \epsilon^I-\frac{1}{4 \sqrt{2}} \Gamma^{a b} F_{a b} \Gamma_M \varepsilon^{I J} \epsilon_J = 0\,.
\ee
We will not go into details regarding the solutions, but see \cite{Gauntlett:1998kc,Lu:1998nu,Duff:1986hr,Heydeman:2020hhw,Freedman:2012zz}. In the case of finite temperature there are no solutions. In the case of the extremal BH with zero temperature, there are four independent Killing spinors that make the extremal solution half-BPS. In the zero temperature limit, the solution develops an infinitely long AdS$_2 \times S^2$ throat. Using a standard coordinate change we can zoom in on the throat 
\be
ds^2 = \frac{\ell_2^2}{z^2} (-dt^2 + dz^2) + \ell_2^2 d \Omega_2^2,
\ee
where $\ell_2=\sqrt{G_N} Q$ is the AdS$_2$ and sphere radius. The above background has eight independent Killing spinor solutions, and so there is an enhancement of supersymmetry when zooming into the near-horizon region and discarding the flat space asymptotics \cite{Kallosh:1997qw,Claus:1998yw,Boonstra:1998yu,Heydeman:2020hhw}.

The eight Killing spinors give us the super-isometry algebra, which generates the symmetry group of the solution. It was shown \cite{Kallosh:1997qw,Claus:1998yw,Boonstra:1998yu,Heydeman:2020hhw} that the Killing spinors of AdS$_2\times S^2$ generate the supergroup $PSU(1,1|2)$, with bosonic subgroup $SL(2,\mathbb{R}) \times SU(2)$. The bosonic subgroup can be seen to be the isometry group of AdS$_2 \times S^2$.

\subsection{The super-Schwarzian}
\label{subsec:super-Schw}

In \cite{Heydeman:2020hhw}, one-loop quantum corrections were analyzed for near-extremal reissner-Nordstr\"om black holes in Euclidean $\mathcal{N}=2$ supergravity. It was found that at very low temperatures, the dominant fluctuations of the fields localized in the near-horizon AdS$_2\times S^2$ region were governed by an $\mathcal{N}=4$ super-Schwarzian theory. Schematically, we have
\be \label{eqn:Schwarzian_PI}
\int \mathcal{D}(\delta g) \mathcal{D}(\delta A) \mathcal{D}(\delta \Psi) \exp \lr{-S_{\mathcal{N}=2}} \sim \int \frac{\mathcal{D}f \mathcal{D}g \mathcal{D}\eta \mathcal{D} \ol \eta}{PSU(1,1|2)} \exp \lr{-S_{\mathcal{N}=4}^{\t{Sch}}}\,,
\ee
where $\delta g, \delta A, \delta \Psi$ are fluctuations of the metric, gauge field, and gravitini around the classical background of the Euclidean RN solution. On the right we have the $\mathcal{N}=4$ super-Schwarzian action $S_{\mathcal{N}=4}^{\t{Sch}}$ with bosonic part 
\be
S_{\mathcal{N}=4}^{\t{Sch.}}= -\Ebrk^{-1} \int_0^\beta d \tau \lr{\t{Sch}(f,\tau) + \Tr (g^{-1} \partial_\tau g )^2 + \t{fermions}  } \,,
\ee
where the light modes of the metric in supergravity become the bosonic modes $f,g$ with $g\in SU(2)$ while the light modes of the gravitini become the complex fermions $\eta, \ol \eta$. The scale for the action is set by the breakdown scale \eqref{eqn:Ebrk} of the corresponding BH, and the dimensionless parameter $\beta \Ebrk$ determines when quantum effects become important.

\paragraph{Symmetries and states.}  The symmetries of the super-Schwarzian were analyzed in detail in \cite{Heydeman:2020hhw}, and we only highlight the most important points. The theory has a $PSU(1,1|2)$ gauge symmetry which acts on the fields $(f,g,\eta,\ol \eta)$ which we mod out by in \eqref{eqn:Schwarzian_PI}. In gravity variables, this symmetry generates identical field configurations under diffeomorphisms and originates from the super-isometry group of the near-horizon AdS$_2\times S^2$ geometry.

The more interesting symmetry group is the $\mathcal{N}=4$ super-Poincare group with algebra
\begin{gather} \label{eqn:N=4SPC_algebra}
\{\ol Q_p, Q^q  \}= \delta_p^q H, \qquad \{ Q^p ,Q^q \}=0, \qquad \{ \ol Q_p , \ol Q_q \}=0 \,,\\
[J_i, J_j] = i \epsilon_{i j k} J_k, \qquad [J^2,J_i]=0, \qquad [Q^p,J_i]=\frac{1}{2}(\sigma_i)^{p}_{~q} Q^{q}\,, \qquad [\ol Q_p,J_i]=-\frac{1}{2}\ol Q_{q}(\sigma_i)^{q}_{~p} \nn
\end{gather}
where we have complex supercharges labelled by $p,q=1,2$, and $J_i$ are angular momenta generators on the $S^2$. Importantly, the states of the theory are organized into irreps of $\mathcal{N}=4$ super-Poincare, not of $PSU(1,1|2)$. States will be labelled by an energy, angular momenta, and  axial angular momenta $|E^j_{j_z}\rb$.

To construct the multiplets We can take an initial state of definite energy and angular momentum $j$ to be annihilated by all $Q^i|j\rb=0$. The multiplet is spanned by $|j\rb, \ol Q_1 |j\rb, \ol Q_2 |j\rb, \ol Q_1 \ol Q_2 | j\rb$. Using the algebra it can be found that $[J^2,  \ol Q_1 \ol Q_2 ]=0$ and so the last state in the multiplet also has spin $j$. The middle states have spin $\frac{1}{2}\otimes j = j-\frac{1}{2}\oplus j+\frac{1}{2}$. It's convenient to shift $j$ down by half and label the multiplets according to the largest value of angular momentum:
\begin{gather}
\mathbf{j}= j \oplus 2 (j-\frac{1}{2}) \oplus (j-1)\,. \\
\mathbf{\frac{1}{2}} = \frac{1}{2} \oplus 2(0)\,. 
\end{gather}
The $\mathbf{\frac{1}{2}}$ multiplet is special and has one fewer state since $\ol Q_1 |j\rb, \ol Q_2 |j\rb$ are linear combinations of the spin $j=\frac{1}{2}$ state with different axial angular momenta. There are also special BPS multiplets of arbitrary spin and energy that are annihilated by all supercharges $Q^i|\t{BPS}\rb=0=\ol Q_i|\t{BPS}\rb$. In our case, we will only have a BPS state with $j=0$ and $E=0$. For later convenience, we will introduce notation to further distinguish states in the multiplet
\be \label{eqn:schw_multiplet}
\mathbf{j} = |E^j_{m_1} , H \rb \oplus |E^{j-\frac{1}{2}}_{m_2}, \Psi \rb \oplus |E^{j-\frac{1}{2}}_{m_3}, \chi \rb \oplus|E^{j-1}_{m_4}, L \rb\,.
\ee
These states should be thought of as one-sided black hole states of energy $E$ above extremality, angular momentum $j$ indicated by the superscript, and axial angular momentum $m\in \{-j_{\t{max}},\ldots,j_{\t{max}} \}$ where $j_{\t{max}}$ differs between states in the multiplet. In the case of the $\mathbf{\frac{1}{2}}$ multiplet the $|E,L\rb$ state does not exist. We will label the BPS state by $|\t{BPS}\rb$ since in our case it only exists for $E=j=0$.

\paragraph{Thermal partition function.}
We will only need the final result for the thermal partition function, which can be written \cite{Heydeman:2020hhw}
\be
\begin{aligned}
Z(\beta, \alpha) =&  \sum_{\mathbf{j}} \chi_j(\alpha) \rho_{\t{BPS}}(\mathbf{j})+\sum_{\mathbf{j} \geq \frac{1}{2}} \int_{E_0(\mathbf{j})}^\infty d E e^{-\beta E}\left(\chi_j(\alpha)+2 \chi_{j-\frac{1}{2}}(\alpha)+\chi_{j-1}(\alpha)\right) \rho_{\mathbf{j}}(E), \\
E_0(\mathbf{j})=&\frac{{j}^2}{2}\Ebrk
\end{aligned}
\ee
with $\chi_j(\a)=\sum_{m=-j}^j e^{4\pi i \a m}$ the $SU(2)$ character, and $\a$ is a chemical potential for the angular momentum. Here and throughout the rest of the paper, we have used $\mathbf j$ to denote quantities that depend on the supermultiplet that a state belongs to and $j$ to denote the spin of states within a given supermultiplet. In each term in the sum above, we have a sum over spins starting with a maximum spin $\mathbf{j} = j$. We are working in conventions where $\chi_{-\frac{1}{2}}=0$ to automatically include the $\mathbf{j}=\frac{1}{2}$ multiplet in the above. The density of states is 
\begin{align}
\label{eqn:dos}
\rho_{\text {BPS},\mathbf{j}}(E) & =e^{S_0} \delta_{\mathbf{j}, 0} \delta(E)\\
\rho_{\mathbf{j}}(E) & =\Ebrk\frac{e^{S_0} \mathbf{j} }{2 \pi^2  E^2} \sinh \left(2 \pi \sqrt{2 \Ebrk^{-1}\left(E-E_0(\mathbf{j})\right)}\right) \Theta\left(E-E_0(\mathbf{j})\right), \qquad  \mathbf{j} \geq \frac{1}{2}\,.
\end{align}
The multiplet with the highest spin $j$ has a gap from the ground state of $E_0(\mathbf{j})$. That means that states of angular momentum $j$ begin at $E_0(\mathbf{j}) = \frac{\mathbf{j}^2}{2} \Ebrk$. In the $j=0$ sector there are BPS states at $E=0$ followed by a gap in the spectrum until $E=\frac{1}{8}\Ebrk$ where additional $j=0$ states from the $\mathbf{\frac{1}{2}}$ multiplet begin to appear.

\subsection{Symmetries in the near-horizon and the spectrum of matter fields}
\label{subsec:symm-NHR}

We will couple an $\mathcal{N}=2$ hypermultiplet to our supergravity theory. The hypermultiplet consists of four real scalar fields $q^i$ with $i=1,\ldots, 4$, and two four-component Majorana Fermions $\zeta_I$ labelled by $I=1,2$. The action for the multiplet is
\begin{equation}
\begin{aligned} \label{eqn:I_Sugra_hyper}
    S_{\t{hyp.}} = \frac{1}{8 \pi G_N}\int d^4 x \sqrt{-g} \lr{-\frac{1}{2} h_{i j}(q) \nabla_N q^i \nabla^N q^j + i \bar{\zeta}^I \Gamma^M D_M \zeta_I + \frac{\varepsilon_{IJ}}{2} \bar{\zeta}^I \Gamma^{MN}(F_{MN}+i\star F_{MN}\Gamma_5) \zeta^J }\,.
\end{aligned}
\end{equation}
The fields are not charged under the photon in the graviton multiplet, but the fermions couple to the field strength through the last term, and so we do not have a free fermion action when the field strength is turned on.\footnote{The fermions are left-handed if they have an upper index $P_L \zeta^I = \zeta^I$ and  right handed if lower index.} The four scalar fields have a non-trivial metric $h_{i j}(q)$ on scalar moduli space, which perturbatively around the scalar vacuum is flat $h_{i j}(\delta q) = \delta_{i j} + \mathcal{O}(\delta q^2)$. We will only consider perturbations around the vacuum, so we effectively have four free massless fields. The supersymmetry transformations for the hypermultiplet can be found in \cite{Freedman:2012zz}.

We are interested in this theory on the near-horizon AdS$_2 \times S^2$ background of the extremal BH. As explained previously, this background has a $PSU(1,1|2)$ super-isometry group, and so the Hilbert space of the matter fields can be decomposed into unitary irreducible representations of this supergroup. See \cite{Lee:1999yu,Michelson:1999kn,Aharony:1999ti} for more on the representation theory.

The states are labelled by quantum numbers $(\D,j)$  of the bosonic subgroup $SL(2,\mathbb{R})\times SU(2)$, with $\D$ the scaling dimension in AdS$_2$ and $j$ the total angular momentum. The scalar and fermion fields in the hypermultiplet can be expanded in spherical harmonics, and the resulting Kaluza-Klein modes are organized into multiplets. It turns out that the single particle states fall into short multiplets with four states
\begin{gather}
\textbf{k} \equiv (k+1,k-1) \oplus 2 (k+\frac{1}{2},k-\frac{1}{2}) \oplus (k,k)\,, \qquad k \geq 1\,,\\
\mathbf{\frac{1}{2}} \equiv (\frac{1}{2},\frac{1}{2}) \oplus 2 (1,0)\,,
\end{gather}
with the $\mathbf{\frac{1}{2}}$ having one fewer state. The single-particle hypermultiplet Hilbert space decompose into these irreps according to \cite{Lee:1999yu}
\be \label{eqn:hypermultiplet}
\mathcal{H}^{\t{hyp.}}_{\t{single-particle}} = \bigoplus_{k \in \mathbb{Z}+\frac{1}{2}} 2 (\mathbf{k})\,,
\ee
We have multiplicity two for each half-integer irrep.\footnote{The multiplicity comes from four scalars and two fermions.} The most important multiplet for our purposes is $\mathbf{\frac{1}{2}}$. This comes from the s-wave reduction of the fermion giving $\D=\frac{1}{2}$ and $j=\frac{1}{2}$, and the s-wave of the scalar which has $\D=1$ and $j=0$. For completeness, the single-particle graviton multiplet organizes as \cite{Michelson:1999kn}
\be
\mathcal{H}^{\t{grav.}}_{\t{single-particle}} = \bigoplus_{k \in \mathbb{Z}} 2 (\mathbf{k})\,.
\ee

The above discussion is for QFT in a fixed background. When we couple the QFT to gravity through the Schwarzian theory, the spacetime symmetry group changes. It changes since the Schwarzian breaks some of the asymptotic symmetries of AdS$_2\times S^2$, with the resulting symmetry algebra becoming $\mathcal{N}=4$ super-Poincaré as we discussed previously. Thus, the full quantum gravity Hilbert space now falls into irreps of this $\mathcal{N}=4$ algebra.

\paragraph{Semiclassical Hawking radiation.} 
Hawking radiation of the black hole in supergravity does not greatly differ at the semiclassical level from that in pure GR.\footnote{For a recent discussion of Hawking radiation and greybody factors for a variety of BHs see \cite{Arbey:2021jif,Arbey:2021yke}.} 
For example, a minimally coupled massless scalar field in the s-wave sector has energy flux from Hawking radiation 
\be \label{eqn:dedt_semiclassical}
\frac{d E}{d t}= \frac{1}{2\pi} \int_0^\infty d \omega \omega \frac{4 (r_+ \omega)^2}{e^{\beta \omega}-1}\,,
\ee
where $\beta$ is the inverse temperature, and the numerator $4(r_+ \omega)^2$ is the greybody factor for the scalar to tunnel through the BH effective potential. There is also radiation into propagating gravitons, photons, and spin-$\frac{3}{2}$ gravitini, where the gravitini are a new feature of supergravity. All of these modes would have a semiclassical flux similar to \eqref{eqn:dedt_semiclassical} - \eqref{eqn:dedt_fermion_semiclassical}, with different greybody factors. See \cite{Brown:2024ajk} for more discussion. 

For simplicity, we will only study Hawking radiation into fields in the hypermultiplet, which gives the dominant decay channel since the greybody factors for the hypermultiplet are smaller due to angular momentum considerations. The dominant decay channel will thus be through four massless scalars in the s-wave sector \eqref{eqn:dedt_semiclassical}, which do not carry away any angular momentum from the black hole, and two s-wave fermions which carry away spin half. These are precisely the $\mathbf{\frac{1}{2}}$ states in \eqref{eqn:hypermultiplet}, with $\lr{\frac{1}{2},\frac{1}{2}}$ the s-wave fermion and $(1,0)$ the s-wave scalar. The semiclassical flux for a single s-wave fermion mode $\lr{\frac{1}{2},\frac{1}{2}}$ is
\be \label{eqn:dedt_fermion_semiclassical}
\frac{d E}{d t}= \frac{1}{2\pi} \int_0^\infty d \omega \omega \frac{4 (r_+ \omega)^2}{e^{\beta \omega}+1}\,.
\ee
To get the full flux from fermions and scalars we should multiply the rates by the degeneracy of modes. Note that the greybody factor in the numerator $4(r_+ \omega)^2$ is identical to the case of the scalar. The fermion we are considering is not free due to extra terms \eqref{eqn:I_Sugra}, and we calculate the greybody factor in appendix \ref{sec:greybody}.

\section{Correlators in the $\mathcal{N}=4$ super-Schwarzian}
\label{eq:correlators-in-super-Schw}

Euclidean two-point functions of matter fields in the near-horizon throat of the black hole are equivalent to bi-local correlators evaluated in the super-Schwarzian theory. These correlators are difficult to evaluate directly. However, there is a complicated field redefinition of the Schwarzian theory that turns it into a super-Liouville Quantum mechanics (LQM) that makes the evaluation of these correlators manageable. This field redefinition is not explicitly known except in the simplest bosonic case \cite{Mertens:2017mtv,Lin:2022zxd}, but the resulting LQM can be guessed based on symmetry principles as was done for the cases $\mathcal{N}=1,2$ in \cite{Lin:2022zxd,Lin:2022rzw}, also see \cite{Fan:2021wsb,Belaey:2024dde}. 
The super-Schwarzian theory has $\mathcal{N}=4$ supersymmetry, so it is natural to guess that the associated LQM also has $\mathcal{N}=4$ supersymmetry.
The calculations in such LQM quickly become unmanageable, and we include a \texttt{Mathematica} notebook in the source file that automates the calculations.

\subsection{The $\mathcal N=4$ super-Liouville quantum mechanics}
 We now summarize the details of the theory. We have four bosonic fields $\ell, g$ with the matrix valued field $g \in SU(2)$ and eight complex fermionic fields given by left and right fermions $\psi_{l}^p, \psi_{r}^p$ and their barred counterparts $\bar \psi_{l,p}, \bar \psi_{r,p}$ with $p=1,2$ a fermion flavor index. The Lagrangian is given by
\be
I=\int \mathrm{d} u\lr{\frac{1}{4} \dot{\ell}^2 - e^{-\ell} - \frac{1}{4} - \frac{1}{2}\Tr \lr{(g^{-1} \dot g)^2} +i \bar{\psi}_{r,p} \dot{\psi}_r^{~p}+i \bar{\psi}_{l,p} \dot{\psi}_l^{~p} 
 - i e^{-\frac{\ell}{2}} \bar{\psi}_{r,q} (g^{-1})^q_{~p} \psi_l^p 
+ i e^{-\frac{\ell}{2}} \bar{\psi}_{l,q} (g)^q_{~p} \psi_r^p }\,,
\ee
where the dot denotes a derivative with respect to $u$. The kinetic term for $g$ is the $SU(2)$ Hamiltonian describing the motion of a particle on the $SU(2)$ group manifold. The particle is additionally coupled to the length $\ell$ and to fermions to make the theory supersymmetric. Canonically quantizing the above theory, the Hamiltonian and fermion commutation relations become
\begin{equation}\label{eq:Hn=4}
    H = -\partial_{\ell}^2  + \frac{1}{4} + e^{-\ell} +  H^{\rm SU(2)} 
 + i e^{-\frac{\ell}{2}} \bar{\psi}_{r,q} (g^{-1})^q_{~p} \psi_l^p 
- i e^{-\frac{\ell}{2}} \bar{\psi}_{l,q} (g)^q_{~p} \psi_r^p\,,
\end{equation}
\be
\{\psi_{l}^p, \ol \psi_{l,q} \} = \delta^p_{q}, \qquad \{\psi_{l}^p, \psi_{l}^q \} = 0=\{\ol \psi_{l,p}, \ol \psi_{l,q} \}\,, \qquad (\psi_l^p)^\dag \equiv \ol \psi_{l,p}\,,
\ee
These relations also apply for right fermions, and all left and right fermions mutually anti-commute in the obvious way $\{\psi_l, \psi_r \} = 0 =\{\psi_l, \ol \psi_r \}$. The $SU(2)$ Hamiltonian has standard form $H^{\rm SU(2)} = \sum_{i=1}^3 \mathcal{J}_{l,i}^2 = \sum_{i=1}^3 \mathcal{J}_{r,i}^2$ where the left/right derivatives have action $\mathcal{J}_{l,i} g = -\frac{1}{2} \sigma_i \cdot g$ and $\mathcal{J}_{r,i} g = -\frac{1}{2} g \cdot \sigma_i $ with $\sigma_i$ the Pauli matrices.\footnote{The conventions we follow in this paper is that the Pauli matrices satisfy $[\sigma_i, \sigma_j] = i \epsilon_{i j k} \sigma_k$, the Wigner D-matrices are defined as in Mathematica 14, and the matrix element is given by $g=e^{-\frac{i}{2} \alpha \sigma_3 }e^{-\frac{i}{2} \beta \sigma_2 } e^{-\frac{i}{2} \gamma \sigma_3 }$ where $\alpha, \beta, \gamma$ are the Euler angles. \label{footnote:paulimatrices}} The derivatives satisfy $ [\mathcal{J}_{l,i},\mathcal{J}_{l,i}]=i \epsilon_{ijk} \mathcal{J}_{l,k},\quad  [\mathcal{J}_{r,i},\mathcal{J}_{r,i}]=-i \epsilon_{ijk} \mathcal{J}_{r,k}\,$. The eigenfunctions are given by the Wigner-D matrices $H^{\rm SU(2)} D^j_{m n}(g) = j (j+1) D^j_{m n}(g)$ \cite{Chu:1994hm}. 

This system has $\mathcal{N}=4$ supersymmetry. There are eight total supercharges with four left supercharges $\ol Q_{l,i} , Q_{l}^i$ and four right charges $\ol Q_{r,i} , Q_{r}^i$ with $i=1,2$. We write out all supercharges in appendix \ref{app:supercharges}, as an example, we have
\begin{gather}
Q_{l}^1 = i \psla ( \partial_\ell - 
  \mathcal{J}_{l,3}) - i\pslb \mathcal{J}_{l,-} + e^{-\frac{\ell}{2}} g^1_{~q} \psi^q_r  -  \frac{i}{2} \psla [\pslb, \bar{\psi}_{l,2}]\,,\\ 
\ol{Q}_{l,1} = i \bar{\psi}_{l,1} (\partial_\ell + \mathcal{J}_{l,3}) +  i\bar{\psi}_{l,2} \mathcal{J}_{l,+} + e^{-\frac{\ell}{2}} \bps_{r,q} (g^{-1})^{q}_{~1} -   \frac{i}{2} \bar{\psi}_{l,1} [\bar{\psi}_{l,2} ,{\psi}_{l}^2]\,, \nn
\end{gather}
where we defined raising and lowering operators $\mathcal{J}_{\pm} = \mathcal{J}_{1} \pm i \mathcal{J}_{2}$ with action $\mathcal{J}_{l,\pm } g = -\frac{1}{2} (\sigma_1 \pm \sigma_2) \cdot g$ and $\mathcal{J}_{r,\pm } g = -\frac{1}{2} g \cdot (\sigma_1 \pm \sigma_2)$ in the derivative representation. The Hamiltonian is given by
\be
\{Q_{l}^i,\ol{Q}_{l,j}\}=\delta^i_j H = \{Q_{r}^i,\ol{Q}_{r,j}\}\,.
\ee
Left and right supercharges mutually commute $\{Q_l, Q_r \}=\{Q_l, \ol Q_r \}=\{Q_r, \ol Q_l \}=0$.

\paragraph{$\mathbf{SU(2)}$ Symmetry.}
The theory has an $SU(2)_r \times SU(2)_l$ R-symmetry
\begin{gather}
SU(2)_r: \qquad \psi_r^p \to h^p_{~q} \psi_r^q, \quad  \bar{\psi}_{r,p} \to \bar{\psi}_{r,q} (h^{-1})^q_{~p}, \quad g \to g \cdot h^{-1}\,, \\
SU(2)_l: \qquad \psi_l^p \to h^p_{~q} \psi_l^q, \quad \bar{\psi}_{l,p} \to  \bar{\psi}_{l,q} (h^{-1})^q_{~p}, \quad g \to h \cdot g\,,
\end{gather}
with $h \in SU(2)$. This symmetry leaves $\ol \psi_r g^{-1} \psi_l$ and $\ol \psi_l g \psi_r$ invariant. The supercharges transform as $Q_{l} \to h \cdot Q_{l}$ and $Q_r \to h \cdot Q_r$ under the left/right transformations and leave the superalgebra invariant. The conserved charges are
\begin{gather}
        J_{l,i}=  \mathcal{J}_{l,i} +  \frac{1}{2}\bar{\psi}_{l,q} (\sigma_i)_{~p}^{q} \psi_l^p\,, \quad 
        J_{r,i} = -  \mathcal{J}_{r,i} + \frac{1}{2} \bar{\psi}_{r,q} (\sigma_i)_{~p}^{q} \psi_r^p\\
        [J_i, J_j] = i \epsilon_{i j k} J_k\,. 
\end{gather} 
There are both left and right Casimirs $J^2_{l}, J^2_{r}$ for total angular momentum, along with an additional element of the algebra which commutes with the Casimir, which we denote by $J_{z}\equiv J_{3} $ 
\be
J_{l}^2 \equiv \sum_{i=1}^3 J_{l,i}^2\,, \qquad [J_{l}^2,J_{l,z}]=0\,,
\ee
with identical expressions for $J_r$. In the two-sided black hole context, left/right Casimirs label the angular momentum of the left/right entangled black hole states. We also have raising/lowering operators given by $J_{\pm}^{l} = J_{l,1} \pm i J_{l,2}$. These operators raise/lower the left $J_{z}^l$ eigenvalue by one unit, with the same operator with $l\to r$ raising/lowering $J_z^r$. We have the commutators with the supercharge
\begin{gather}
[Q_l^p,J_{l,i}]=\frac{1}{2}(\sigma_i)^{p}_{~q} Q_l^{q}\,, \qquad [\ol Q^l_p,J_{l,i}]=-\frac{1}{2}\ol Q^l_{q}(\sigma_i)^{q}_{~p}\,, \\
[Q_r^p,J_{r,i}]=\frac{1}{2}(\sigma_i)^{p}_{~q} Q_r^{q}\,, \qquad [\ol Q^r_p,J_{r,i}]=-\frac{1}{2}\ol Q^r_{q}(\sigma_i)^{q}_{~p}\,,
\end{gather}
Left and right charges and angular momenta operators mutually commute $[Q_l, J_{r}]=0$ since $Q_l$ only acts on the left and vice-versa.

\subsection{Two-sided black hole states}\label{sec:states}
We now examine the states of the theory. A maximally commuting set of operators are given by the Hamiltonian $H$, the left/right $J^2_{l,r}$, and left/right $J_z^{l,r}$. Since we are interested in studying two sided TFD black hole states, we will restrict to states with equal left/right total angular momentum, and opposite axial angular momenta
\be
|\Psi^{E,j}_{m,m}\rb , \qquad J_z^l = m = -J_z^r\,, \qquad J^2_l = J^2_r=j(j+1)\,,
\ee
where the upper index indicates total angular momentum. To construct supermultiplets, we must choose an initial state annihilated by half of the supercharges.\footnote{Choosing an initial state not annihilated by any supercharges instead generates a reducible multiplet.} Suppose our initial state $|\Psi\rb$ is annihilated by all $Q$'s, we can build the full multiplet by acting with all left/right charges
\be
|\Psi\rb, \ol Q_{r,a} |\Psi\rb, \ol Q_{l,a} |\Psi\rb, \ol Q_{l,a} \ol Q_{r,b} |\Psi\rb , \ldots, \ol Q_{r,a} \ol Q_{r,b} \ol Q_{l,c} \ol Q_{l,d} |\Psi\rb\,.
\ee
However, only a subset of these states describe two-sided TFD black holes, since $\ol Q_{l,r}$  raises/lowers the total angular momentum on the left/right. We have to act with both a left and right supercharge to satisfy the TFD condition, which picks out four states, with total spin $\mathbf{j} = j \oplus 2 (j-\frac{1}{2}) \oplus j-1$. We will label these states in analogy to the Schwarzian theory \eqref{eqn:schw_multiplet}
\be
\mathbf{j} \t{ Multiplet}: \qquad |H^{E,j}\rb \oplus |\Psi^{E,j-\frac{1}{2}}\rb \oplus |\chi^{E,j-\frac{1}{2}}\rb \oplus |L^{E,j-1}\rb\,,
\ee
where each state labels the family of states with all possible values of axial angular momenta satisfying $J_{z}^l = - J_{z}^r$ and total angular momentum indicated. These states should be thought of as two-sided entangled states of the Schwarzian theory \eqref{eqn:schw_multiplet}. It is sometimes convenient to shift the spin of the $|\Psi\rb$ state by half, as should be clear from the notation. 

We will now construct the multiplet.
We choose a fermionic ground state
\be
\psi_{l,i} |\Omega\rb = \ol \psi_{r,i} |\Omega\rb = 0\,. 
\ee

\begin{figure}[t!]
    \centering

    \begin{subfigure}[b]{0.48\textwidth}
        \centering
        \includegraphics[width=1\linewidth]{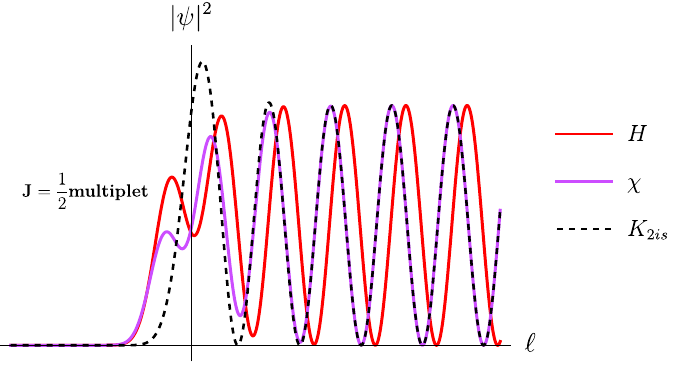}
    \end{subfigure}
    \hfill
    \begin{subfigure}[b]{0.48\textwidth}
        \centering
\includegraphics[width=1\linewidth]{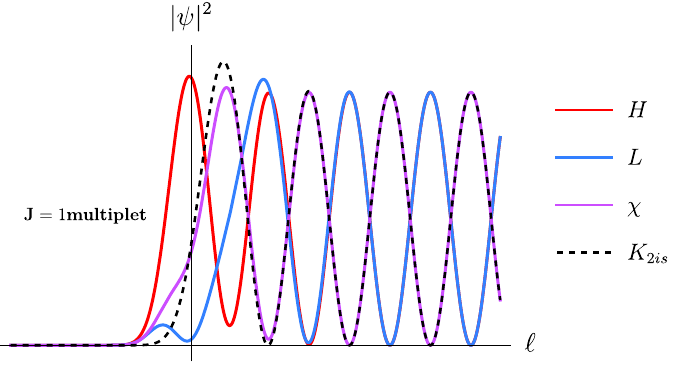}
    \end{subfigure}
    \vspace{0.5cm }
    \begin{subfigure}[b]{0.48\textwidth}
        \centering
       \includegraphics[width=1\linewidth]{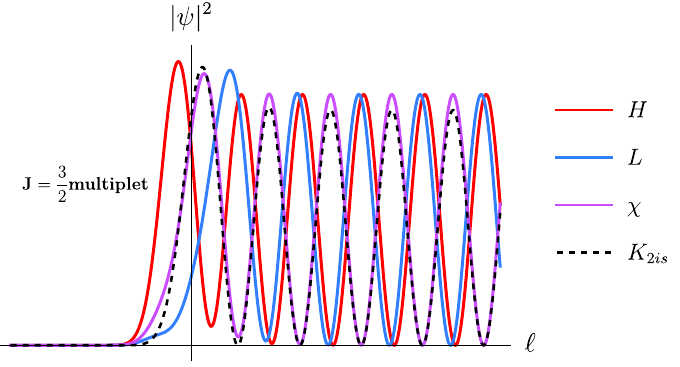}
    \end{subfigure}
    \hfill
    \begin{subfigure}[b]{0.48\textwidth}
        \centering
\includegraphics[width=0.8\linewidth]{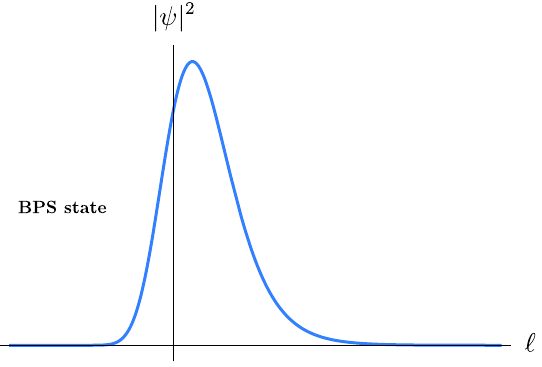}
    \end{subfigure}
    \caption{Plot of the energy eigenfunctions in the $\ell$-basis for the different states ($\ket{H}$, $\ket{L}$, $\ket{\chi}$, $\ket{\psi}, \ket{\t{BPS}}$) that can exist in the supermultiplets in $\mathcal N=2$ supergravity. The wavefunctions are shown when their energy is $E=\Ebrk$ (except the BPS state with $E=0$), a regime in which we expect quantum corrections to be large. The result can be compared to the energy eigenfunction of JT gravity in non-supersymmetric theories, shown by the dotted black curve. Compared to this eigenfunction, some of the wavefunctions in the supersymmetric have greater support for lower values of $\ell$, which implies a growth in the two-point function and explains why the quantum-corrected flux shown in Figure \ref{fig:dedt_micro} and \ref{fig:dedt_psi} is greater than the semiclassical estimate in the appropriate frequency range. }
    \label{fig:wavefunc}
\end{figure}
This choice is slightly unnatural but will simplify the eigenstates we are interested in.\footnote{This specification of the fermionic vacuum is identical to taking the more standard definition of the vacuum $|\tilde \Omega \rb$ which satisfies $\psi_{l,i} |\tilde \Omega\rb =0= \psi_{r,i} |\tilde \Omega\rb$, and taking the fermionic state $|\Omega \rb \equiv \ol \psi_r^2 \ol \psi_r^1 |\tilde \Omega \rb$ to be our new vacuum. The spectrum of the Hamiltonian doesn't change under different choices of fermionic vacuum, but eigenstates will have slightly different expressions.} It is annihilated by ${Q}_{l,i} | \Omega \rangle = \ol{Q}_{r,j} |\Omega\rangle$. 
The simplest energy eigenstates are found by making the ansatz $h(\ell)D^j_{m m}(g) \psi_{r}^1 \psi_{r}^2 |\Omega\rangle$ where $D^j$ is a Wigner D-matrix and solving for the function $h(\ell)$. However, it turns out that these states do not give the TFD state for a two-sided black hole, see the discussion around \eqref{eqn:TFD_condition}. Imposing this additional condition out-of-time-order, we will guess the correct ansatz:
\begin{equation}
    |\Psi \rangle = h_a(\ell) D(g) + h_b^{(ij)}(\ell) D^{ij}(g) \bar{\psi}_{l,i} \psi_{r,j} + \bar{\psi}_{l,1}\bar{\psi}_{l,2}\psi_{r,1}\psi_{r,2}  h_c(\ell)\,,
\end{equation}
In the above $h_i(\ell)$ are general functions of the geodesic distance and $D, D^{i j}$ are general functions of the $SU(2)$ field. To solve for the eigenstate, we will consider a highest weight state with angular momentum $j$ and equal but opposite axial angular momenta
\begin{gather}
H |\Psi^{E,j}_{j,j} \rb = E |\Psi^{E,j}_{j,j} \rb, \qquad \mathcal{J}^2 | \Psi^{E,j}_{j,j} \rb = j^2 | \Psi^{E,j}_{j,j} \rb\,, \nn \\ J_z^l | \Psi^{E,j}_{j,j} \rb = j | \Psi^{E,j}_{j,j} \rb, \qquad J_z^r | \Psi^{E,j}_{j,j} \rb = -j | \Psi^{E,j}_{j,j} \rb ,  \qquad J_+^l | \Psi^{E,j}_{j,j} \rb = 0, \qquad J_-^r | \Psi^{E,j}_{j,j} \rb = 0\,.
\end{gather}

\paragraph{Solve for the  $|\Psi\rb$ State.}
The equations are complicated to solve directly, and the solution is not unique. 
This is natural to expect because in supersymmetric theory, one often encounters different states with the same bosonic charges. To uniquely specify at least one state, one needs to exploit the supercharges. 
As a starting point, we seek the ``analog" of the fermionic ground state and require that 
\begin{equation}
    Q_{l,i}|\Psi^{E,j}_{j,j}\rangle = \bar{Q}_{r,i}|\Psi^{E,j}_{j,j}\rangle = 0.
\end{equation}
Then one gets a unique solution (up to an overall factor) of the $|\Psi\rangle$ state 
\be \label{eqn:Psi_state}
\begin{aligned}
 |\Psi^{E,j}_{j,j} \rangle =& D^{j}_{j j}(g) \big(h_3(\ell) + h_1(\ell) \bar{\psi}_{l,1}\bar{\psi}_{l,2}\psi_{r,1}\psi_{r,2}   \big) + \Theta \lr{j \geq \frac{1}{2}} h_4(\ell)  D^{j-\frac{1}{2}}_{j-\frac{1}{2},j-\frac{1}{2}}(g) \ol \psi_l^1 \psi_r^1 \\
 &+ h_2(\ell)~ \ol \psi_l^T \cdot \begin{pmatrix}
    \frac{1}{2j+1} D^{j+\frac{1}{2}}_{j-\frac{1}{2},j-\frac{1}{2}} (g) & \frac{1}{\sqrt{2j+1}}D^{j+\frac{1}{2}}_{j-\frac{1}{2},j+\frac{1}{2}} (g)\\
     \frac{1}{\sqrt{2j+1}}D^{j+\frac{1}{2}}_{j+\frac{1}{2},j-\frac{1}{2}} (g) & D^{j+\frac{1}{2}}_{j+\frac{1}{2},j+\frac{1}{2}} (g)
 \end{pmatrix} \cdot {\psi}_r 
\end{aligned}
\ee
where $D^j_{m n}(g)$ are the Wigner D-matrices and
\begin{equation}
\begin{aligned}
    & h_1(\ell)= e^{-\ell} h(\ell) \equiv 2 e^{-\ell} K_{2 i s}(2e^{-\frac{\ell}{2}}) , \quad  h_2(\ell)= -\frac{i}{2} e^{-\frac{\ell}{2}} (2 h^{\prime}(\ell) - (2j+1) h(\ell)) , \\
    & h_4(\ell)= -i \frac{2j}{2j+1} e^{-\frac{\ell}{2}} (2 h^{\prime}(\ell) + (2j+1) h(\ell)), \quad h_3(\ell) = h^{\prime\prime}(\ell) -\frac{(2j+1)^2}{4}h(\ell)
\end{aligned}
\end{equation}
The state exists for all non-negative half integer values of angular momenta $j=0,\frac{1}{2},1,\ldots$ with energies $E=s^2+(j+\frac{1}{2})^2 \geq (j+\frac{1}{2})^2$ with $s$ a positive parameter labelling the energy. 

\paragraph{Inner product and normalization.}
We specify the normalization of $|\Psi\rangle$, and other states of course, as follows. 
The inner product for the length and $SU(2)$ part of the wavefunction will be given by integrating over all lengths and over the $SU(2)$ group with the normalized Haar measure $\lb \psi | \chi \rb = \int d g \int_{\infty}^\infty d \ell \psi(\ell,g)^* \chi(\ell,g)$. An important integral is
\be
4\int_{-\infty}^\infty d \ell K^*_{2i s'}(\ell) K_{2is}(\ell) = \delta(s-s') \frac{ 2\pi^2 E^2}{ s \sinh (2\pi s)}\,, \
\ee
where $E=s^2$. The normalized Haar measure satisfies $\int d g = 1$, and an important identity is
\be
\int d g (D^{ j'}_{m',n'}(g))^* D^{ j}_{m,n}(g)\equiv \frac{1}{8\pi^2} \int d \alpha d \beta \sin \beta d\gamma (D^{ j'}_{m',n'}(g))^* D^{ j}_{m,n}(g) = \frac{\delta_{j j'} \delta_{m m'} \delta_{n n'}}{2 j+1}\,.
\ee
where in the second equality the angles are Euler angles and our conventions for the matrix $g$ are in footnote \ref{footnote:paulimatrices}. The state is delta function normalized with the appropriate density of states
\be \label{eqn:inner_prod_answer}
\lb \Psi_{m',n'}^{E',j'} | \Psi_{m,n}^{E,j} \rb = \delta_{j j'} \delta_{m' m} \delta_{n' n} \delta(s-s') \frac{\pi^2 E^2 }{ (2j+1) 2 s \sinh (2\pi s)}\,,
\ee
where again $E=s^2 + (j+\frac{1}{2})^2$.

\subsubsection*{Building the supermultiplet.} We get the other TFD states in the supermultiplet by consecutively acting with one left and one right supercharge supercharge on the initial state $|\Psi^{E,j-\frac{1}{2}}\rb$\footnote{In our earlier discussion we used conventions where the multiplet was built using an initial state that was annihilated by all $Q$'s, but we have chosen a slightly different ansatz so now it is annihilated by half $Q$'s and half $\ol Q$'s.}
\begin{gather}
      |H^{E,j}_{j,j} \rangle = \frac{i}{E} \ol Q_{l,1} Q_{r,1} |\Psi^{E,j-\frac{1}{2}}_{j-\frac{1}{2},j-\frac{1}{2}} \rb, \nn \\
      |\chi^{E,j-\frac{1}{2}}_{j-\frac{1}{2},j-\frac{1}{2}} \rangle = -\frac{1}{E^2}  \ol Q_{l,1} \ol Q_{l,2} Q_{r,1} Q_{r,2} |\Psi^{E,j-\frac{1}{2}}_{j-\frac{1}{2},j-\frac{1}{2}} \rb, \\
      |L^{E,j-1}_{j-1,j-1} \rangle = \frac{i}{E} \Big(\ol Q_{l,2} Q_{r,2} -\frac{\mathcal{J}_{l,-}}{2j+1} \ol Q_{l,1}Q_{r,2} + \frac{\mathcal{J}_{r,+}}{2j+1} \ol Q_{l,2}Q_{r,1}  - \frac{\mathcal{J}_{l,-}\mathcal{J}_{r,+}}{(2j+1)^2} Q_{l,1}Q_{r,1} \Big)|\Psi^{E,j-\frac{1}{2}}_{j-\frac{1}{2},j-\frac{1}{2}} \rb.  \nn   
\end{gather}
The labels on the state clearly indicate the quantum numbers, and we have introduced $E^{-1}$ to keep the normalization of all states the same as \eqref{eqn:inner_prod_answer} with $j$ shifted down
\begin{align}
\lb \Psi^{E,j-\frac{1}{2}}_{m,m} | \Psi^{E,j-\frac{1}{2}}_{m,m} \rb &= \lb H^{E,j}_{m,m} | H^{E,j}_{m,m} \rb = \lb \chi^{E,j-\frac{1}{2}}_{m,m} | \chi^{E,j-\frac{1}{2}}_{m,m} \rb = \lb L^{E,j-1}_{m,m} | L^{E,j-1}_{m,m} \rb\,, \nn\\
&= \delta(s-s') \frac{\pi^2 E^2 }{ 4 j s \sinh (2\pi s)}\,, \qquad E=s^2+j^2\,,
\end{align}
with axial momenta equal between bra and ket, but of arbitrary value. The $L$ state is slightly more complicated than the others and we do not write out it's full expression.\footnote{
Let us explain the $|L\rb$ state since it is more complicated. Acting with $\overline{Q}_{l,2}\overline{Q}_{r,1} |\Psi\rb$ we get a state with $J_{z}^l=j-1 = -J_{z}^r$. However, this is a linear combination of states with different total angular momentum but identical axial angular momentum $\overline{Q}_{l,2}\overline{Q}_{r,1} |\Psi\rb \propto a |j-1,j-1\rb+ b|j,j-1\rb$ with $a,b$ some coefficients. To isolate the portion of the state with angular momentum $j-1$ we must subtract off the $|j,j-1\rb$ component of the state. This can be identified by acting with lowering operators $J^l_{-} J^r_{+} |H^j_{j,j}\rb$ and taking the inner product to subtract the overlap with $|H^j_{j-1,j-1}\rb$. We have given the final answer for this subtraction. The full expression for the $L$ state can be found in the Mathematica notebook.}

\paragraph{Explicit $|\chi\rb$ State.} 
For $|\chi\rangle$ state we have
\be
\begin{aligned}
 |\chi^{E,j}_{j,j} \rangle =& D^{j}_{j j}(g) \big(h_3(\ell) + h_1(\ell) \bar{\psi}_{l,1}\bar{\psi}_{l,2}\psi_{r,1}\psi_{r,2}   \big) + \Theta \lr{j \geq \frac{1}{2}} h_4(\ell)  D^{j-\frac{1}{2}}_{j-\frac{1}{2},j-\frac{1}{2}} \ol \psi_l^1 \psi_r^1 \\
 &+ h_2(\ell) \ol \psi_l^T \cdot \begin{pmatrix}
    \frac{1}{2j+1} D^{j+\frac{1}{2}}_{j-\frac{1}{2},j-\frac{1}{2}} & \frac{1}{\sqrt{2j+1}}D^{j+\frac{1}{2}}_{j-\frac{1}{2},j+\frac{1}{2}}\\
     \frac{1}{\sqrt{2j+1}}D^{j+\frac{1}{2}}_{j+\frac{1}{2},j-\frac{1}{2}} & D^{j+\frac{1}{2}}_{j+\frac{1}{2},j+\frac{1}{2}}
 \end{pmatrix} \cdot {\psi}_r 
\end{aligned}
\ee
where 
\begin{equation}
\begin{aligned}
    & h_3(\ell)= e^{-\ell} h_1(\ell) \equiv - 2 e^{-\ell} K_{2 i s}(2e^{-\frac{\ell}{2}}) , \quad  h_2(\ell)= \frac{i}{2} e^{-\frac{\ell}{2}} (2 h^{\prime}(\ell) - (2j+1) h(\ell)) , \\
    & h_4(\ell)= i \frac{2j}{2j+1} e^{-\frac{\ell}{2}} (2 h^{\prime}(\ell) + (2j+1) h(\ell)), \quad h_1(\ell) = h^{\prime\prime}(\ell) -\frac{(2j+1)^2}{4}h(\ell)
\end{aligned}
\end{equation}
As defined the energy of this state is $E=s^2 + (j+\frac{1}{2})^2 \geq (j+\frac{1}{2})^2$.

\paragraph{Explicit $|H\rb$ State.}
This state only exists for half integers $j \geq \frac{1}{2}$
\be
\begin{aligned}
 |H^{E,j}_{j,j} \rangle =& D^{j}_{j j}(g) \big(h_3(\ell) + h_1(\ell) \bar{\psi}_{l,1}\bar{\psi}_{l,2}\psi_{r,1}\psi_{r,2}   \big) + h_4(\ell)  D^{j-\frac{1}{2}}_{j-\frac{1}{2},j-\frac{1}{2}} \ol \psi_l^1 \psi_r^1 \\
 &+ h_2(\ell) \ol \psi_l^T \cdot \begin{pmatrix}
    \frac{1}{2j+1} D^{j+\frac{1}{2}}_{j-\frac{1}{2},j-\frac{1}{2}} & \frac{1}{\sqrt{2j+1}}D^{j+\frac{1}{2}}_{j-\frac{1}{2},j+\frac{1}{2}}\\
     \frac{1}{\sqrt{2j+1}}D^{j+\frac{1}{2}}_{j+\frac{1}{2},j-\frac{1}{2}} & D^{j+\frac{1}{2}}_{j+\frac{1}{2},j+\frac{1}{2}}
 \end{pmatrix} \cdot {\psi}_r 
\end{aligned}
\ee
where 
\begin{equation}
\begin{aligned}
    & h_2(\ell)=  e^{-\ell} h(\ell) \equiv 2 i e^{-\ell} K_{2 i s}(2e^{-\frac{\ell}{2}}) , \quad   h_1(\ell) = - h_3(\ell) = e^{-\frac{\ell}{2}} ( h^{\prime}(\ell) + j h(\ell)), \\
     & h_4(\ell) = h^{\prime\prime}(\ell) +2 j h^{\prime}(\ell)+j^2 h(\ell) - \frac{1}{2j+1} e^{-\ell}h(\ell)
\end{aligned}
\end{equation}
The energy is $E=s^2 + j^2 \geq j^2$.

\paragraph{Changing axial angular momentum.}
The above states have the largest allowed value for axial angular momenta for each spin $j$ with $J^l_{z} = j = - J^r_{z}$. There are $2j+1$ other states with the other possible values of axial momenta. To get these we act with left/right lowering/raising operators on the highest weight state
\be
|\Phi^{E,j}_{j-n,j-n}\rb \equiv c_{j-\frac{1}{2},n} (J_-^l J_+^r)^n |\Phi^{s,j}_{j_z,j_z}\rb\, \qquad \t{for} \quad \Phi = \{ \Psi^{j-\frac{1}{2}}, H^j, L^{j-1}, \chi^{j-\frac{1}{2}} \}\,,
\ee
where it's understood that the value $j,j_z$ should be set to the highest allowed angular momentum value for each state. The operator $J_-^l$ lowers the left axial angular momentum by one unit while $J_+^r$ raises the right momenta by one unit, and $c_{j,n} = \lr{n! (2j)!/(2j-n)!}^{-1}$ fixes the normalization after acting with lowering operators such that the norm of the state is unchanged.

\paragraph{$\mathbf{j}=\frac{1}{2}$ multiplet.} This is a special case where the multiplet is short. It is given by
\be
|\Psi^{E,j=0}_{0,0}\rb\,, \qquad |H^{E,j=\frac{1}{2}}_{\frac{1}{2},\frac{1}{2}}\rb\,, \qquad |\chi^{E,j=0}_{0,0}\rb\,. 
\ee
The middle $j=1/2$ state has $J_{z}^l=1/2=-J^r_{z}$, with the other value obtained by acting with $J_{l,-} J_{r,+}$. These states have a minimum energy $E=s^2 + (\frac{1}{2})^2 \geq \frac{1}{4}$.

\paragraph{$\mathbf{j}=0$ multiplet.} There is a special BPS multiplet with only a single state. It must satisfy the condition $Q|\text{BPS} \rangle=0=\ol{Q}|\text{BPS} \rangle$ for all $Q$ and can be explicitly solved for. However, it turns out that it is also an analytic continuation of the $j=0$ state $|L^{E,0}_{0,0}\rangle $ in the $\J=1$ multiplet if we set $E=0$ by taking $s\to i$
\be
|\text{BPS} \rangle \equiv \lim_{s\to i}  |L^{s,0}_{0,0}\rb\,.
\ee
Following either approach, the wavefunction can be worked out 
\begin{equation}
 |\t{BPS}\rangle =  2 e^{-\ell} K_1(2e^{-\ell/2}) (\bar\psi_{l,1}\bar\psi_{l,2} \psi_{r,1}\psi_{r,2}-1 ) + 2 i  e^{-\ell} K_0 (2e^{-\ell/2}) \bar{\psi}_l\cdot g \cdot \psi_r \,.
\end{equation} 
It has been normalized $\lb \text{BPS} | \text{BPS} \rangle = 1$.

\subsection*{Thermo-field Double state.} 
We are now ready to use the CRT invariant states we constructed to build the TFD. There are infinitely many maximally entangled states that can serve as the TFD \cite{Cottrell:2018ash}. This ambiguity is fixed by making a choice of anti-unitary CRT operator $\Theta$ which defines an associated maximally entangled state $|\t{TFD}\rb = \sum_n e^{-\beta E_n/2} |n\rb_L \otimes \Theta | n \rb_R$. This state satisfies the condition $(\mathcal{O}_L - \Theta \mathcal{O}_R^\dag \Theta^\dag) | \t{TFD} \rb_{\beta \to 0} = 0$ where $\mathcal{O}_L$ is an operator that acts on the left subsystem and $\mathcal{O}_R$ is the corresponding operator acting on the right. Instead of making an explicit choice of $\Theta$ we will define the TFD by demanding\footnote{In higher dimensional QFT the CRT condition imposes that $\Th \psi(t,x_1,x_2,\ldots) \Th^\dag = (\Gamma^0 \Gamma^1)^* \psi^*(-t,-x_1,x_2,\ldots)$ where $\psi^*$ on the operator is the Hilbert space adjoint and spinor indices are not acted on, and the gamma matrices satisfy $\{\Gamma^i, \Gamma^j \}=2\eta^{i j}$. In $0+1 d$ QM with single component spinors we have $\G^0=\pm i$ where the sign is up to conventions. We choose conventions such that $\th \psi^\dag \th^\dag = - i \psi$, which gives us the extra factor of $i$ in the TFD condition. See \cite{Harlow:2025notes,Witten:2025ayw} for recent discussion on CRT action on spinors.}
\be \label{eqn:TFD_condition}
(\psi_l^a + i \psi_r^a) |\t{TFD}\rb_{\beta \to \infty} =0=(\ol \psi_l^a + i \ol \psi_r^a) |\t{TFD}\rb_{\beta \to \infty}\,,
\ee
which is the natural boundary condition that fermions satisfy on the thermal circle as it shrinks to zero size. We have made various implicit choices in the preceding construction so that this condition is satisfied by the ansatz
\begin{align} \label{eqn:TFD}
| \t{TFD}\rb = \mathcal{A}_0 |\text{BPS} \rangle +\sum_{\mathbf{j} =\frac{1}{2},1,\frac{3}{2},\ldots}^\infty \sum_{m=-j_{\t{max}}}^{j_{\t{max}}}\int_0^\infty d s e^{-\frac{\beta}{2} E_{s,j}} \mathcal{A}_{s,j} \lr{ | \Psi^{E,j-\frac{1}{2}}_{m m} \rb + |H^{E,j}_{m m}\rb + |L^{E,j-1}_{m m}\rb + |\chi^{E,j-\frac{1}{2}}_{m m}\rb }\,,
\end{align}
where $E_{s,j}=s^2+j^2$, $j_{\t{max}}$ is the maximal value of axial angular momentum and differs between the states $H,\Psi,\chi,L$ in the same multiplet. We are also using the definition that $|L^{j=-\frac{1}{2}} \rb \equiv 0$ so that we automatically include the short multiplet.

We have to determine the coefficients $\mathcal{A}_{s,j}$ to match the Schwarzian theory, by normalizing the TFD state as 
\be
\lb \t{TFD} | \t{TFD} \rb = Z^{\mathcal{N}=4}_{\t{Schw.}}(\beta)
\ee
Using the inner product \eqref{eqn:inner_prod_answer}, we can immediately see that the correct choices of coefficients are
\be
\mathcal{A}_0 = e^{S_0/2}, \qquad  \mathcal{A}_{s,j} = \frac{2 s j e^{S_0/2}}{\pi^2E^2} \sinh(2\pi s)\,,
\ee
where the second piece is the density of states.

\subsection{The two-point function } \label{sec:3.4_2pt}

We now consider two-sided correlators with the operator $e^{-\Delta \ell}$. Since $[J^2, e^{-\Delta \ell}]=0=[J_z, e^{-\Delta \ell}]$ the correlator vanishes between states of different $j,j_z$. The two-point function is defined between two general states by 
\be
\lb \Psi_{m',m'}^{E', j'} | e^{-\D \ell}|\chi_{m,m}^{E, j}\rb \equiv \int d g \int_{-\infty}^\infty d \ell e^{-\D \ell} \lb \Psi^{E',j'}_{m',m'} | \ell, g\rb \lb \ell, g | \chi^{E,j}_{m,m} \rb
\ee
where $\lb \ell, g | \Psi \rb$ is the more formal version of \eqref{eqn:Psi_state}. The basic integral is
\be
4\int d \ell e^{-\Delta \ell} K_{2 i s}(2 e^{-\ell/2}) K_{2 i s'}(2 e^{-\ell/2}) = \frac{\Gamma(\Delta \pm i s \pm i s')}{\Gamma(2\Delta)}  \equiv \Gamma^\Delta_{s,s'} \,,
\ee
where the variable $s$ is related to the energy of different states through,
\begin{align} \label{eqn:s_definition}
    &|\Psi^{E,j}_{m,m}\rb \implies s^2 = E-\lr{j+\frac{1}{2}}^2\,,\\
    &|\chi^{E,j}_{m,m}\rb \implies s^2 = E-\lr{j+\frac{1}{2}}^2\,, \\
    &|H^{E,j}_{m,m}\rb \implies s^2 = E-j^2\,, \\
    &|L^{E,j}_{m,m}\rb \implies s^2 = E-\lr{j+1}^2\,.
\end{align}
In writing the above expressions for the correlators, we must use the variable $s$ defined above for the respective states for the expression to be simple. For $j=0$ states ($|H\rb$ cannot have spin zero and doesn't appear), we find  the following correlators:
\be \label{eqn:psi_psi} 
\lb \Psi_{m',m'}^{E', 0} | e^{-\D \ell}|\Psi_{m,m}^{E, 0}\rb = \lb \chi_{m',m'}^{E', 0} | e^{-\D \ell}|\chi_{m,m}^{E, 0}\rb =\delta_{mm^{\prime}}\big(E E^{\prime} \G^{\Delta}_{s,s'} + \Delta(\Delta+1) \G^{\Delta+1}_{s,s'})\,,
\ee

\be \label{eqn:psi_chi}
\lb \Psi_{m',m'}^{E', 0} | e^{-\D \ell}|\chi_{m,m}^{E, 0}\rb = \delta_{mm^{\prime}} \Delta(\Delta+1) \G^{\Delta+1}_{s,s'}\,,
\ee

\be \label{eqn:L_L}
\lb L_{m',m'}^{E', 0} | e^{-\D \ell}|L_{m,m}^{E, 0}\rb  = \delta_{m,m'}\frac{\Delta(E+E^{\prime}-1+\Delta^2)(E+E^{\prime}-1+(\Delta-1)^2)+(1-2\Delta)E E^{\prime}}{2(2\Delta+1)}\Gamma_{s,s'}^{\Delta}\,,
\ee

\be \label{eqn:L_BPS}
    \lb \t{BPS} | e^{-\D \ell}|L_{m,m}^{E, j}\rb = \delta_{j,0} \delta_{m,0}\frac{\Delta}{2\Delta+1}(E-1+(\Delta-1)^2)(E-1+\Delta^2) \G^{\Delta}_{s,s'=i}\,,
\ee

\be \label{eqn:psi_L}
\lb \Psi_{m',m'}^{E', 0} | e^{-\D \ell}|L_{m,m}^{E, 0}\rb  =  \lb \chi_{m',m'}^{E', 0} | e^{-\D \ell}|L_{m,m}^{E, 0}\rb = \delta_{m,m'}  \frac{E^{\prime}(\Delta+1)+\Delta  (E-1+\Delta^2 )}{2\Delta+1} \G^{\Delta+\frac{1}{2}}_{s,s'}\,,
\ee

\be \label{eqn:psi_bps}
\lb \Psi_{m',m'}^{E, 0} | e^{-\D \ell}|\mathrm{BPS} \rb   =  \lb \chi_{m',m'}^{E, 0} | e^{-\D \ell}|\mathrm{BPS}\rb  = \delta_{m',0} \frac{E(\Delta+1)+\Delta(\Delta^2 -1 )  }{2\Delta+1} \G^{\Delta+\frac{1}{2}}_{s,s'=i}\,,
\ee

\be \label{eqn:bps_bps}
\lb \text{BPS} |e^{-\D \ell} | \text{BPS} \rangle = \frac{\Delta^2 \left(\Gamma(\Delta) \Gamma(\Delta+2)\right)^2}{ \Gamma(2\Delta+2)}\,.
\ee

\subsection*{Two-point function in TFD}
Using the above formulas, we can also calculate the two-point function in the thermal TFD state \eqref{eqn:TFD}
\be \label{eqn:thermal_2pt}
\lb \cO(\tau) \cO(0) \rb_\beta = {}_{\tau}\lb \t{TFD}| e^{-\Delta \ell} | \t{TFD} \rb_{\beta-\tau}\,,
\ee
where the operators have scaling dimension $\Delta$. Since the operator is spinless, one significant simplification will be that the inner product vanishes between states building the TFD with different $j,j_z$. The full expression will be very complex, containing fourteen independent terms. However, the two-point functions in appendix \ref{app:spinless_ops} fully determine this correlator. It would be interesting to study it further in order to understand how a black hole that starts out in the canonical ensemble evolves with time \cite{Biggs:2025nzs}.

\section{The spectrum of Hawking radiation in $\mathcal N=2$ supergravity} \label{sec:4}

We now discuss the spectrum of Hawking radiation from a near-BPS black hole in $\mathcal{N}=2$ supergravity minimally coupled to a single hypermultiplet. The matter content of the hypermultiplet consists of two Weyl fermions and two complex scalar fields. In addition, we have propagating gravitons, spin-$\frac{3}{2}$ gravitini, and photons from the gauge field that the black hole is charged under. 

In this section, we will, for simplicity, only consider the Hawking radiation from the hypermultiplet. The s-wave modes in the hypermultiplet carry the minimum amount of angular momentum, so at low temperatures, this is the sector that will dominate the Hawking radiation since the potential barrier of the black hole will be smallest for the lowest values of angular momenta, allowing most of the radiation to escape to infinity. 

\paragraph{How to transition to BPS states.} We first clarify in which sense a near-BPS state can emit Hawking radiation and transition into a BPS state in asymptotically flat spacetimes. After all, BPS states are exact eigenstates of the Hamiltonian: they do not evolve under time evolution, and cannot be transitioned into. The key point is that the initial state is a near-BPS black hole with ADM energy $M > Q$ (BPS black holes instead have $M=Q$). Because the ADM energy is measured with respect to conformal infinity, the ADM energy does not change as the black hole evaporates, and instead, the energy lost by the black hole is transferred into radiation at null infinity. Eventually, the black hole stops emitting radiation, and we have a spacetime described by a black hole with its emitted radiation at null infinity. We denote this final configuration of the black hole as ``BPS". The entire spacetime still has ADM energy $M=Q + (M-Q) > Q$ and so strictly speaking is not BPS, with the correct interpretation that $M-Q$ of the energy is carried by Hawking radiation at null infinity while the rest is carried by the black hole.\footnote{To be more specific, the integral of the matter stress tensor (component along the outgoing null directions) along future null infinity will be given by $M-Q$, while the Bondi mass will be given by $Q$ as $u\to \infty$ where $u$ is one of the Bondi-Sachs coordinates that parametrizes the retarded time along outgoing light rays.
}

\subsection{Fermi's golden rule}
We follow the general discussion in \cite{Brown:2024ajk} to explain how Hawking radiation from the black hole can be calculated using the Schwarzian two-point function. 

Consider a near-BPS reissner-Nordstr\"om black hole in $d=4$ $\mathcal{N}=2$ supergravity. Far from the black hole, we have asymptotically flat space, while in the near-horizon region, the spacetime topology becomes approximately AdS$_2 \times S^2$. Within the framework of AdS/CFT, the Schwarzian theory describing the BH should be thought of as living at the boundary of AdS embedded within the larger asymptotically flat spacetime. 

Hawking radiation escapes from the AdS$_2$ throat and reaches asymptotic infinity, and so we must understand how to properly couple the Schwarzian theory to free QFT in asymptotically flat space. The argument for the correct coupling is simple. Imagine a classical scalar field $\phi_{\t{cl.}}$ in the full spacetime that solves the equations of motion. In the AdS/CFT correspondence, a classical field in the full spacetime will have a non-normalizable component in the AdS$_2$ region of the spacetime, and will deform the boundary CFT by a source term
\be \label{eqn:deformed_schw_action}
I = I_{\t{sch.}} + \int d t\,  \phi_{\t{cl}} (t) \mathcal{O}(t),
\ee
where $\mathcal{O}$ is a CFT operator with scaling dimension such that the total operator is marginal. The preceding discussion can be promoted from classical sources to operators by making the replacement $\phi_{\t{cl.}} \to \hat{\phi}$ since for a free field we have the operator equation $(\nabla^2 - m^2)\hat{\phi}=0$. Now $\hat \phi(t)$ acts on the canonically quantized Fock Hilbert space built around the reissner-Nordstr\"om black hole background, and using the field equations $\hat \phi(t)$ can be expressed in terms of creation and annihilation operators that create modes of frequency $\omega$ at infinity. Once we have coupled the Schwarzian to the free-field Hilbert space, we have effectively deformed our Hamiltonian by an interaction term
\be \label{eqn:H_I}
H_I = \hat{\phi} \mathcal{O}\,.
\ee
With this interaction Hamiltonian, we can use standard QM perturbation theory to compute transition rates from an initial BH state $|E_i\rb$ to a final state $|E_f,\omega\rb$ where $\omega$ is a particular mode of frequency $\omega$ at infinity. Using Fermi's Golden rule, this transition rate is given by
\be \label{eqn:FGR}
\Gamma_{i \to f} = 2 \pi\, |\langle E_f, \omega| \cO \hspace{.035cm}\hat{\phi}|E_i \rangle|^2 \,\delta(E_i - \omega - E_f).
\ee
If we start in a maximally mixed initial state characterized by some macroscopic charges, and we want to calculate the transition rate into an arbitrary final state, we would have
\be
\Gamma_{\t{spon.}} = \frac{1}{|\t{initial}|}\sum_{\substack{\t{final}}} \sum_{\substack{\t{initial}}} \Gamma_{i \to f}\,,
\ee
where we average over initial states with the specified macroscopic charges, but sum over all possible final states.

\paragraph{The Hilbert space.}
Let us elaborate on the structure of the black hole Hilbert space for initial/final states used in Fermi's golden rule. The one-sided black hole Hilbert space decomposes into supermultiplets and can be labeled according to
\be
\mathcal{H}_{\t{BH}} = e^{S_0}|\t{BPS}\rb \oplus \sum_{\mathbf{j} \geq \frac{1}{2}} \sum_{m_i=-j_{\t{max}}}^{j_{\t{max}}} \mathop \int_{E_0(\mathbf{j})}^\infty d E  \rho_{\mathbf{j}}(E) \lr{|E^j_{m_1} , H \rb \oplus |E^{j-\frac{1}{2}}_{m_2}, \Psi \rb \oplus |E^{j-\frac{1}{2}}_{m_3}, \chi \rb \oplus|E^{j-1}_{m_4}, L \rb }\,,
\ee
There are $e^{S_0}$ BPS states with $M=Q$ and $j=0$, and then a continuum of states after an energy gap in each supermultiplet. The upper index gives the angular momentum of the state $j$, the lower index the axial angular momentum $m_i$; as before, the range of $m_i$ is given by the spin within each state within the $\mathbf j$ supermultiplet. The extra label in the set $\Phi = \{ H, \Psi, \chi, L \}$ distinguishes states with identical quantum numbers across different multiplets, as an example we have four distinct states with zero angular momentum $|E,\Psi \rb, |E,\chi \rb, |E, L \rb, |\t{BPS}\rb$ across three distinct supermultiplets. When calculating matrix elements between states, we must keep these additional labels to distinguish which particular BH state we are dealing with.

The Hilbert space of the radiation is the free field Fock space. In the simplest case, without angular momentum in the single particle sector, we have
\be \label{eqn:H_rad}
\mathcal{H}_{\t{rad}} = |0\rb \oplus \int_0^\infty d \omega |\omega\rb \oplus \mathcal{H}_{\t{multi-part.}}\,, \qquad |\omega\rb = a^\dag |0\rb, \qquad [a_\omega, a^\dag_{\omega'}]=\delta(\omega-\omega')\,,
\ee
where $\lb \omega | \omega' \rb=\delta(\omega-\omega')$, and $|0\rb$ is the free field vacuum and $|\omega\rb$ is a single outgoing mode at future infinity. We have included the multi-particle Hilbert space for completeness. The full Hilbert space that enters Fermi's golden rule is the tensor product $\mathcal{H} = \mathcal{H}_{\t{BH}} \otimes \mathcal{H}_{\t{rad}}$.

\subsection{Massless Scalar}
We first consider the radiation of the massless scalar. We restrict ourselves to the s-wave sector, which is the dominant emission channel for Hawking radiation. In \cite{Brown:2024ajk}, the field equations were solved on a reissner-Nordstr\"om background, and the source coupling to the Schwarzian was found to be
\be
 \hat{\phi}(t) = \mathcal{N} \int_0^\infty d \omega \hspace{.035cm}\sqrt{r_+^2 \omega} \hspace{.035cm} (a_{\omega} + {a^\dag_{\omega}})\,.
\ee
This gives the interaction Hamiltonian \eqref{eqn:H_I},
\be
H_I = \mathcal{N} \cO \int_0^\infty d \omega \hspace{.035cm}\sqrt{r_+^2 \omega} \hspace{.035cm} (a_{\omega} + {a^\dag_{\omega}}), \qquad \mathcal{N}^2 = \frac{1}{\pi^2}\,,
\ee
where $\mathcal{O}$ is a spinless operator in the Schwarzian theory with scaling dimension $\D=1$. We can now use \eqref{eqn:FGR} to calculate transition rates between different initial and final states, and as a consequence, the rate of emission of Hawking radiation into the s-wave of the scalar. General transition rates are quite complicated, so we will consider examples where the initial state of the BH has zero spin $j=0$. There are a variety of initial states that we will consider in sequence.

\paragraph{Initial State $|E, \Psi\rb$:} The initial state has $j=0$ and is in the supermultiplet with highest spin $\mathbf{j}=\frac{1}{2}$. In this case the possible final states are $|E_f, \Psi\rb, |E_f, \chi\rb, |E_f, L\rb, |\t{BPS}\rb$ tensored with a mode $|\omega\rb$ in the radiation Hilbert space. Summing over all possible final states, we get
\begin{align}
\Gamma_{\t{spon.}} = &\int_0^\infty d \omega \int_{E_0(\mathbf{\frac{1}{2}})}^\infty d E_f \rho_{\mathbf{j_f}=\frac{1}{2}}(E_f) \delta(E_i - \omega - E_f) \times 2 \pi\, \lr{|\langle E_f, \Psi ,\omega| \cO \hspace{.035cm}\hat{\phi}|E_i, \Psi \rangle|^2 \,+ |\langle E_f, \chi ,\omega| \cO \hspace{.035cm}\hat{\phi}|E_i, \Psi \rangle|^2} \nn \\
&+ \int_0^\infty d \omega \int_{E_0(\mathbf{1})}^\infty d E_f \rho_{\mathbf{j_f}=1}(E_f) \delta(E_i - \omega - E_f)\times 2\pi |\langle E_f, L ,\omega| \cO \hspace{.035cm}\hat{\phi}|E_i, \Psi \rangle|^2 \nn \\
&+ \underbrace{\int_0^\infty d \omega \int_{0}^\infty d E_f \rho_{\t{BPS},\mathbf{j_f}}(E_f) \delta(E_i - \omega - E_f) \times 2\pi |\langle \t{BPS},\omega| \cO \hspace{.035cm}\hat{\phi}|E_i, \Psi \rangle|^2}_{\t{transition from near-BPS $\to$ BPS state}}
\end{align}
In the first line, we use the density of states $\rho_{\mathbf{j_f}=\frac{1}{2}}$ since the final states fall into the supermultiplet with largest spin $\mathbf{j_f}=\frac{1}{2}$, while in the second line the final $|L\rb$ state falls into the $\mathbf{j_f}=1$ multiplet. The last line consists of transitions to BPS states which only have support at $\mathbf{j_f}=0$.

\paragraph{Matrix elements from LQM.} We now explain how to evaluate the one-sided correlators appearing in the rate. Using the interaction Hamiltonian, we can evaluate the radiation inner product and use the result from LQM 
\begin{align} \label{eqn:matrix_element_example}
|\langle E_f, \Psi ,\omega| \cO \hspace{.035cm}\hat{\phi}|E_i, \Psi \rangle|^2 &= \mathcal{N}^2 r_+^2 \omega \times |\lb E_f, \Psi | \mathcal{O} | E_i, \Psi \rb|^2 \nn \\
&= \mathcal{N}^2 r_+^2 \omega \times \lb \Psi^{E_f, j=0} | \mathcal{P}_\Delta | \Psi^{E_i, j=0} \rb  \nn \\
&= \mathcal{N}^2 r_+^2 \omega \times \big(E_f E_i \G^{\Delta}_{s_i,s_f} + \Delta(\Delta+1) \G^{\Delta+1}_{s_i,s_f})
\end{align}
In the first line, we evaluated the radiation Hilbert space inner product $|\lb \omega | \hat{\phi} | 0 \rb|^2 = \mathcal{N}^2 r_+^2 \omega$. In the second line, we turned the modulus squared of the one-sided correlator into the two-sided correlator in LQM. In the last line, we used the LQM correlator \eqref{eqn:psi_psi}. The variables $s_i, s_f$ are related to the initial energies \eqref{eqn:s_definition} through $E_{i,f}=s_{i,f}^2 + (1/2)^2$. The same steps apply for the other matrix elements for which we list the final answers 
\begin{align}
|\langle E_f, \chi ,\omega| \cO \hspace{.035cm}\hat{\phi}_0|E_i, \Psi \rangle|^2 &= \mathcal{N}^2 r_+^2 \omega \times  \D(\D+1)\G^{\Delta+1}_{s_i,s_f}\,,
\end{align}
using \eqref{eqn:psi_chi} with the same $s_i,s_f$ as above. The other correlator is
\begin{align}
|\langle E_f, L ,\omega| \cO \hspace{.035cm}\hat{\phi}|E_i, \Psi \rangle|^2 &= \mathcal{N}^2 r_+^2 \omega \times |\lb E_f, L | \mathcal{O} | E_i, \Psi \rb|^2 \nn \\
&= \mathcal{N}^2 r_+^2 \omega \times \lb L^{E_f, j=0} | \mathcal{P}_\Delta | \Psi^{E_i, j=0} \rb  \nn \\
&= \mathcal{N}^2 r_+^2 \omega \times \frac{E_i (\D+1)+\D(E_f -1+\D^2) }{2\D + 1} \G^{\D+\frac{1}{2}}_{s_i,s^L_f}, \qquad E_f=(s_f^L)^2 + 1^2\,\,,
\end{align}
where we have used \eqref{eqn:psi_L}. In this case, $s^L_f$ has a different relation to the final energy since the $L$ state is in the supermultiplet $\mathbf{j_f}=1$. The BPS state transition is
\be
|\langle \t{BPS}, \omega| \cO \hspace{.035cm}\hat{\phi}|E_i, \Psi \rangle|^2 = \mathcal{N}^2 r_+^2 \omega \times \frac{E_i (\D+1)+(\D^2-1) }{2\D + 1} \G^{\D}_{s_i,s_f=i}
\ee
To calculate the spontaneous emission rate, we also need the densities of states from \eqref{eqn:dos},
\be
\rho_{\t{BPS},\mathbf{j_f}}(E)= e^{S_0} \delta_{\mathbf{j_f},0} \delta(E)\,, \qquad \rho_{\mathbf{j_f}}(E)=\frac{e^{S_0} \mathbf{j_f} }{ \pi^2  E^2} \sinh \left(2 \pi \sqrt{E-E_0(\mathbf{j_f})}\right) \Theta\left(E-E_0(\mathbf{j_f})\right), \qquad E_0=\mathbf{j_f}^2\,.
\ee
Putting everything together, the total rate is
\begin{align}
\Gamma_{\t{spon.}} = &2\pi \mathcal{N}^2 r_+^2 \int_0^\infty d \omega \rho_{\mathbf{j_f}=\frac{1}{2}}(E_f) \times \omega (E_f E_i \Gamma^\D_{s_i,s_f}+2\D(\D+1) \Gamma^{\D+1}_{s_i,s_f} ) \nn \\
&+ 2\pi \mathcal{N}^2 r_+^2 \int_0^\infty d \omega \rho_{\mathbf{j_f}=1}(E_f) \times \omega \frac{E_i (\D+1)+\D(E_f -1+\D^2) }{2\D + 1} \G^{\D+\frac{1}{2}}_{s_i,s^L_f} \nn\\
&+ 2\pi \mathcal{N}^2 r_+^2 \underbrace{\int_0^\infty d \omega \rho_{\t{BPS},\mathbf{j_f}}(E_f) \times \omega \frac{E_i (\D+1)+(\D^2-1) }{2\D + 1} \G^{\D}_{s_i,s_f=i} }_{\t{transition from near-BPS $\to$ BPS state} }\\
\t{where we have used,}  & \qquad E_f=E_i-\omega,\qquad  E_i=s_i^2 + \frac{1}{4}, \qquad E_f=(s^L_f)^2 + 1 = s_f^2 + \frac{1}{4} \,. \nn
\end{align}
The final energy is always $E_f=E_i-\omega$ from the delta function constraint. We have left the expression in terms of $E_f$ for simplicity. The relation between the $s_f$ variable and $E_f$ changes line by line depending on the multiplet the final state is in. In the final line, we have listed all the parameters in the rate and how they are related to the final energy. We always have $E_i=s_i^2+1/4$ for this initial state. The density of states naturally cuts off the integral over frequencies when there don't exist black holes that can be transitioned into. 

The energy flux is given by multiplying the integrand by the energy $\omega$ of the emitted mode
\begin{align}
\frac{d E_{|\Psi\rb}}{d t} = &2\pi \mathcal{N}^2 r_+^2 \underbrace{\int_0^\infty d \omega \rho_{\mathbf{j_f}=\frac{1}{2}}(E_f) \times \omega^2 (E_f E_i \Gamma^\D_{s_i,s_f}+2\D(\D+1) \Gamma^{\D+1}_{s_i,s_f} )}_{|\Psi\rb \to |\Psi\rb, |\chi\rb ~~\t{transitions}} \nn \\
&+ 2\pi \mathcal{N}^2 r_+^2 \underbrace{\int_0^\infty d \omega \rho_{\mathbf{j_f}=1}(E_f) \times \omega^2 \frac{E_i (\D+1)+\D(E_f -1+\D^2) }{2\D + 1} \G^{\D+\frac{1}{2}}_{s_i,s^L_f} }_{|\Psi\rb \to |L\rb ~~\t{transitions}} \nn\\
&+ 2\pi \mathcal{N}^2 r_+^2 \underbrace{\int_0^\infty d \omega \rho_{\t{BPS},\mathbf{j_f}}(E_f) \times \omega^2 \frac{E_i (\D+1)+(\D^2-1) }{2\D + 1} \G^{\D}_{s_i,s_f=i} }_{\t{transition from near-BPS $\to $ BPS state} } \\
\t{where we have used,} & \qquad E_f=E_i-\omega,\qquad  E_i=s_i^2 + \frac{1}{4}, \qquad E_f=(s^L_f)^2 + 1 = s_f^2 + \frac{1}{4} \,. \nn
\end{align}

\paragraph{Initial State $|E, \chi\rb$:} This case is identical to the one discussed above since the correlators between $|E,\chi\rb$ and all other states are identical to the correlators for $|E,\psi\rb$ (see \eqref{eqn:psi_chi},\eqref{eqn:psi_psi}, \eqref{eqn:psi_L}). We immediately have the final answer for the flux
\be
\frac{d E_{|\chi\rb}}{d t} =\frac{d E_{|\Psi\rb}}{d t}\,.
\ee
\paragraph{Initial State $|E, L \rb$:} The initial state has $j=0$ and is in the supermultiplet with highest spin $j=1$. The final states are $|E_f, \Psi\rb, |E_f, \chi\rb, |E_f, L\rb, |\t{BPS}\rb$ tensored with a mode $|\omega\rb$ in the radiation Hilbert space. Summing over all possible final states and inserting a factor of frequency to directly get the energy flux
\begin{align}
\frac{d E_{|L\rb}}{d t} = &\int_0^\infty d \omega \omega \int_{E_0(\mathbf{\frac{1}{2}})}^\infty d E_f \rho_{\mathbf{j_f}=\frac{1}{2}}(E_f) \delta(E_i - \omega - E_f) \times 2 \pi\, \lr{|\langle E_f, \Psi ,\omega| \cO \hspace{.035cm}\hat{\phi}|E_i, L \rangle|^2 \,+ |\langle E_f, \chi ,\omega| \cO \hspace{.035cm}\hat{\phi}|E_i, L \rangle|^2} \nn \\
&+ \int_0^\infty d \omega \omega \int_{E_0(\mathbf 1)}^\infty d E_f \rho_{\mathbf{j_f}=1}(E_f) \delta(E_i - \omega - E_f)\times 2\pi |\langle E_f, L ,\omega| \cO \hspace{.035cm}\hat{\phi}|E_i, L \rangle|^2 \nn \\
&+ \underbrace{\int_0^\infty d \omega \omega \int_{0}^\infty d E_f \rho_{\t{BPS}}(\mathbf{j_f},E_f) \delta(E_i - \omega - E_f) \times 2\pi |\langle \t{BPS},\omega| \cO \hspace{.035cm}\hat{\phi}|E_i, L \rangle|^2}_{\t{transition from near-BPS $\to$ BPS state}}
\end{align}
Following the same steps as before, and using \eqref{eqn:psi_L}, \eqref{eqn:L_L}, \eqref{eqn:L_BPS} we find
\begin{align}
&\frac{d E_{|L\rb}}{d t} = 2\pi \mathcal{N}^2 r_+^2 \underbrace{\int_0^\infty d \omega \rho_{\mathbf{j_f}=\frac{1}{2}}(E_f) \times  \omega^2 \frac{E_f (\D+1)+\D(E_i -1+\D^2) }{2\D + 1} \G^{\D+\frac{1}{2}}_{s_f,s^L_i} }_{|L\rb \to |\Psi\rb, |\chi\rb}\nn \\
+ 2\pi \mathcal{N}^2 r_+^2 &\underbrace{\int_0^\infty d \omega \rho_{\mathbf{j_f}=1}(E_f) \times \frac{\D (E_i+E_f - 1 +\D^2)(E_i+E_f - 1 +(\D-1)^2)+(1-2\D)E_i E_f}{2(2\D + 1)} \G^{\D}_{s^L_f,s^L_i}}_{|L\rb \to |L\rb} \nn\\
+ 2\pi \mathcal{N}^2 r_+^2 &\underbrace{\int_0^\infty d \omega \rho_{\t{BPS}}(\mathbf{j_f},E_f) \times \omega^2 \frac{\D(E_i-1+(\D-1)^2)(E_i-1+\D^2)}{2\D + 1} \G^{\D}_{s^L_i,s_f=i} }_{\t{transition from near-BPS $\to$ BPS state} }\\
\t{where we have used, } \quad & \qquad E_f=E_i-\omega,\qquad  E_i=(s^L_i)^2 + 1, \qquad E_f=(s^L_f)^2 + 1 = s_f^2 + \frac{1}{4}\,. \nn
\end{align}

\paragraph{Microcanonical ensemble.} In the microcanonical ensemble, we should average over the possible initial states with zero angular momentum, we have
\begin{gather} \label{eqn:dedt_microcanonical}
\left.\frac{d E}{d t}\right\rvert_{E_i} = f_{|\Psi\rb} \frac{d E_{|\Psi\rb}}{d t} + f_{|\chi\rb} \frac{d E_{|\chi\rb}}{d t} + f_{|L\rb} \frac{d E_{|L\rb}}{d t}  \,,\\
f_{|\Psi\rb} = f_{|\chi\rb} = \frac{\rho_{\mathbf j=\frac{1}{2}}(E_i)}{\rho_{\mathbf j=\frac{1}{2}}(E_i)+\rho_{\mathbf j=1}(E_i)}, \qquad f_{|L\rb} =\frac{\rho_{\mathbf j=1}(E_i)}{\rho_{\mathbf j=\frac{1}{2}}(E_i)+\rho_{\mathbf j=1}(E_i)}\,,
\end{gather}
where, as a reminder, the initial state cannot be of type $H$ since there are no such states with zero angular momentum. The fractions $f$ represent the relative number of states $|\Psi\rb, |\chi\rb, |L\rb$ in the microcanonical window $E_i$.\footnote{Fermi's golden rule into a particular final state $|f\rb$ for an initial maximally mixed state in a microcanonical window with $N$ states is given by $\Gamma=\frac{1}{N} \sum_{i=1}^N |\lb f| \mathcal{O}| i \rb|^2$. If the states $|i\rb$ can be further subdivided into groups with different (on average) matrix elements with the final state, then we arrive at the version of FGI used in the main text.} We can write the flux in a compact way using the Liouville quantum mechanics correlators
\begin{align}
\left.\frac{d E}{d t}\right\rvert_{E_i} &= \sum_{\mathbf{j_f}=0,\frac{1}{2},1}\int_0^\infty d \omega  \rho_{\mathbf{j_f}} (E_f) \times \frac{2\pi \mathcal{N}^2}{3} r_+^2 \omega^2 \times f_{|\Phi_i\rb} \sum_{\Phi_f} \sum_{\Phi_i=\Psi,\chi,L}  \lb \Phi_f^{E_f, j=0} | e^{-\D \ell} | \Phi_i^{E_i, j=0} \rb \,,\\
\t{where we have used:} &\qquad E_f = E_i-\omega\,.
\end{align}
The final states $\Phi_f\in\{\t{BPS}, \Psi, \chi, H, L \}$ all have zero angular momentum but are spread across multiplets with the highest angular momentum $j_f=j_{\t{max}}$. The correlators can be read off from section \ref{sec:3.4_2pt}. The microcanonical flux can be calculated from the formulas that follow.

\paragraph{Quantum Flux for $\D=1$.} We write out the full quantum fluxes explicitly for $\D=1$ which is the case for the massless scalar
\begin{figure}
    \centering
    \includegraphics[width=1\linewidth]{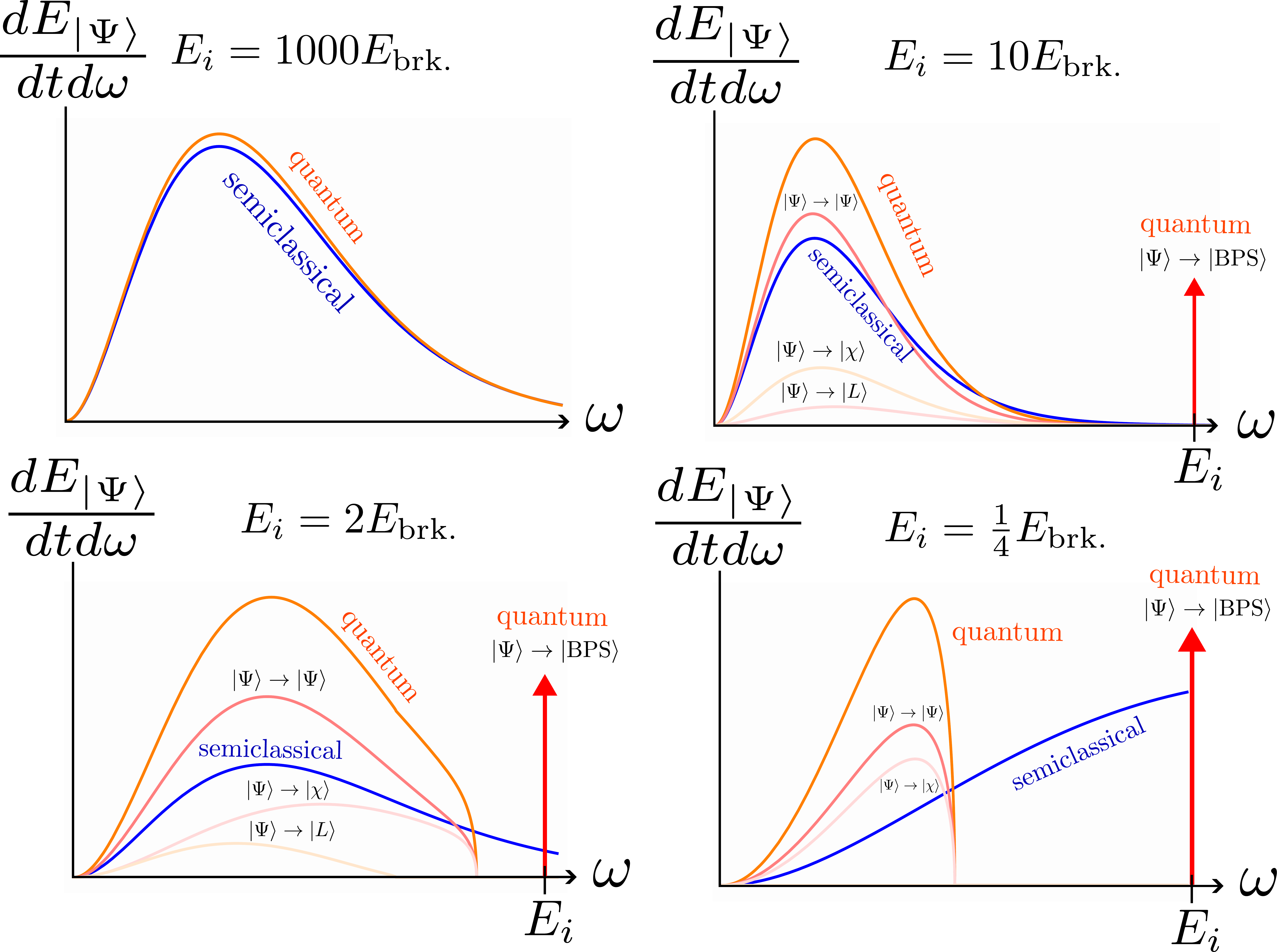}
    \caption{Comparison of the \textcolor{blue}{semiclassical prediction} vs. \textcolor{orange}{quantum corrected} Hawking radiation into a massless scalar field. The energy flux is plotted for an initial energy eigenstate $|\Psi\rb$ of the black hole with energy $E_i$ above extremality with zero spin $j=0$. \textbf{Upper Left:} At large energies $E_i$ above extremality, the flux approach the semiclassical answer. There is an exponentially small probability for the BH to emit all of its energy in one particle and transition into a BPS state; because of this, we have not included this in the plot, and we have cut off the plot at smaller energies. \textbf{Upper Right:} At intermediate energies, we once again find large deviations from the semiclassical answer. We have plotted the various decay channels available through scalar emission (\textcolor{RedOrange}{light red}, \textcolor{YellowOrange}{light orange}, \textcolor{pink}{pink}). The \textcolor{orange}{quantum curve} is the sum of the three possible emission channels. The transition into \textcolor{red}{BPS states} is now significant. \textbf{Lower Left:} At low energies $E_i \sim \Ebrk$, there are very large deviations from the semiclassical answer. The flux into \textcolor{red}{BPS states} is now comparable to the flux into near-BPS states (given by the \textcolor{orange}{quantum curve}). \textbf{Lower Right:} At very low energies, many emission channels are completely missing since the final states do not exist. The multiplet containing $|L\rb$ has minimum energy $E_0(\mathbf{J}=1)=\Ebrk/2$ and so is missing in the plot.}
    \label{fig:dedt_psi}
\end{figure}

\begin{align} \label{dedtscalar_psi}
\frac{d E_{|\Psi\rb}}{d t} = & \underbrace{2\pi \mathcal{N}^2 \int_0^\infty d \omega \omega \frac{(r_+ \omega)^2(3 E_i^2 + E_i (8-3 \omega) +2\omega(\omega-2))\sinh \lr{2\pi\sqrt{E_i-\omega-\frac{1}{4}}}}{3 (E_i-\omega)^2 \lr{\cosh \lr{2\pi\sqrt{E_i-\frac{1}{4}}} - \cosh \lr{2\pi\sqrt{E_i-\omega-\frac{1}{4}}} }  } }_{|\Psi\rb \to |\Psi\rb, |\chi\rb} 
 \Theta\lr{E_i-\omega-\frac{1}{4}}\nn  \\
&+\underbrace{2\pi \mathcal{N}^2 \int_0^\infty d \omega  \frac{(r_+ \omega)^2(3 E_i - \omega)(E_i + \omega + \omega^2) \sinh \lr{2\pi\sqrt{E_i-\omega-1}}}{3 (E_i-\omega)^2 \lr{\cosh \lr{2\pi\sqrt{E_i-\frac{1}{4}}} + \cosh \lr{2\pi\sqrt{E_i-\omega-1}} }  }}_{|\Psi\rb \to |L\rb} 
 \Theta\lr{E_i-\omega-1}\nn\\
&+\underbrace{ \frac{2\pi^3 \mathcal{N}^2}{3} r_+^2 E_i^4(E_i+2) \sech^2 \lr{\pi \sqrt{E_i-\frac{1}{4}}} }_{\t{transition from near-BPS $\to$ BPS state} } \Theta\lr{E_i-\frac{1}{4}}\,.
\end{align}
As a reminder, we are working in units where $\Ebrk=2$. To restore units, factors of energy $\Ebrk/2=1$ should be inserted wherever the units don't agree. The first two lines are near-BPS$\to$near-BPS transitions while the last are BPS transitions. The theta functions come from the density of states and ensure the black hole cannot radiate more energy than it has to begin with. 

In figure \ref{fig:dedt_psi}, we plot this flux and analyze the various emission channels across a range of energies. There are a number of interesting features in the full quantum answer:
\bi
\item At large energies above the breakdown scale $E_i \gg \Ebrk$, the quantum answer approaches the semiclassical flux computed by QFT in a fixed background.
\item A near-BPS BH can transition to a BPS BH by emitting all of its energy above extremality. This is the red delta function peak in figure \ref{fig:dedt_psi}.
\item There are multiple decay channels between different states within and between supermultiplets. Since supermultiplets begin at an energy gap $E_0(J)$, decay channels cease to exist when the final state must have energy below the gap.
\ei

A similar formula can be found for an initial $|L\rb$ state
\begin{align} \label{dedtscalar_L}
\frac{d E_{|L\rb}}{d t} = & \underbrace{2\pi \mathcal{N}^2 \int_0^\infty d \omega \frac{(r_+ \omega)^2(3 E_i - 2 \omega)(E_i + \omega (\omega-2)) \sinh \lr{2\pi\sqrt{E_i-\omega-\frac{1}{4}}}}{3 (E_i-\omega)^2 \lr{\cosh \lr{2\pi\sqrt{E_i-1}} + \cosh \lr{2\pi\sqrt{E_i-\omega-\frac{1}{4}}} }  } }_{|L\rb \to |\Psi\rb, |\chi\rb} 
 \Theta\lr{E_i-\omega-\frac{1}{4}}\nn  \\
&+\underbrace{2\pi \mathcal{N}^2 \int_0^\infty d \omega  \omega \frac{(r_+ \omega)^2(3 E_i^2 - 2 E_i + \omega(1-3 E_i)+\omega^2) \sinh \lr{2\pi\sqrt{E_i-\omega-1}}}{3 (E_i-\omega)^2 \lr{\cosh \lr{2\pi\sqrt{E_i-1}} - \cosh \lr{2\pi\sqrt{E_i-\omega-1}} }  }}_{|L\rb \to |L\rb} 
 \Theta\lr{E_i-\omega-1}\nn\\
&+\underbrace{ \frac{2\pi^3 \mathcal{N}^2}{3} r_+^2 E_i^4(E_i-1) \csch^2 \lr{\pi \sqrt{E_i-1}} }_{\t{transition from near-BPS $\to$ BPS state} } \Theta\lr{E_i-\frac{1}{4}}\,.
\end{align}

\paragraph{Semi-Classical limit.}
We can take semiclassical limits of the above formula by taking $E_i/ \Ebrk\to \infty$ while keeping $\beta \omega \sim \omega / \sqrt{\Ebrk E_i}$ fixed. In this limit, all of the above fluxes reduce to the expected semiclassical answer given by a scalar field on a fixed reissner-Nordstr\"om background at an inverse temperature determined by the energy above extremality
\be
\lim_{\substack{E_i \to \infty } } \frac{d E_{|\Psi\rb, |\chi\rb, |L\rb}}{d t} = \frac{1}{2\pi} \int_0^
\infty d \omega \omega \frac{4 (r_+ \omega)^2 }{e^{\beta \omega}-1}\,, \qquad \beta = \sqrt{\frac{2\pi^2 }{\Ebrk E_i}}\,,
\ee
where we have restored units by reintroducing $\Ebrk$ which is set to $\Ebrk=2$ elsewhere. This also ensures that a black hole in the microcanonical ensemble has the same semiclassical flux at large energies. The numerator $4(r_+ \omega)^2$ is the semiclassical greybody factor for a massless scalar field.

\paragraph{Semi-Classical limit: dominant transitions.} It is interesting to understand which transitions between black hole microstates dominate in the semiclassical limit. In the case of an initial $|\Psi\rb$ state, the transitions that dominate are
\be
\lim_{\substack{E_i \to \infty } } \frac{d E_{|\Psi\rb}}{d t} = \frac{d E_{|\Psi\rb \to |\Psi\rb}}{d t} + \frac{d E_{|\Psi\rb \to |\chi\rb}}{d t} = \frac{1}{2\pi} \int_0^
\infty d \omega \omega \frac{4 (r_+ \omega)^2 }{e^{\beta \omega}-1}\,,
\ee
which is the first line of \eqref{dedtscalar_psi}. The transition to $|L\rb$ states in the second line \eqref{dedtscalar_psi} is suppressed by a power of $\Ebrk \ll \omega_{\t{typical}}$ in the semiclassical limit 
\be
\lim_{E_i \to \infty} \frac{d E_{|\Psi\rb \to |L\rb}}{d t} = \frac{1}{2\pi} \int_0^
\infty d \omega \frac{\Ebrk}{2} \frac{4 (r_+ \omega)^2 }{e^{\beta \omega}+1} \ll \lim_{E_i \to \infty} \frac{d E_{|\Psi\rb \to |\Psi\rb, |\chi\rb}}{d t}\,.
\ee
along with a flipped sign in the denominator. Since $|\Psi\rb$ and $|\chi\rb$ states are identical for scalars, the same results apply to them. For an initial $|L\rb$ state, the story is reversed, with emission into other $L$ states dominating
\be
\lim_{\substack{E_i \to \infty } } \frac{d E_{|L\rb}}{d t} = \frac{d E_{|L\rb \to |L\rb}}{d t} = \frac{1}{2\pi} \int_0^
\infty d \omega \omega \frac{4 (r_+ \omega)^2 }{e^{\beta \omega}-1}
\ee
while transitions into $|\Psi\rb, |\chi\rb$ states are highly suppressed. It is interesting that black hole states favor particular decay channels even at large energies.

\paragraph{Energy flux into BPS states.}

\begin{figure}
    \centering
    \includegraphics[width=.5\linewidth]{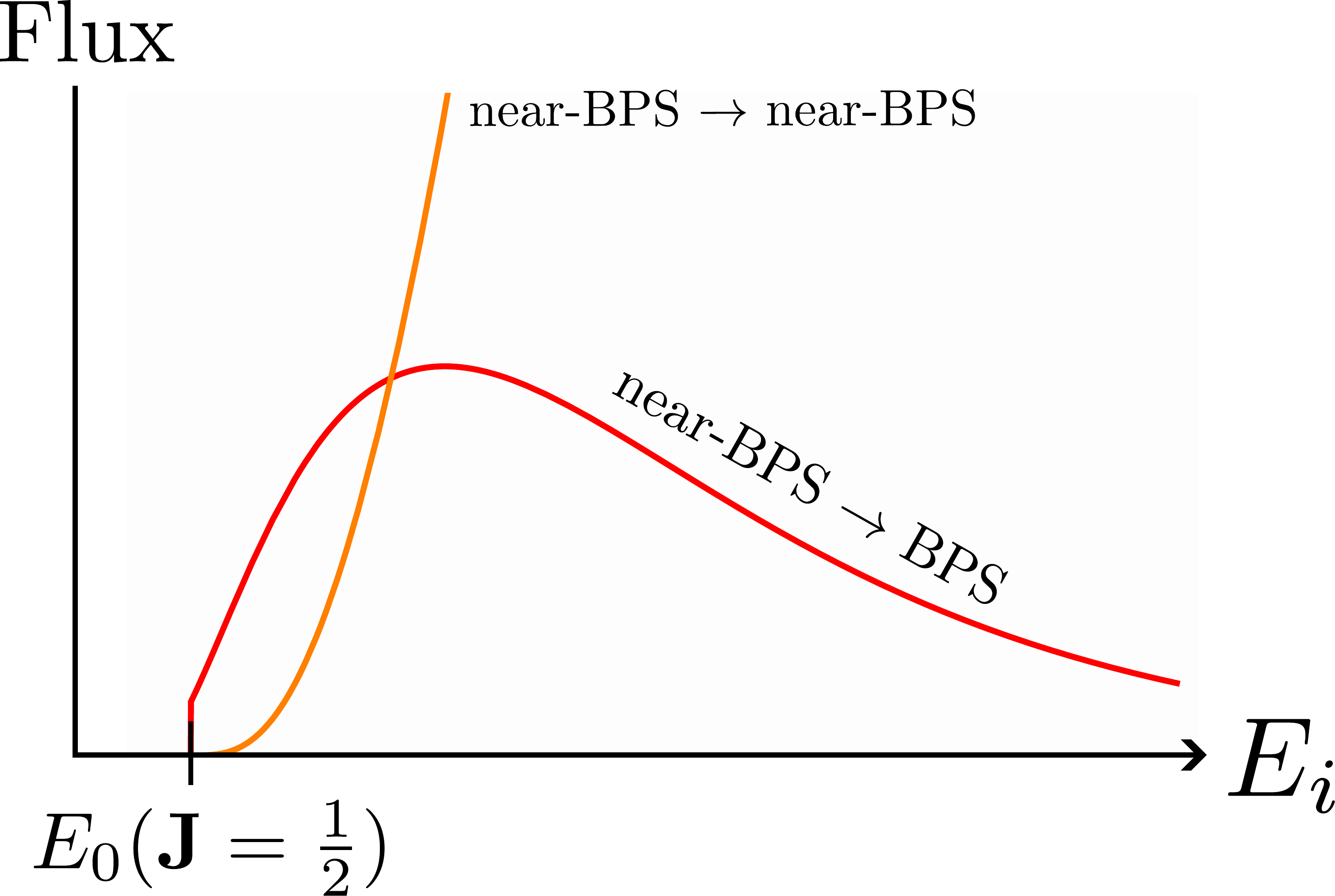}
    \caption{Comparison of the integrated flux $\frac{d E_{|\Psi\rb}}{d t}$ from an initial state $|\Psi\rb$ with  energy $E_i$ into \textcolor{red}{BPS states} and \textcolor{orange}{near-BPS states} $|\Psi\rb, |\chi\rb, |L\rb$. The flux into BPS states dominates at very low energies $E_i \sim E_0(\mathbf{J} = \frac{1}{2}) = \Ebrk/8$ since there are very few near-BPS states to transition into. The crossover point can be numerically found, $E_i \simeq .7 \Ebrk$. The BH state $|\Psi\rb$ does not exist in the spectrum for $E_i \leq E_0(\frac{1}{2})$, so there is a sharp cutoff. The total flux emitted is the sum of the two curves, and at large energies, the flux behaves semiclassically and is dominated by emission into \textcolor{orange}{near-BPS states}. The flux into BPS states very quickly decays with energy since the probability to emit a single particle carrying all of the excess energy of the BH exponentially decays with $E_i/\Ebrk$.}
    \label{fig:BPS_flux_transition}
\end{figure}

A BH can emit all of its energy above extremality and transition into a BPS state. As expected, for states in all supermultiplets, transitions into BPS states are exponentially suppressed at large energies
\be
\lim_{E_i \to \infty}\frac{d E_{|\Psi,\chi,L\rb \to |\t{BPS}\rb}}{d t} \sim e^{-2\pi \sqrt{2E_i/\Ebrk}}\,.
\ee
However, at sufficiently low energies $E_i$, the transition into BPS states begins to dominate because the number of near-BPS states begins to decrease. The details of when the flux is dominated by transitions into BPS states depends on the initial state. In figure \ref{fig:BPS_flux_transition} we plot the total flux into BPS and near-BPS states, and find the numerical transition to be at $E_i \simeq .7 \Ebrk$.

\subsection{Massless fermion}
We now consider massless fermion radiation. Fermions provide the dominant decay channel for the BH to lose angular momentum since scalars that carry angular momentum have larger greybody factor suppression. We focus on the dominant decay channel, which is given by fermions that carry away spin-$1/2$ angular momentum. The fermionic creation and annihilation operators are
\be
\{b_{\omega,m}, b_{\omega',m'}^\dag \} = \delta(\omega-\omega') \delta_{m m'}
\ee
where $m=\pm\frac{1}{2}$ labels the axial angular momentum. These operators act on an asymptotic fermionic Hilbert space analogous to \eqref{eqn:H_rad} and create fermionic plane waves at infinity. We will write out a general interaction Hamiltonian similar to the case of photons \cite{Brown:2024ajk}
\be
H_I=\mathcal{N} \sum_{m_\gamma= \pm \frac{1}{2}} \mathcal{O}_{j, m_\gamma} \int_0^\infty d \omega (r_{+} \omega) \left(b_{\omega, m_\gamma}+b_{\omega, m_\gamma}^{\dagger}\right), \qquad \mathcal{N}^2= \frac{4}{\pi^2}\,.
\ee
In the above $\mathcal{O}_{j,m_\gamma}$ is a spinning operator with $j=\frac{1}{2}$ and $\D=\frac{1}{2}$, and we have input the prefactor to match the semiclassical limit. In this case, the spinning operator $\mathcal{O}_{j,m_\gamma}$ will induce transitions between BH microstates with different angular momenta in the Schwarzian theory. We will not exhaustively study all aspects of fermion emission, specializing to a simple case that displays all of the interesting features. 

We will choose our initial state to be $|E^{j=\frac{1}{2}}_m, \Psi\rb$ in the $\mathbf{j}=1$ supermultiplet, where we will average over initial axial momenta. Since we start with a BH with $j=1/2$ and emit a spin-$1/2$ particle, the only possible final states have $j_f=0,1$. The possible final states at $j=0$ are $|E_f^{j=0}, \Psi\rb, |E_f^{j=0}, \chi\rb, |E_f^{j=0}, L\rb, |\t{BPS}\rb$. The possible final states at $j_f=1$ consist of $|E_f^{j=1}, \Psi\rb, |E_f^{j=1}, \chi\rb, |E_f^{j=1}, L\rb, |E_f^{j=1}, H\rb$ across different supermultiplets. These states are tensored with a mode $|\omega\rb$ in the fermion radiation Hilbert space. Summing over all possible final states, we get
\begin{align}\label{eqn:Gammaferm1}
\Gamma_{\t{spon.}}^{\ket{\Psi,j=\frac{1}{2}}} = & \frac{1}{2}\sum_{m_i=\pm \frac{1}{2}}\sum_{m_f=\pm 1,0} \sum_{\Phi \in \{\Psi, \chi, L, H \} } \int_0^\infty d \omega \int_{E_0(\mathbf{j_\Phi})}^\infty d E_f \rho_{\mathbf{j_\Phi}}(E_f) \delta(E_i - \omega - E_f) \nonumber \\ &\qquad \qquad \qquad \qquad  \qquad \qquad   \qquad  \qquad \qquad 
 \times 2 \pi\,|\langle E_{f,m_f}^{j=1}, \Phi ,\omega| H_I |E_{i,m_i}^{j=\frac{1}{2}}, \Psi \rangle|^2  \nn \\
&+ \frac{1}{2} \sum_{m_i=\pm \frac{1}{2}} \sum_{\Phi \in \{\Psi, \chi, L \} } \int_0^\infty d \omega \int_{E_0(j_\Phi)}^\infty d E_f \rho_{j_\Phi}(E_f) \delta(E_i - \omega - E_f) \times 2 \pi\, |\langle E_f^{j=0}, \Phi ,\omega| H_I | E_{i,m_i}^{j=\frac{1}{2}}, \Psi \rangle|^2  \nn \\
&+ \frac{1}{2} \sum_{m_i=\pm \frac{1}{2}}\underbrace{\int_0^\infty d \omega \int_{0}^\infty d E_f \rho_{\t{BPS}}(\mathbf{j_f},E_f) \delta(E_i - \omega - E_f) \times 2\pi |\langle \t{BPS},\omega| H_I | E_{i,m_i}^{j=\frac{1}{2}}, \Psi \rangle|^2}_{\t{transition from near-BPS $\to$ BPS state}}
\end{align}
The energy flux is obtained by inserting an extra factor of energy in the above formulas $\int d \omega \to \int d \omega \omega$. The operator that is inserted in LQM variables for the spinor is $e^{-\frac{\ell}{2}} D^{\frac{1}{2}}_{m_p,m_p}$ from the discussion around \eqref{eqn:hypermultiplet}. This operator is very special, it is a half-BPS operator since half of the supercharges commute with it \eqref{eqn:selection_rules_commutators}, and therefore it has additional selection rules for transitions \eqref{eqn:selection_rules}. In the first line of \eqref{eqn:Gammaferm1}, only the transitions $\Psi \to \Psi$ and $\Psi \to H$ are allowed, while in the second line, only $\Psi \to \Psi$ and $\Psi \to L$ are allowed. Interestingly, transitions into BPS states are forbidden. Taking this into account, the flux is
\begin{align}\label{eqn:Gammaferm2}
\frac{d E_{|\Psi,j=\frac{1}{2}\rb}}{d t} = & \frac{1}{2}\sum_{m_i=\pm \frac{1}{2}}\sum_{m_f=\pm 1,0} \int_0^\infty d \omega  \omega \int_{E_0( \mathbf{j}=\frac{3}{2})}^\infty d E_f \rho_{\mathbf{j}=\frac{3}{2}}(E_f) \delta(E_i - \omega - E_f) \times 2 \pi\,|\langle E_{f,m_f}^{j=1}, \Psi ,\omega| H_I |E_{i,m_i}^{j=\frac{1}{2}}, \Psi \rangle|^2  \nn\\  &+ \frac{1}{2}\sum_{m_i=\pm \frac{1}{2}}\sum_{m_f=\pm 1,0} \int_0^\infty d \omega \omega \int_{E_0(\mathbf{j}=1)}^\infty d E_f \rho_{\mathbf{j}=1}(E_f) \delta(E_i - \omega - E_f) \times 2 \pi\,|\langle E_{f,m_f}^{j=1}, H ,\omega| H_I |E_{i,m_i}^{j=\frac{1}{2}}, \Psi \rangle|^2  \nn \\
&+ \frac{1}{2} \sum_{m_i=\pm \frac{1}{2}}\int_0^\infty d \omega  \omega \int_{E_0(\mathbf{j}=\frac{1}{2})}^\infty d E_f \rho_{\mathbf{j}=\frac{1}{2}}(E_f) \delta(E_i - \omega - E_f) \times 2 \pi\, |\langle E_f^{j=0}, \Psi, \omega| H_I | E_{i,m_i}^{j=\frac{1}{2}}, \Psi \rangle|^2  \nn \\
&+ \frac{1}{2} \sum_{m_i=\pm \frac{1}{2}}\int_0^\infty d \omega  \omega \int_{E_0(\mathbf{j}=1)}^\infty d E_f \rho_{\mathbf{j}=1}(E_f) \delta(E_i - \omega - E_f) \times 2 \pi\, |\langle E_f^{j=0}, L, \omega| H_I | E_{i,m_i}^{j=\frac{1}{2}}, \Psi \rangle|^2  \nn
\end{align}
The first two lines have a transition to a higher angular momentum state. The third and fourth lines have a transition to $j=0$ near-BPS states

\begin{figure}
    \centering
    \includegraphics[width=1\linewidth]{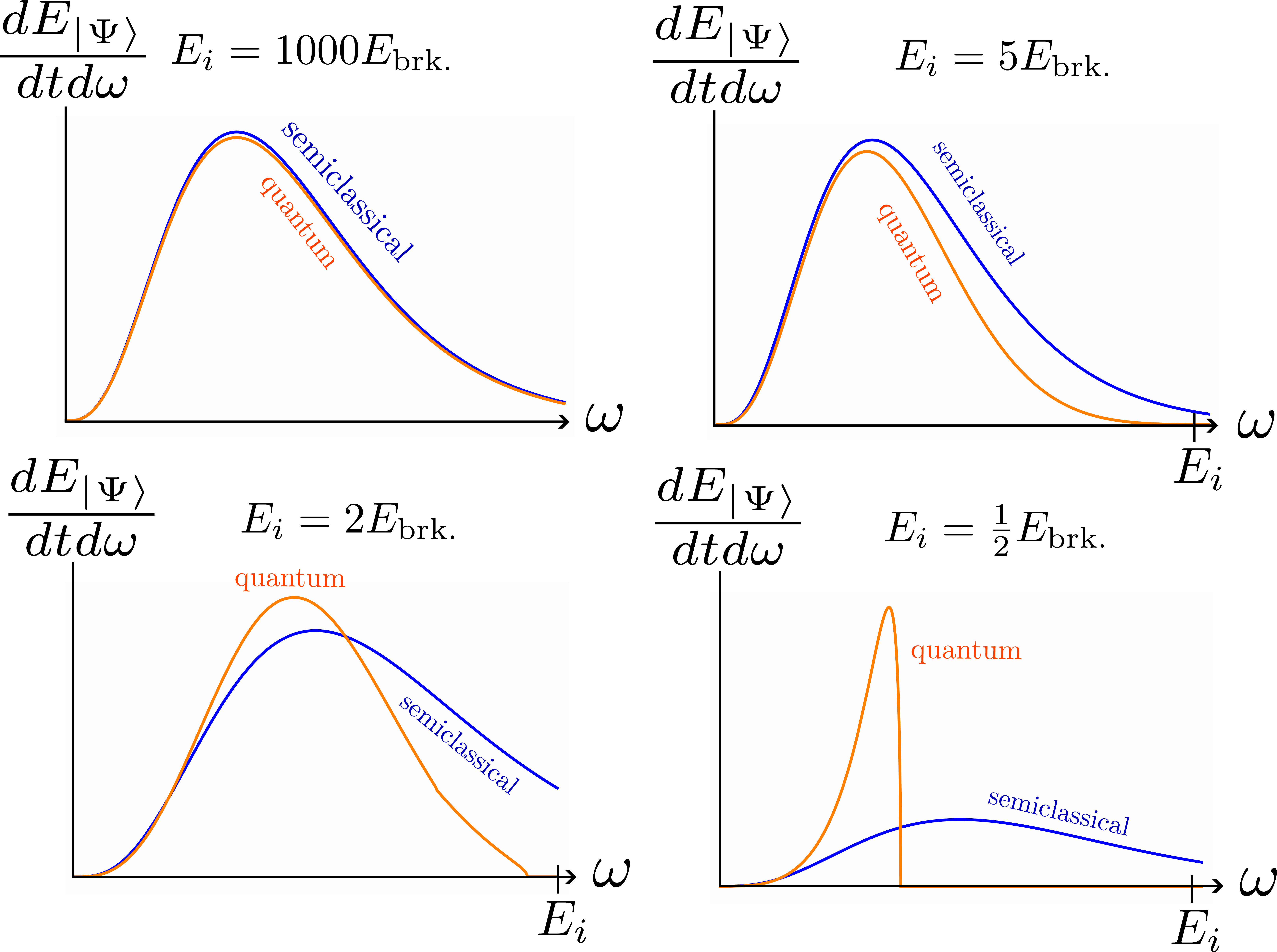}
    \caption{Comparison of the \textcolor{blue}{semiclassical prediction} vs. \textcolor{orange}{quantum corrected} Hawking radiation into a spin half fermion. The energy flux is plotted for an initial energy eigenstate $|\Psi\rb$ of the black hole with energy $E_i$ above extremality with spin $j=\frac{1}{2}$. \textbf{Upper Left:} At large energies $E_i$ above extremality, the flux approach the semiclassical answer. \textbf{Upper Right/Lower plots:} As we decrease the initial energy, we start to see large deviations from the semiclassical answer, just as in the case of scalar emission. Surprisingly, the deviations are less severe than in the case of the scalar in figure \ref{fig:dedt_psi}. One new feature for fermion radiation is that there are no transitions into BPS states. This is explained by the spin-$1/2$ fermion channel being special, since the corresponding operator is BPS in LQM variables. All of the other features described in the caption of figure \ref{fig:dedt_psi} also appear here.}
    \label{fig:dedt_psi_fermion}
\end{figure}

Using the formulas in appendix \ref{sec:spinhalf_2pt}, and summing over quantum numbers using identities presented there, the above becomes surprisingly simple
\begin{align}
&\frac{d E_{|\Psi,j=\frac{1}{2}\rb}}{d t} =  \frac{3\pi \mathcal{N}^2 }{8} \int_0^\infty d \omega \omega \frac{ (r_+\omega)^2}{E_i-\omega} \times \underbrace{\left( \frac{2 E_i \sinh \lr{2\pi \sqrt{E_i-\omega -\frac{9}{4}}} \Theta(E_i-\omega-\frac{9}{4}) }{\cosh \lr{2\pi \sqrt{E_i-1}}+\cosh \lr{2\pi \sqrt{E_i-\omega-\frac{1}{4}}} }  \right.}_{|\Psi^{j=\frac{1}{2}}\rb \to |\Psi^{j=1}\rb} \nn\\ 
&+\underbrace{(1+\frac{1}{3}) \frac{2 \omega \sinh \lr{2\pi \sqrt{E_i-\omega -1}} \Theta(E_i-\omega-1) }{\cosh \lr{2\pi \sqrt{E_i-1}}-\cosh \lr{2\pi \sqrt{E_i-\omega-1}} } }_{|\Psi^{j=\frac{1}{2}}\rb \to |H^{j=1}\rb, |L^{j=0}\rb} +  \underbrace{\left. \frac{1}{3}\frac{2 E_i \sinh \lr{2\pi \sqrt{E_i-\omega -\frac{1}{4}}} \Theta(E_i-\omega-\frac{1}{4}) }{\cosh \lr{2\pi \sqrt{E_i-1}}+\cosh \lr{2\pi \sqrt{E_i-\omega-\frac{1}{4}}} } \right)}_{|\Psi^{j=\frac{1}{2}}\rb \to |\Psi^{j=0}\rb} \nn\,.
\end{align}
The channels each term originates from is indicated in the equation. In the second line, transitions into $L,\,H$ states are identical up to a prefactor of $\frac{1}{3}$. At large energies $E_i$, the above approaches the semiclassical answer \eqref{eqn:dedt_fermion_semiclassical}.\footnote{The semiclassical answer \eqref{eqn:dedt_fermion_semiclassical} must be multiplied by a factor of two since there, the answer was given for a single polarization of a spin half particle, and here we include both polarizations.} We plot the flux into fermions in figure \ref{fig:dedt_psi_fermion}.

\subsection{The evaporation history}

We now describe the evaporation history of an initially charged, rotating black hole in supergravity coupled to a single hypermultiplet consisting of four scalars and two fermions. The black hole will radiate into all possible emission channels: gravitons, photons, spin-$\frac{3}{2}$ gravitini, along with the scalars and fermions in the hypermultiplet. For simplicity, we will focus on radiation into fields in the hypermultiplet, which gives the dominant decay channel since the greybody factors for the hypermultiplet are significantly smaller due to angular momentum considerations, and so the majority of the flux will go into the lowest spin modes of the scalars and fermions.

\paragraph{Step 1: Losing angular momentum.} A charged, rotating black hole will gradually lose its angular momentum due to fermion emission. The dominant loss will come from losing $\frac{1}{2}$ a unit of angular momentum, so the black hole will alternate between being bosonic and fermionic. Other channels for angular momentum loss like the emission gravitons or gravitinos are suppressed by much larger greybody factors. Furthermore, s-wave scalar emission keeps the angular momentum constant but reduces energy. The emission of s-wave scalars and s-wave fermions occurs at roughly the same rate since the classical greybody factors are the same. Thus, as the black hole loses its angular momentum it will also lose energy to scalar and fermion emissions. 

Black hole states with angular momentum $j$ at fixed charge $Q$ only start to exist in the spectrum above the extremality bound
\be \label{eqn:energy_gap}
E=M_{\t{ext}}+\frac{j^2}{2}\Ebrk\,,
\ee
where $M_{\t{ext}}$ is the mass for the extremal solution $M=Q$. As energy is lost into radiation, we must necessarily find ourselves in states with smaller values of angular momentum whose energy must be greater than \eqref{eqn:energy_gap}.\footnote{The $j=0$ states in \eqref{eqn:energy_gap} above correspond to BPS states, and so near-BPS $j=0$ states start in the $j=\frac{1}{2}$ multiplet.}

\paragraph{Step 2: The quantum regime.} As the black hole loses energy into the deeply quantum regime, we end up with states highly concentrated in the $\mathbf j =\frac{1}2$ multiplet. These states either have $j=0$ or $j=\frac{1}{2}$. The only possibility is to transition to the same $\mathbf j =\frac{1}2$  multiplet or to the BPS state. This is shown in the plot. Eventually, all states end up in the BPS state, and the probability $P_{\t{BPS}}(t)$ for that to occur with a given initial state $|E^{j=0},\Psi\rb$ with $E\approx \Ebrk$ is given in figure \ref{fig:statesevolution}. In this figure, we assume we only have transitions between zero angular momentum states in the $\mathbf j =\frac{1}2$ multiplet.

\paragraph{Evolution of BH state.} We now explain how the state of the BH evolves when starting from a microcanonical ensemble (see \cite{Biggs:2025nzs}, which studied the evolution of the non-supersymmetric case). With a finite number of states labelled by indices $a,b$, and transition rates $\Gamma_{a\to b}$, the probability to be in a state $P_a(t)$ evolves with time according to
\be
\frac{d P_a(t)}{d t} = \sum_{b} P_b(t) \Gamma_{b \to a} - \sum_{b } P_{a}(t) \Gamma_{a \to b}\,,
\ee
The continuous version of the above is obtained by replacing the sums by integrals over the density of final states\footnote{As a reminder the density of final states for the radiation has flat measure as explained in \cite{Brown:2024ajk}.}, and replacing the probabilities by probability densities per unit energy, which we will denote by $\rho_{\Phi_a}(E) P_{\Phi_a}(E,t)$
\begin{equation}\label{eqn:prob_evolution}
\begin{aligned}
 &\frac{d P_{\Phi_a}(E,t)}{d t} = \sum_{\Phi_b\in \{\Psi,\chi,H,L,\mathrm{BPS} \}} \int_0^\infty d \omega \int d E' \rho_{\Phi_b}(E') P_{\Phi_b}(E',t) \Gamma_{{\Phi_b} \to {\Phi_a}} \\  
&- \int_0^\infty  d \omega \int d E' \rho_{\Phi_b}(E') P_{\Phi_a}(E,t) \Gamma_{{\Phi_a}\to {\Phi_b}}\,. \nn\\
 &\text{where } \quad \Gamma_{{\Phi_a}\rightarrow {\Phi_b}}(E \to E') = | \lb E, {\Phi_a}| \mathcal{O}\hat{\phi} | E', {\Phi_b}, \omega \rb|^2, \qquad E-E'=\omega\,. \nn
\end{aligned}
\end{equation} 
The transition rates are given by Fermi's golden rule \eqref{eqn:FGR}, which kills the integral over $\omega$. The indices ${\Phi_a},{\Phi_b}$ will now label the possible states $\{ \Psi, \chi, H, L, \t{BPS}\}$. The density of states $\rho_{\Phi_b}$ depends on the supermultiplet the final state is in, and evaluating $\Gamma_{{\Phi_a} \to {\Phi_b}}$ in LQM was explained around \eqref{eqn:matrix_element_example}. In $\rho_{{\Phi_a}}(E)$ we also include the BPS contribution as $\rho_{\t{BPS}}(E)=\delta(E)$. The total probability to be in any state is given by $\sum_{\Phi_a} \int d E \rho_{\Phi_a}(E) P_{\Phi_a}(E,t)=1$ which can be checked from the evolution equation \eqref{eqn:prob_evolution} and time reversal $\Gamma_{{\Phi_a} \to {\Phi_b}}= \Gamma_{{\Phi_b} \to {\Phi_a}}$.

\begin{figure}
    \centering
    \hspace*{-1cm}
    \includegraphics[width=0.5\linewidth]{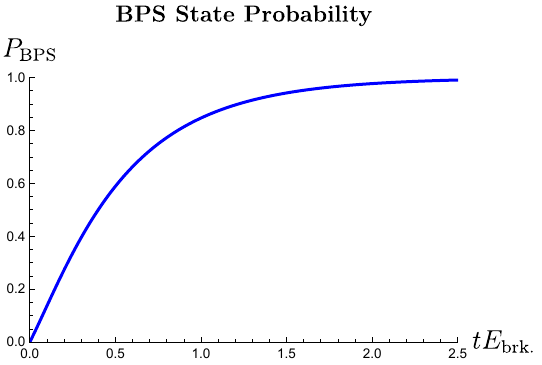}
    \caption{We plot the probability to be in the BPS state with time if we start in an initial state $|E_i^{j=0},\Psi\rb$ with $E_i=\Ebrk$. The evolution probability for all channels is plotted in figure \ref{fig:statesevolution}, and here we only plot the probability to be in the BPS state.}
    \label{fig:bpsevolution}
\end{figure}

In the special case where we are at initial energies where only the $\mathbf{j}=\frac{1}{2}$ multiplet survives, and we focus on states with $j=0$, the possible final states simplify
\begin{align}
\frac{d P_{\Phi_a}(E,t)}{d t}=\sum_{{\Phi_b}\in \{|\Psi\rangle,|\chi\rangle,|\mathrm{BPS}\rb \}}  &\int_{E^{\prime}>E} d E^\prime \rho_{\Phi_b}(E') P_{\Phi_b}(E',t) \Gamma_{{\Phi_b}\to {\Phi_a}}(E^\prime \rightarrow E) \\
&- \int_{E>E'} dE' \rho_{\Phi_b}(E') P_{\Phi_a}(E',t) \Gamma_{{\Phi_a}\rightarrow {\Phi_b}}(E \to E') \,. 
\end{align} 
The correlators can be found near \eqref{eqn:matrix_element_example}, and the evolution is plotted in figure \ref{fig:statesevolution}.
We also give the plot for BPS probability evolution with time in \ref{fig:bpsevolution}. Notice that the black hole has a high probability (say $>95\%$) in the BPS state in a polynomial amount of time, with the time scale $\mathcal{O}(t E_{\rm brk.})\sim \mathcal{O}(1)$. 

\subsection{The absorption cross-section: transparent black holes}

So far in the paper, we have analyzed how black holes lose energy through spontaneous emission. We will now analyze how black holes absorb radiation in a scattering experiment where we send an incoming planar wave. Like any other scattering experiment, the absorption cross-section measures the effective size of the black hole that is seen by such plane waves at a given frequency. 
The absorption cross-section of the non-supersymmetric case was studied by \cite{Emparan:2025sao,Biggs:2025nzs}. Here, we shall repeat the analysis for the supersymmetric black hole. Imagine sending in a large classical wave with a large amplitude. The wave reflects off the BH, and we can measure the outgoing flux. The difference in fluxes arises through three processes:
\be \label{eqn:ingoing_flux}
(\t{in flux})-(\t{out flux}) = (\t{absorbed flux}) - (\t{stim. emission}) - (\t{spon. emission})\,,
\ee
where some of the ingoing flux is absorbed (and not reflected back to infinity), the presence of the wave induces stimulated emission, which sends additional flux to infinity, and there is spontaneous emission of Hawking radiation. In the limit that the classical wave is very large, the first two channels are much larger than the Hawking radiation, so we can neglect it. The ratio 
\be
P_{\t{abs}}(\omega)=\frac{(\t{in flux})-(\t{out flux})}{(\t{in flux})}\,,
\ee
is equivalently known as the greybody factor, the transmission probability, and the absorption probability. Coupling the Schwarzian to a classical source wave, we can compute the first two quantities on the right of \eqref{eqn:ingoing_flux}. An incoming classical wave induces a time dependent coupling in the Schwarzian action \eqref{eqn:deformed_schw_action} with $\phi_0(t) = \mathcal{N} (N_\omega r_+^2 \omega)^{1/2}  e^{-i \omega t}$ with $N_\omega$ the amplitude of the wave at infinity, where the precise prefactor was found in \cite{Brown:2024ajk}, see also \cite{Emparan:2025sao}. With this coupling, the absorption and stimulated emission rates per unit frequency are given by Fermi's Golden rule \eqref{eqn:FGR}, which, after integrating over final states, gives
\begin{gather}
\Gamma_{\t{abs}}=2 \pi\, |\langle E_i+\omega | \cO \hspace{.035cm}\phi_0|E_i \rangle|^2 \, \rho(E_i+\omega)\,,\\
\Gamma_{\t{stim. emis.}}=2 \pi\, |\langle E_i-\omega | \cO \hspace{.035cm}\phi_0|E_i \rangle|^2 \,\rho(E_i-\omega)\,.
\end{gather}
Normalizing by the ingoing flux, the greybody factor is
\be
P_{\t{abs}}(\omega)= \frac{2\pi}{ N_\omega} (\Gamma_{\t{abs.}}-\Gamma_{\t{stim. emis.}})\,.
\ee
It can be checked that in the semiclassical limit $E_i \gg \Ebrk$ for an initial $|E_i^{j=0},\Psi\rb$ state, the above reduces to $4(r_+ \omega)^2$, which is the expected greybody factor. The greybody factor goes to zero as $\omega \to 0$ since the probability to tunnel through the BH effective potential goes to zero. 

An interesting quantity that measures the size of the black hole as seen by an incoming plane wave is the absorption cross-section, which is related to the absorption probability for an s-wave through \cite{Das:1996we,Gubser:1996zp}
\be
\sigma_{\t{abs}}(\omega) = \frac{\pi}{\omega^2} P_{\t{abs}}(\omega)\,.
\ee
In our case we will imagine starting with an initial state $|E_i^{j=0}, \Psi\rb$, and absorption of the s-wave scalar can take us to $\Psi, \chi, L$, while stimulated emission can take us to $\Psi, \chi, L, \t{BPS}$. We have
\begin{align}
\sigma_{\t{abs}}(\omega) = \frac{\pi^3 \mathcal{N}^2  r_+^2}{\omega} &\left( \sum_{\Phi_f\in\{\Psi, \chi, L \}} \rho_{\mathbf{j_{\Phi_f}}}(E_i+\omega)  \underbrace{|\langle E_i+\omega, \Phi | \cO |E_i, \Psi \rangle|^2}_{\t{absorption}} \right. \nn\\ &- \left. \sum_{\Phi_f\in\{\Psi, \chi, L, \t{BPS} \}} \rho_{\mathbf{j_{\Phi_f}}}(E_i-\omega)  \underbrace{|\langle E_i-\omega, \Phi | \cO |E_i, \Psi \rangle|^2}_{\t{stimulated emission}} \right)\,.
\end{align}
In the semiclassical limit $E_i \gg \Ebrk$ with low frequency radiation $\omega \to 0$ we recover
\be
\lim_{\substack{E_i \gg \Ebrk\\ \omega \to 0}}\sigma_{\t{abs}}(\omega) = 4\pi r_+^2= A_{\t{H}}\,.
\ee
This is the expected semiclassical absorption cross section \cite{Das:1996we}.

It's also intriguing to write the cross-section for scattering a wave on an extremal BH in the BPS state. In this case, we can only have absorption, with no stimulated emission
\be
\sigma_{\t{abs}}^{\t{BPS}}(\omega) = \frac{\pi^3 \mathcal{N}^2 r_+^2}{\omega} \sum_{\Phi_f\in\{\Psi, \chi, L \}} \rho_{\mathbf{j_{\Phi_f}}}(E_i+\omega)  \underbrace{|\langle E_i+\omega, \Phi | \cO | \t{BPS} \rangle|^2}_{\t{absorption}} \,.
\ee
We plot the cross section for scattering off both near-BPS and BPS states in figure \ref{fig:absorption}. The plots have many interesting features depending on the initial state of the black hole:
\bi
\item Near-BPS state: the cross section has resonances that show up as discontinuities in the derivative of the cross section whenever new states can be transitioned into at a particular frequency, or when stimulated emission into a set of states cannot occur due to gaps in the spectrum. For example, the red curve in figure \ref{fig:absorption} that is associated to a black hole with $E_i = .425 \Ebrk$ we see three resonances: the one at lowest $\omega$ is associated to a new absorption channel $\ket{\Psi} \to \ket{L}$, the next resonance is associated to losing the stimulated emission channels $\ket{\Psi} \to \ket{\Psi}$ and $\ket{\Psi} \to \ket{\chi}$ while the final resonance at $\omega = E_i$ is associated to the new stimulated emission channel into BPS states.
\item BPS state: an initial BPS black hole is transparent to incoming radiation with frequencies $\omega < \frac{1}{8}\Ebrk.$ because the radiation can not be absorbed or emitted by the BPS state at frequency $\omega$. Even for BPS states we also see resonances associated to new absorption channels: the first at $\omega=\Ebrk/8$ is associated to the appearance of the $\ket{\rm BPS} \to \ket{\Psi}$ or $\ket{\chi}$ transition, while the second at $\omega=\Ebrk/2$ is associated to the appearance of $\ket{\rm BPS} \to \ket{L}$.
\ei

\section{Discussion}
In this work, we have analyzed how one-loop quantum gravity corrections modify the spectrum of Hawking radiation and other related properties of near-BPS reissner-Nordstr\"om black holes in $\mathcal{N}=2$ supergravity in asymptotically flat spacetimes. 

We have found that these black holes behave very differently from their non-supersymmetric cousins in standard general relativity. Importantly, the spectrum consists of a large number of supersymmetric black holes with $M=Q$, followed by a gap in the spectrum to the near-BPS states beginning at $M=Q+\frac{\Ebrk}{8}$. These features heavily correct the results from semiclassical QFT in a fixed background and drastically alter the evaporation of such black holes. We highlight the most important points:
\bi
\item Near-BPS black holes can transition to BPS states with $M=Q$ by emitting all of their excess energy into a single quanta. This gives a sharp spectral line in the Hawking radiation. See figures \ref{fig:dedt_micro}, \ref{fig:dedt_psi}, \ref{fig:BPS_flux_transition}. 
\item Large quantum effects make it easier for Hawking radiation to penetrate the BH effective potential and escape to infinity, indicated by the corrected quantum flux rising above the semiclassical answer. See figures \ref{fig:dedt_micro}, \ref{fig:dedt_psi}, \ref{fig:dedt_psi_fermion}.
\item We have studied how an initial near-BPS state evolves in time, until it has radiated down to a BPS state. See figure \ref{fig:statesevolution}.
\item BPS black holes are transparent/invisible to very low frequency $\omega < \frac{\Ebrk}{8}$ probe radiation. See figure \ref{fig:absorption}.
\item To analyze the above problem, we solved $\mathcal{N}=4$ Liouville quantum mechanics and obtained the states and two-point functions of the theory. These are equivalent to the two-point functions in the $\mathcal{N}=4$ Schwarzian theory.
\ei

\subsection*{Future directions} 

\paragraph{Deviations from the thermal ensemble.} Consider a black hole that begins in an initial thermal state far above extremality, as opposed to the microcanonical initial states considered in this work. As it approaches extremality, large quantum corrections to the spectrum and to the radiation will cause the state to evolve to a very non-thermal distribution. In the case of non-supersymmetric charged black holes, the deviations of the state away from thermality were studied in \cite{Biggs:2025nzs}, where it was found that there is an attractor state (which is a function of only initial energy and amount of time spent evaporating) that the black hole evolves towards. This attractor state deviates from the naive thermal/microcanonical states of the same energy.\footnote{In \cite{Mohan:2024rtn} corrections to the radiation rate were also studied if the non-supersymmetric black hole spectrum was modified to instead be given by various non-perturbative completions of JT gravity.} It would be interesting to understand the attractor state for near-BPS black holes in supergravity, where we found that quantum effects seem more severe than in the case of non-supersymmetric black holes. This would partially entail understanding the thermal two-point function in the thermal ensemble better, which we have implicitly solved for around \eqref{eqn:thermal_2pt} but did not analyze in detail. 

\paragraph{BPS Chaos and higher-point correlators.} Signatures of chaos in BPS black hole states have been recently studied in \cite{Lin:2022rzw,Lin:2022zxd,Boruch:2023trc,Chen:2024oqv,Chang:2022mjp,Chang:2023zqk,Chang:2024lxt,Chang:2025rqy}. It would be interesting to better understand the chaotic properties of BPS black holes described by the $\mathcal{N}=4$ Schwarzian. We have computed the two-point function between BPS states and found features similar to the $\mathcal{N}=2$ calculation of \cite{Lin:2022rzw,Lin:2022zxd}. In both cases, the length of the TFD as diagnosed by the two-point function is finite rather than infinite, as expected from the semiclassical geometry. It would be interesting to compute higher point correlators in the $\mathcal{N}=4$ theory, to compute the eigenvalue spectrum of simple operators coupled to the  $\mathcal{N}=4$ theory. On general grounds, we expect higher-point functions that include crossings between the possible Wick contractions of the boundary operators to be suppressed in $e^{-\#\Delta}$. The only surviving diagrams would include  Wick contractions that do not cross, and with our results, this would give a semi-circle distribution for the eigenvalue of simple operators, just as in the $\mathcal N=2$ theory \cite{Lin:2022rzw,Lin:2022zxd}. This semicircle distribution typically indicates that the matrix elements of simple operators are drawn from a Gaussian ensemble.

\paragraph{The evaporation of black holes in supergravity in AdS.} We can repeat the calculations above for near-BPS black holes in AdS once we put absorbing boundary conditions for the outgoing Hawking radiation. There are two cases to consider.  First, we can analyze the cases in which the $\mathcal{N}=4$ super-Schwarzian is important.  In \cite{Heydeman:2020hhw} it was argued that the $\mathcal{N}=4$ Schwarzian describes the near-horizon dynamics of near-BPS black holes in supergravity on AdS$_3 \times S^3$. This supergravity theory is the holographic dual to the 2d SCFT arising from the low-energy limit of the D1-D5 system. To understand the evaporation of black holes in this theory, one can use the correlators in the $\mathcal N=4$ theory derived above. Secondly, we can analyze the cases in which the $\mathcal{N}=2$ super-Schwarzian is relevant, which is the more generic case for black holes in AdS. For example, such an EFT describes the quantum fluctuations for near-BPS black holes in AdS$_4$ \cite{Heydeman:2024ezi}, and AdS$_5$ \cite{Boruch:2022tno}. Given that the correlators in the $\mathcal{N}=2$ super-Schwarzian were previously derived in \cite{Lin:2022rzw, Lin:2022zxd} computing the Hawking radiation fluxes in these theories should now be a straightforward exercise.

\paragraph{Alternative descriptions of near-horizon dynamics.}
In this paper, we employed $\mathcal{N}=4$ Liouville quantum mechanics to study the near-horizon dynamics of near-BPS black holes described by $\mathcal{N}=4$ super-JT gravity. As established in JT with $\mathcal{N}=0,1,2$ supersymmetry, JT gravity admits multiple equivalent formulations: the Schwarzian description \cite{Kitaev:2017awl,Maldacena:2016upp,Engelsoy:2016xyb,Jensen:2016pah,Mertens:2017mtv,Forste:2017kwy,Stanford:2017thb,Forste:2017apw}, BF theory description\cite{Mertens:2018fds,Blommaert:2018iqz,Iliesiu:2019xuh,Iliesiu:2019lfc,Kapec:2019ecr,Belaey:2023jtr}, and the particle-on-group-manifold picture \cite{Yang:2018gdb,Blommaert:2018oro,Lin:2022zxd,Belaey:2024dde}. See \cite{Mertens:2022irh,Turiaci:2024cad} for recent reviews. Each description emphasizes distinct physical or mathematical features of the model. A systematic exploration of this web of formulations for the $\mathcal{N}=4$ case would be interesting, and we intend to elucidate these relationships and their implications in forthcoming work \cite{ongoing}.

\section*{Acknowledgments}
We thank Jan Boruch, Gary Horowitz, Maciej Kolanowski, Don Marolf, Geoff Penington, and Stephen Shenker for illuminating discussions. We also thank Anna Biggs and Roberto Emparan for useful comments about the draft. MU was supported in part by grant NSF PHY-2309135 to the Kavli Institute for Theoretical Physics (KITP), and by grants from the Simons Foundation (Grant Number 994312, DG), (216179, LB). LVI was supported by the Department of Energy, Office of Science, Office of High Energy Physics under the DOE Early Career Award DE-SC0025522. GL  was partly supported by the Department of Energy, Office of Science, Office of High Energy Physics under QuantISED Award DE-SC0019380. This work was also performed in part at the Aspen Center for Physics, which is supported by National Science Foundation grant PHY-2210452.

\appendix

\section{Greybody factors: spinors}
\label{sec:greybody}

In this section we calculate greybody factors for spinors on the extremal reissner-Nordstr\"om background. See appendix A of \cite{Brown:2024ajk} for the case of other particles. A field propagating on a BH background experiences an effective potential barrier. Hawking radiation is generated at the horizon of the BH, and as it propagates outwards it is partly reflected by the barrier and only a small fraction manages to escape to infinity. The fraction that propagates through the barrier is known as the greybody factor. The greybody factor cannot be calculated in full generality, but in limiting cases such as for a very cold BH with very low frequency Hawking radiation it can be evaluated analytically by approximately solving wave equations.

\subsection{Decoupling the Dirac equation}\label{sec:minimalfermgrey1}

Solving the Dirac equation on black hole backgrounds has a long history. See \cite{Teukolsky:1973ha,Press:1973zz,Page:1976jj,chandrasekhar1976transformation} for just a few papers. One subtlety with spinors compared to other particles is that different spinor components become coupled in the equation, but through various field redefinitions, they can be decoupled. In the case of Kerr-Newman  see\cite{Page:1976jj,Page:1977um,Hartman:2009nz,Cvetic:2009jn}, see \cite{Arbey:2021jif,Arbey:2021yke} for RN. For the wave equation on a general rotating background, see \cite{Teukolsky:1973ha,Press:1973zz}. We first review how to decouple it before solving for the greybody factor.

The massless Dirac Lagrangian on a curved manifold, and the equations of motion are
\begin{gather}
\mathcal{L} = i \int d^4 x \sqrt{g} \ol{\Psi} \g^\mu \nabla_\mu \Psi\,, \quad \gamma^\mu \nabla_\mu  \Psi = 0\,,\\
    \g^\mu(x) = e_a^{~\mu}(x) \g^a, \qquad g_{\mu \nu} = e_\mu^{~a} e_{\nu}^{~b} \eta_{a b}  ,\qquad \{\g^a,\g^b \} = 2 \eta^{a b}\,,\qquad 
    \nabla_\mu \Psi = \lr{\partial_\mu + \frac{1}{4} \omega_\mu^{~a b} \g_{a b}} \Psi, \qquad \nn
\end{gather}
with $\g_a$ the flat space gamma matrices $\{\g_a, \g_b \}=2\eta_{a b}$, $\g_{a b} = \frac{1}{2}[\g_a,\g_b]$, and $\omega_\mu^{a b}$ the spin connection. For a spherically symmetric black hole, we have the metric, tetrad, and spin connection
\begin{gather}
ds^2 = -f(r) dt^2 + \frac{dr^2}{f(r)}+r^2(d\theta^2 + \sin^2(\theta) d \phi^2), 
\\  e_t^{~0} = \sqrt{f}, \quad e_r^{~0}=\frac{1}{\sqrt{f}},\quad e_\theta^{~2} = r, \quad e_\phi^{~3} = r \sin \theta\,, \\
\omega_t^{01} = -\frac{f^{\prime}(r)}{2}, \quad \omega_{\theta}^{12}= \sqrt{f(r)}, \quad \omega_{\phi}^{13}= \sqrt{f(r)}\sin\theta, \quad \omega_{\phi}^{23}=\cos\theta\,.
\end{gather}
With indices $a,b=0,1,2,3$ flat space indices. In the case of the massless spinor, we take the Weyl representation of the gamma matrices
\begin{equation}
\begin{aligned}
       &\gamma^a = i \begin{pmatrix}
        0 & \sigma^a \\ \bar{\sigma}^a & 0
    \end{pmatrix}, \quad \text{with }  \sigma^a=(\mathbf{1},\vec{\sigma}) \,, \qquad  \bar \sigma^a=(\mathbf{1},-\vec{\sigma})\,.
\end{aligned}
\end{equation}
This choice decouples the left-handed from the right-handed Weyl fermions in the Dirac equation $\Psi=(\psi_L, \psi_R)^T$. The components of the Weyl fermion are still coupled, and the equation for the top two components of the Weyl fermion $\psi_L=(\psi_1, \psi_2)^T$ becomes
\begin{equation} \label{eq:diraceqn_A8}
   \left[ - i \begin{pmatrix}
        \partial_t &  f (\partial_r + r^{-1}) + \frac{1}{4}f^{\prime} \\ f (\partial_r + r^{-1}) + \frac{1}{4}f^{\prime} & \partial_t
    \end{pmatrix} +  \frac{\sqrt{f} }{r} \begin{pmatrix}
     - i \csc\theta \partial_\phi  & -  \partial_\theta - \frac{ \cot\theta}{2}  \\ 
            \partial_\theta + \frac{ \cot\theta}{2} & i   \csc\theta  \partial_\phi
    \end{pmatrix} \right]
    \begin{pmatrix}
        \psi_1\\ \psi_2
    \end{pmatrix} = 0 \,.
\end{equation}
There is a similar equation for the two components of the right-handed fermion. 
The angular part can be diagonalized through functions defined by
\begin{equation}
    \begin{pmatrix}
     - i \csc\theta \partial_\phi  & -  \partial_\theta - \frac{ \cot\theta}{2}  \\ 
            \partial_\theta + \frac{ \cot\theta}{2} & i   \csc\theta  \partial_\phi
    \end{pmatrix} \begin{pmatrix}
        \Theta_1(\theta,\phi) \\ \Theta_2(\theta,\phi)
    \end{pmatrix} =  \Big(j+\frac{1}{2}\Big)
    \begin{pmatrix}
        \Theta_1(\theta,\phi) \\ -\Theta_2(\theta,\phi)
    \end{pmatrix}\,.
\end{equation}
One can assume that $\Theta_i(\theta,\phi)\sim e^{i m \phi} f_i(\theta)$ and the equations becomes a pair of coupled set of ODEs for the $\theta$ variable.It turns out that there are two independent solution to this equation, one of the satisfies $\Theta_1=\Theta_2\equiv \Theta_+ $ and the other satisfies $\Theta_1=-\Theta_2\equiv \Theta_-$. Furthermore, Requiring that $f_i(\theta)$ is regular at $\theta=0,\pi$ fixes half-integer $m$ with  $-j\leq m\leq j$.\footnote{As a simple example, when $j=\frac{1}{2}$, the two solutions are $\Lambda_1=\Lambda_2=\cos\frac{\theta}{2}e^{\pm i\frac{\phi}{2}}$ and $\Lambda_1=-\Lambda_2=\sin\frac{\theta}{2}e^{\pm i\frac{\phi}{2}}$.} 
Importantly, these are not spinor spherical harmonics. Expanding the spinor using the ansatz
\be
\begin{pmatrix}
        \psi_1
        \\ \psi_2
\end{pmatrix} = \begin{pmatrix}
        \eta_+(r)\Theta_+(\theta)+\eta_-(r)\Theta_-(\theta) \\
        \eta_+(r)\Theta_+(\theta)-\eta_-(r)\Theta_-(\theta)
    \end{pmatrix} e^{i\omega t}\,,
\ee
and, using \eqref{eq:diraceqn_A8}, we get a coupled set of linear differential equations
\begin{equation}
    \begin{pmatrix}
        \omega + (j+\frac{1}{2})\frac{\sqrt{f}}{r} &  - i f (\partial_r + r^{-1}) - \frac{i}{4}f^{\prime} \\ - i f (\partial_r + r^{-1}) - \frac{i}{4}f^{\prime} & \omega -  (j+\frac{1}{2}) \frac{\sqrt{f}}{r} 
    \end{pmatrix} . \begin{pmatrix}
        \eta_+(r)\Lambda_+(\theta)+\eta_-(r)\Lambda_-(\theta) \\
        \eta_+(r)\Lambda_+(\theta)-\eta_-(r)\Lambda_-(\theta)
    \end{pmatrix} e^{i\omega t} = 0\,.
\end{equation}
The two coupled first-order equations can be combined to get the decoupled second-order equations:
\begin{equation}\label{eqn:dirac2nd1}
    16rf\left((3f + rf')\eta_\pm' + rf\eta_\pm''\right)+  \left(16 f^2 + r^2 (4 \omega + i f')^2 - 4f\left(\frac{1}{2}(1+2j)^2 \pm 4ir\omega - r(5f' + rf'')\right)\right) \eta_\pm(r)  = 0\,.
\end{equation}
Making a field redefinition $\Phi_\pm = r^{-3/2}\eta_\pm $
\begin{equation}
    \Big(\partial_{r^*}\partial_{r^*}+ \omega^2\Big) \Phi_\pm \pm \frac{1}{2}i \omega r^2\Big(\frac{f}{r^2}\Big)^\prime \Phi_\pm + \frac{1}{2}\Big(\frac{1}{2}+j\Big)^2 \Phi_\pm + D(r) \Phi_\pm = 0\,.
\end{equation}
with $\partial_*\equiv\partial_{r^*} = f(r)\partial_r$ the tortoise coordinate and $D(r)$ some function of $r$ of which its explicit form is not illuminating or necessary.

As mentioned earlier, this equation is not particularly easy to solve because of the factor of $i\omega$. The Chandrasekhar transformation\cite{chandrasekhar1976transformation} transforms it into a more standard Schrodinger-type equation with a real potential. We closely follow\cite{Arbey:2021jif,Arbey:2021yke}, the transformation is
$\tilde{\Phi}_\pm=\Big(\frac{r^2}{f}\Big)^{\frac{s}{2}}\Phi_\pm$ with $s$ the spin of the particle (here it is $\frac{1}{2}$ for fermion) such that 
\begin{equation}\label{eqn:phitilde}
    \Lambda^2 \tilde{\Phi}_\pm + P \Lambda_{\mp} \tilde{\Phi}_{\pm} -Q \tilde{\Phi}_{\pm} = 0, \quad \text{with } \Lambda_\mp= \partial_*\pm i \omega ,\quad  P= s\partial_*\log\frac{f}{r^2}, \qquad  Q=\frac{(2j+1)^2}{8}\frac{f}{r^2}-\frac{1}{2} f \left(r^2 \left( \frac{f}{r^2}  \right)^{\prime} \right)^{\prime} \,.
\end{equation}
Note that $\Lambda^2=\Lambda_+ \Lambda_-= \partial_*\partial_*+ \omega^2$. Doing another field redefinition
\begin{equation}
    \tilde{\Phi}_{\pm}= (2j+1)\sqrt{\frac{f}{r^2}} Z_{\pm} + \Lambda_{\pm} Z_{\pm} \,,
\end{equation}
we get a rather complicated third order differential equation. 
But magically, this complicated equation is actually solved by $Z_{\pm}$ satisfying the following standard Schrodinger-type equation
\begin{equation} \label{eqn:wave_eqn}
   \Big(\partial^2_*+ \omega^2-V_{\pm}(r)\Big) Z_{\pm} = 0 
\end{equation}
where the effective potential is given by
\begin{equation}
V_{\pm}\left(r\right)=\Big(j+\frac{1}{2}\Big)^2\frac{f}{r^2} \pm \Big(j+\frac{1}{2}\Big)\partial_*\left(\sqrt{\frac{f}{r^2}}\right)\,.
\end{equation}
This equation is easier, than \eqref{eqn:dirac2nd1}, to solve  for obtaining the greybody factors.

\subsection{Greybody factor: minimally coupled spinor}\label{sec:minimalfermgrey}
We now solve the wave equation \eqref{eqn:wave_eqn}, which can be written in the form
\be\label{eqn:fermionpotential}
f^2 Z''(r)+f f' Z'(r)+ \lr{\omega^2 - \frac{(2j+1)\sqrt{f}((2j+1)\sqrt{f}\mp 2f \pm r f')} {4 r^2}}Z(r)=0
\ee
The half integer $j=\frac{1}{2}, \frac{3}{2},\frac{5}{2},\ldots$ labels the total angular momentum of the field. The reason there are two equations is the total angular momentum comes from the tensor product of the orbital angular momentum $l$ and the intrinsic spin, giving $j = l \otimes \frac{1}{2}= (l+\frac{1}{2})\oplus (l-\frac{1}{2})$. There are thus two possible ways to realize total $j$ given by $l=j+\frac{1}{2},j-\frac{1}{2}\in \mathbb{Z}$. The two equations correspond to these two possible choices for orbital angular momentum. 

To solve these equations it's easier to introduce $Z=r u(r)$ and multiply everything by $r/f(r)$ giving
\be \label{eqn:spinor_diff_eqn}
r^2 f u''(r)  + (2f + r f') u'(r)  + \frac{r^2 u(r)}{f} \lr{\omega^2 - \frac{(2j+1)((2j+1)f \mp 2 f^{\frac{3}{2}} \pm r \sqrt{f}f')}{4r^2}} + f' u(r)=0
\ee
These equations are a rewriting of the Dirac equation in any spherically symmetric BH. They are simpler to solve for an extremal BH with $r_+ = r_-$, which will give us the greybody factor for a spinor on a $T=0$ background. They can be solved in three partially overlapping regions in the appropriate regions and glued together to get the transmission through the effective potential.

\paragraph{Region I: near-horizon region.}  The near-horizon region is defined by $r-r_{+} \ll r_+$. In this region we set $r=r_+$ in \eqref{eqn:spinor_diff_eqn} except for singular terms $(r-r_+)$
\begin{gather}
(r-r_+)^2 u_1''(r)+2 (r-r_+) u_1'(r)+\left(-j (j+2)-\frac{3}{4}+\frac{r_+^4 \omega ^2}{(r-r_+)^2}\right) u_1(r) =0 \ , \\   
(r-r_+)^2 u_2 ''(r)+2 (r-r_+) u_2'(r)+\left(-j^2+\frac{1}{4}+\frac{r_+^4 \omega ^2}{(r-r_+)^2}\right) u_2 (r) =0 \ ,
\end{gather}

The solutions are
\begin{gather}
\label{eqn:region1spinor}
u^I_1(r) = \frac{i}{\sqrt{2}}\sqrt{\frac{\omega r_+^2 }{r-r_+}} \left(a_1 \Gamma\lr{-j} J_{-j-1}\left(\frac{\omega r_+^2 }{r_+-r}\right)+a_2 \Gamma \left(j+2\right) J_{j+1}\left(\frac{\omega r_+^2}{r_+-r}\right)\right)\,,\\
u^I_2(r) = -\frac{i}{\sqrt{2}}\sqrt{\frac{\omega r_+^2 }{r-r_+}} \left(a_1 \Gamma\lr{-j} J_{-j}\left(\frac{\omega r_+^2 }{r_+-r}\right)-a_2 \Gamma (j) J_{j}\left(\frac{\omega r_+^2}{r_+-r}\right)\right)\,,
\end{gather} 
\paragraph{Region II: barrier.}
The intermediate barrier region is defined to be where $\omega^2$ is smaller than the effective potential, which are the other terms proportional to $u(r)$ in \eqref{eqn:spinor_diff_eqn}. Setting $\omega \to 0$ in \eqref{eqn:spinor_diff_eqn} gives
\begin{gather}
(r-r_+)^2 u_1''(r)+2 (r-r_+) u_1'(r)+\left(-j^2+\frac{1}{4}-\frac{2j r_+}{r}+\frac{(r-2r_+)r_+}{r^2}\right) u_1(r) =0 \ , \\   
(r-r_+)^2 u_2''(r)+2 (r-r_+) u_2'(r)-\frac{1}{4r^2}\left( r^2 (3+4j(2+j)) - 4 (3+2j)r r_+ + 8 r_+^2 \right) u_2(r) =0 \ .
\end{gather}
with solution
\begin{gather} \label{eqn:spinor_II}
    u_1^{II}(r)=\frac{(r-r_+)^{-j-\frac{3}{2}}}{r} \left(b_1 (r-r_+)^{2 j+2}+b_2 \left(\left(2 j^2+3 j+1\right) r^2-2 (j+1) r r_++r_+^2\right)\right)\,,\\
    u_2^{II}(r)=\frac{(r-r_+)^{-j-\frac{1}{2}}}{r} \left(b_1+b_2 \left(2 j^2 r^2+j r (r+2 r_+)+r_+^2\right) (r-r_+)^{2 j} \right)\,.
\end{gather}

\paragraph{Region III: flat space.}
The final region is flat space where we set $f(r)= 1$ with solutions given by spherical Bessel functions
\begin{gather}
u_1^{III}(r)= c_1 j_{j-\frac{1}{2}}(r \omega) + c_2  y_{j-\frac{1}{2}}(r \omega)\,,\\
u_2^{III}(r)= c_1 j_{j+\frac{1}{2}}(r \omega) + c_2  y_{j+\frac{1}{2}}(r \omega)\,,
\end{gather}
These functions behave at large $r$ as
\be \label{eqn:spinor_flatspaceexpansion}
 c_1 j_j(r \omega )+c_2 y_j(r \omega )\to -\frac{1}{2 r \omega} \left( (i c_1 + c_2) e^{i r \omega - \frac{1}{2} i \pi j} +   (-i c_1 + c_2) e^{-i r \omega + \frac{1}{2} i \pi j }\right),
\ee
where the outgoing wave is proportional to $i c_1 + c_2$ while the ingoing wave is proportional to $-i c_1 + c_2$.

\paragraph{Matching conditions:} We can match $a_i,b_i,c_i$ for $u_1$ independently by expanding $u_1^I$ at large $r$ and matching to the small $r$ expansion of $u_1^{II}$. A similar procedure would give different matching conditions for the field $u_2$. Carrying this out, we get
\begin{gather}
    u_1 \t{  matching conditions:} \nn \\ b_1 = -i (-1)^{-j} 2^{j+\frac{1}{2}} a_1 (r_+^2 \omega)^{-j} \omega^{-\frac{1}{2}}, \qquad b_2 = \frac{i (-1)^{j+1} 2^{-j-\frac{3}{2}} a_2 (r_+^2 \omega)^{j+1} \omega^{\frac{1}{2}}}{2j^2+j}\,, \nn\\
    c_1 = \frac{b_1}{\sqrt{\pi}} 2^{j+\frac{1}{2}} \omega^{-j+\frac{1}{2}}\Gamma(j+1),\qquad  c_2 = -\frac{b_2 \sqrt{\pi} }{\Gamma(j)} 2^{-j+\frac{1}{2}} (j+1)(2j+1) \omega^{j+\frac{1}{2}}\,.
\end{gather}
\begin{gather}
    u_2 \t{  matching conditions:} \nn \\ b_1 = -i (-1)^{-j} \frac{ 2^{j-\frac{1}{2}} a_1 j (r_+^2 \omega)^{-j} \sqrt{\omega}\Gamma(-j)}{\Gamma(-j+1)(2j^2+3j+1)} , \qquad b_2 = i(-1)^j \frac{a_2 j (r_+ \omega)^j r_+^2 \sqrt{\omega}\Gamma(j)}{2^{j+\frac{1}{2}}\Gamma(j+1)} \,, \nn\\
    c_1 =  \frac{b_1 2^{j+\frac{3}{2}}}{\sqrt{\pi}\omega^{j+\frac{1}{2}}}  j(2j+1)\Gamma(j+2)  ,\qquad  c_2 = -\frac{b_2 \sqrt{\pi} \omega^{j+\frac{3}{2}}}{2^{j+\frac{1}{2}}\Gamma(j+1)}\,.
\end{gather}

\paragraph{Greybody factor. } To calculate the greybody factor, we must set purely ingoing boundary conditions at the horizon and then calculate the ratio of flux that is reflected back out to infinity from an initial ingoing wave at infinity. To set purely ingoing boundary conditions at the horizon, we must set
\begin{gather}
 \quad \t{ingoing boundary conditions for $u_1$: } 
 \qquad a_1 = a_2 \frac{e^{3\pi i j} \Gamma(j+2)}{\Gamma(-j)}\,, \nn \\   
  \quad \t{ingoing boundary conditions for $u_2$: } 
 \qquad a_1 = a_2 \frac{e^{3\pi i j} \Gamma(j)}{\Gamma(-j)}\,.
\end{gather}
The probability for the mode to be reflected is given by the ratio of outgoing to ingoing fluxes at infinity, using \eqref{eqn:spinor_flatspaceexpansion} the probability for a mode to be reflected and transmitted is consequently given by
\begin{gather}
R = \frac{|c_1-i c_2|^2}{|c_1+i c_2|^2}\,, \qquad P_{\t{abs}}(\omega)\equiv T = 1-\frac{|c_1-i c_2|^2}{|c_1+i c_2|^2}\,.
\end{gather}
The greybody factor at low frequencies $(r_+ \omega) \ll 1$ is given by series expanding the transmission probability
\be
u_1, u_2 \quad \t{greybody factors:} \quad \lim_{(r_+ \omega) \to 0 }  P_{\t{abs}}(\omega) = \frac{\pi^2 (r_+ \omega)^{4 j+2}}{16^{j}\Gamma (j+1)^4}\,.
\ee
this is the greybody factor for the zero temperature BH at small frequencies. It turns out the greybody factors for $u_1$ and $u_2$ are identical for the same value of $j$. We list the first few greybody factors
\be
P_{\t{abs}}(j=\frac{1}{2})= 4 (r_+\omega)^4, \qquad P_{\t{abs}}(j=\frac{3}{2})= \frac{4}{81} (r_+\omega)^8, \qquad P_{\t{abs}}(j=\frac{5}{2})= \frac{4}{50625}  (r_+\omega)^{12}.
\ee
These greybody factors will receive finite temperature corrections by additive terms going as $(r_+ \omega)^{n_1} (\beta \omega)^{n_2}$ with $n_{i}$ positive integers.

\subsection{Greybody factor: hypermultiplet fermion}\label{sec:hyperfermgrey}
The fermion we are interested in comes from the hypermultiplet, and has an additional coupling to the electromagnetic field
\be
\mathcal{L} = i \ol \Psi^I \g^\mu \nabla_\mu \Psi_I + \frac{1}{2} \ol \Psi^I \g^{\mu \nu}(F_{\mu \nu}+i * F_{\mu \nu} \g_5) \Psi_J \epsilon_{I J}
\ee
As a reminder, there are two four component Majorana fermions labelled by $I,J$. The equation of motion after a chiral projection onto the top two components $P_L \Psi_J = \psi_{L,J}$
\be
i P_L \g^\mu  P_L \nabla_\mu \psi_{L,I} + \frac{1}{2} P_L\g^{\mu \nu}P_L (F_{\mu \nu}+i * F_{\mu \nu} \g_5) \psi_{L,J} \epsilon_{I J}=0\,.
\ee
In the equation the gamma matrices are conjugated by $P_L$. This projection of  4$\times$4 component gamma matrices, in the Weyl representation, reduces them to 2$\times$2 sigma matrices  $\sigma^{\mu}\equiv \sigma^a e_a^\mu$, we get
\begin{equation}
    \begin{pmatrix}
        i \sigma^{\mu}\nabla_{\mu} & \frac{1}{2} F_{01}\sigma^{01} \\
        - \frac{1}{2}F_{01}\sigma^{01} & i \sigma^{\mu}\nabla_{\mu} 
    \end{pmatrix}\begin{pmatrix}
        \psi_{L,1}\\ \psi_{L,2}
    \end{pmatrix} = 0\,.
\end{equation}
We have combined the equation into one that acts on a four spinor constructed out of $\psi_{L,I}$. We can diagonalize by taking $\psi_{\pm} = \psi_{1,L}\pm i \psi_{2,L}$ to get the decoupled equations
\begin{equation}
    (i \sigma^{\mu}\nabla_{\mu} \pm   \frac{1}{2} F_{01}\sigma^{01} )\psi_{\pm}=0\,.
\end{equation}
for two components spinors. However, the equations between the two components of say $\psi_+$ are still coupled.

Now using the same separation of variables discussed in the minimally coupled fermion case, we can reduce to a coupled set of first-order equations. To be explicit and simple, in the $j=\frac{1}{2}$ sector they take the form of
\begin{equation}\label{eqn:weyldiraceqn}
\begin{aligned}
    & f \partial_r \psi^{1}_{\pm} -  \Big(i \omega - \sqrt{f} \frac{(r^2 \sqrt{f})^\prime}{r^2} \Big) \psi^1_{\pm} - \frac{\sqrt{f}}{r^2}(r\pm r_+)\psi^2_{\pm}  = 0 \\
     &  f \partial_r \psi^{2}_{\pm} +  \Big(i \omega + \sqrt{f} \frac{(r^2 \sqrt{f})^\prime}{r^2} \Big) \psi^2_{\pm} - \frac{\sqrt{f}}{r^2}(r\pm r_+)\psi^1_{\pm} = 0\,.
\end{aligned}
\end{equation}
We have $\psi_\pm^i$ with the index $i=1,2$ labelling the first or second component of the respective field. From here, one might simply substituting one equation into another and get a second order differential equation schematically in the form \eqref{eqn:phitilde}, and perform the Chandrasekhar transformation to reduce it to the standard Schrodinger form \eqref{eqn:fermionpotential}. 

As a side comment, although such a standard procedure is a bit cumbersome (so we will not perform it fully), the above equations are actually easy to solve in the barrier region. By setting $\omega$ to zero, one finds that\footnote{There are still two independent solutions for $\psi_+$. The two components can be chosen independently.}
\begin{equation}
\begin{aligned}
    \psi_+ & \sim b_{1,+} z^{\frac{1}{2}}  \\
    \psi_- & \sim b_{1,-} z^{\frac{5}{2}} + b_{2,-} z^{-\frac{3}{2}} \\
\end{aligned}
\end{equation}
which follows the pattern of $\psi \sim b_1 z^\Delta + b_2 z^{1-\Delta}$, so that for $\psi_+$ the scaling dimension (near the AdS2 boundary) is 1/2 while for $\psi_-$ the scaling dimension is 5/2, which matches \cite{Lee:1999yu} stating that $\Delta=j,j+2$. 

Instead, one may hope that there is a simpler way to solve the same problem, because the supersymmetry relating bosons and fermions in the action should also induce a relation between the bosonic and fermionic equations of motion.  One can see this by performing the following transformation 
\begin{equation}
    \tilde{\psi}^1= \frac{1}{2}(\psi^1 + \psi^2)\sqrt{\frac{r-r_+}{r}}, \quad \tilde{\psi}^2= -\frac{i}{2}(\psi^1 - \psi^2) \frac{r^2}{r_+^2} \sqrt{\frac{r-r_+}{r}},
\end{equation}
reduce the $\tilde{\psi}^2$ variable, and $\tilde{\psi}^1$ and $\tilde{\psi}^2$ now satisfies (here one should plug in $f=\frac{(r-r_+)^2}{r^2}$)
\begin{equation}\label{eqn:simplifeddiraceq}
     f \partial_r \tilde{\psi}^1 - \frac{r_+^2}{r^2} w \tilde{\psi}^2  - (1\mp 1) \frac{r_+\sqrt{f}}{r^2}\tilde{\psi}^2_{\pm}= 0, \quad  f \partial_r \tilde{\psi}^2_{\pm} + \frac{r^2}{r_+^2} w \tilde{\psi}^1_{\pm} + (1\mp 1) \frac{r_+ \sqrt{f}}{r^2}\tilde{\psi}^2_{\pm} = 0\,.
\end{equation}
Combining these together, we see that 
\begin{equation}
    \partial_r (r^2 f)\partial_r \tilde{\psi}^1_\pm + \omega^2 \frac{r^2}{f} \tilde{\psi}^1_{\pm} - \frac{r_+^2}{r^2}(1\mp 1)(2\mp 1) \tilde{\psi}^1_{\pm}   = 0
\end{equation}
which is precisely the equation of motion for a massless scalar in the $l=0$ or $l=2$ sector.
Based on the supermultiplet analysis, these are the bosonic modes as superpartner of our $\psi_{\pm}$ modes. 

To get the greybody factor, one can work with \eqref{eqn:weyldiraceqn}, following the procedure of the minimally coupled fermion case, and get a greybody factor using the standard matching method. 
On the other hand, it is sufficient to solve the simplified equation \eqref{eqn:simplifeddiraceq} and there is no surprise that the greybody factor is identical to those of scalars. 
In particular, we find  
\begin{equation}
    P^{\t{ferm}}_{\t{abs}}(j=\frac{1}{2},\Delta=\frac{1}{2})=4(r_+ \omega)^2 = P^{\rm scalar}_{\t{abs}}(j=0,\Delta=1) 
\end{equation}
matching the standard greybody factor of the $\ell=0$ scalar.

\section{$\mathcal{N}=4$ LQM Supercharges} \label{app:supercharges}
We define $\mathcal{J}_{\pm} = \mathcal{J}_{1} \pm i \mathcal{J}_{2}\,,$ which applies for both left/right charges. All of the supercharges of LQM are

\begin{equation}
Q_{l}^1 = i \psla ( \partial_\ell - 
 \mathcal{J}^l_3) - i\pslb \mathcal{J}^l_{-} + e^{-\frac{\ell}{2}} g^1_{~q} \psi^q_r  -  \frac{i}{2} \psla [\pslb, \ol{\psi}_{l,2}],
\end{equation}
\begin{equation}
Q_{l}^2 = i \pslb ( \partial_\ell +  \mathcal{J}^l_3) - i\psla \mathcal{J}^l_{+} + e^{-\frac{\ell}{2}}  g^2_{~q} \psi^q_r - \frac{i}{2} \pslb [\psla, \ol{\psi}_{l,1}],
\end{equation}
\begin{equation}
\ol{Q}_{l,1} = i \ol{\psi}_{l,1} ( \partial_\ell + \mathcal{J}^l_3) + i\ol{\psi}_{l,2} \mathcal{J}^l_{+} + e^{-\frac{\ell}{2}} \bps_{r,q} (g^{-1})^{q}_{~1} -   \frac{i}{2} \ol{\psi}_{l,1} [\ol{\psi}_{l,2} ,{\psi}_{l}^2],
\end{equation}
\begin{equation}
\ol{Q}_{l,2} = i {\bps_{l,2} ( \partial_\ell - \mathcal{J}^l_3)} + i{\bps_{l,1} \mathcal{J}^l_-} + {e^{-\frac{\ell}{2}} \bps_{r,q} (g^{-1})^q_{~2}} - \frac{i}{2} \bps_{l,2} [\bps_{l,1}, {\psi}_{l}^1],
\end{equation}
\begin{equation}
   Q_{r}^1 = {i \psra (  \partial_\ell + \mathcal{J}^r_3)} + i {\psrb \mathcal{J}^r_{-}} -  {e^{-\frac{\ell}{2}} (g^{-1})^1_{~q} \psi^q_l } - { \frac{i}{2} \psra [\psrb , \bpsrb]},
\end{equation}
\begin{equation}
     Q_{r}^2 = {i \psrb (  \partial_\ell - \mathcal{J}^r_3)} + i {\psra \mathcal{J}^r_{}} -  {e^{-\frac{\ell}{2}} (g^{-1})^2_{~q} \psi^q_l } - { \frac{i}{2} \psrb [\psra , \bpsra]},
\end{equation}
\begin{equation}
    \ol{Q}_{r,1} = { i \bpsra ( \partial_\ell - \mathcal{J}^r_3)} - i {\bpsrb \mathcal{J}^r_{+}}  {-e^{-\frac{\ell}{2}} \ol{\psi}_{l,q} (g)^q_{~1} }  { - \frac{i}{2} \bpsra [\bpsrb , \psrb]},
\end{equation}
\begin{equation}
    \ol{Q}_{r,2} = {i \bpsrb ( \partial_\ell + \mathcal{J}^r_3)} - i {\bpsra \mathcal{J}^r_{-}}  {-e^{-\frac{\ell}{2}} \ol{\psi}_{l,q} (g)^q_{~2} }  { - \frac{i}{2} \bpsrb [ \bpsra , \psra] },
\end{equation}
and one can check that they satisfy 
\begin{gather}
     \{Q_{l}^i,\ol{Q}_{l,j}\}=\delta^i_j H = \{Q_{r}^i,\ol{Q}_{r,j}\}\\
     \{Q^i, Q^j\}=\{\ol Q_i, \ol Q_j\}=\{Q_l, \ol Q_r \}=0\,,
\end{gather}
where the second line applies to arbitrary left/right pairings of charges.

\section{LQM two-point functions}
We include more two-point functions used in the main text. They are written compactly in terms of a variable $s$. We relist for convenience the relation between $s$ and energy since it varies between states and supermultiplets
\begin{align}
    &|\Psi^{E,j}_{m,m}\rb \implies s^2 = E-\lr{j+\frac{1}{2}}^2\,,\\
    &|\chi^{E,j}_{m,m}\rb \implies s^2 = E-\lr{j+\frac{1}{2}}^2\,, \\
    &|H^{E,j}_{m,m}\rb \implies s^2 = E-j^2\,, \\
    &|L^{E,j}_{m,m}\rb \implies s^2 = E-\lr{j+1}^2\,.
\end{align}
The above values of $s$ must be used in the below correlators.

\subsection{Neutral operator in general spin sector} \label{app:spinless_ops}

For neutral operators, we find the following correlation functions:
\be
    \lb \Psi_{m',m'}^{E', j^{\prime}} | e^{-\Delta \ell}|\Psi_{m,m}^{E, j}\rb  = \lb \chi_{m',m'}^{E', j} | e^{-\Delta \ell}|\chi_{m,m}^{E, j}\rb =\frac{\delta_{jj^{\prime}}\delta_{mm^{\prime}}}{2j+1}\big(E E^{\prime} \G^{\Delta}_{s,s'} + \Delta(\Delta+1) \G^{\Delta+1}_{s,s'}),
\ee

\be
\lb \Psi_{m',m'}^{E', j'} | e^{-\Delta \ell}|\chi_{m,m}^{E, j}\rb =\frac{\delta_{jj^{\prime}}\delta_{mm^{\prime}}}{2j+1}  \Delta(\Delta+1)\G^{\Delta+1}_{s,s'},
\ee

\be
\lb L_{m',m'}^{E', j} | e^{-\Delta \ell}|L_{m,m}^{E, j}\rb  =\frac{\delta_{j,j'}\delta_{m,m'}}{2j+1} \frac{\Delta(E+E^{\prime}-J^2+(\Delta-J)^2)(E+E^{\prime}-J^2+(\Delta+1-J)^2)+\frac{(2\Delta+1-2J)}{-J}E E^{\prime}}{2(2\Delta+1)}\Gamma_{s,s'}^{\Delta},
\ee
with $J=j+1$.

\be
\begin{aligned}
\lb H_{m',m'}^{E', j} | e^{-\Delta \ell}|H_{m,m}^{E, j}\rb  =\frac{\delta_{j,j'}\delta_{m,m'}}{2j+1} \frac{\Delta(E+E^{\prime}-J^2+(\Delta+J)^2)(E+E^{\prime}-J^2+(\Delta+1+J)^2)+\frac{(2\Delta+1+2J)}{J}E E^{\prime}}{2(2\Delta+1)}\Gamma_{s,s'}^{\Delta},
\end{aligned}
\ee
with $J=j$.

\be
\lb \Psi_{m',m'}^{E', j} | e^{-\Delta \ell}|L_{m,m}^{E, j}\rb  =  \lb \chi_{m',m'}^{E', j} | e^{-\Delta \ell}|L_{m,m}^{E, j}\rb = \frac{\delta_{j,j'}\delta_{m,m'} }{2j+1} \Delta 
 \frac{E^{\prime}(\Delta+1)+ \Delta (E + (\Delta-j)^2- (j+1)^2)  }{2\Delta+1} \G^{\Delta+\frac{1}{2}}_{s,s'},
\ee

\be
\lb \Psi_{m',m'}^{E', j} | e^{-\Delta \ell}|H_{m,m}^{E, j}\rb  =  \lb \chi_{m',m'}^{E', j} | e^{-\Delta \ell}|H_{m,m}^{E, j}\rb = \frac{\delta_{j,j'}\delta_{m,m'} }{2j+1} \Delta 
 \frac{E^{\prime}(\Delta+1)+ \Delta (E + (\Delta+j+1)^2- j^2)  }{2\Delta+1} \G^{\Delta+\frac{1}{2}}_{s,s'},
\ee

\begin{equation}
    \lb L_{m',m'}^{E', j} | e^{-\Delta \ell}|H_{m,m}^{E, j}\rb = \frac{\delta_{jj'}\delta_{mm'}}{2j+1}\Delta \frac{(E-E^{\prime}-2j-1+\Delta^2)^2+4\Delta^2(E^{\prime}-(j+1)^2)}{2(2\Delta+1)}\Gamma^{\Delta}_{s,s'}.
\end{equation}

\subsection{Spin-$\frac{1}{2}$ operator in general spin sector} \label{sec:spinhalf_2pt}

4d QFT operators can be expanded in tensor/spinor spherical harmonics on the BH background. A two-point correlator of an operator with total angular momentum $j$, with axial angular momentum $m_p\in \{-j,-j+1,\ldots, j-1,j \}$ on the left and the opposite axial momenta on the right is given in LQM by the operator $e^{-\D \ell} D_{m m}^j(g)$. For a spinor of minimal spin $j=\frac{1}{2}$ and $m_p\in\{-\frac{1}{2},\frac{1}{2}\}$ the operator is $e^{-\D \ell} D^{\frac{1}{2}}_{m_p m_p}(g)$ with $\D=1$ for a minimally coupled fermion in 4d, and $\D=\frac{1}{2}$ for the non-minimally coupled fermion in the hypermultiplet. We compute the matrix element with $\D=\frac{1}{2}$

\be
\lb \Psi_{m',m'}^{E', j'} | e^{-\frac{\ell}{2}} D^{\frac{1}{2}}_{m_p m_p}|\Psi_{m,m}^{E, j}\rb = \lb \chi_{m',m'}^{E', j'} | e^{-\frac{\ell}{2}} D^{\frac{1}{2}}_{m_p m_p}|\chi_{m,m}^{E, j}\rb = \frac{1}{2j'+1}\left(C^{j'm'}_{\frac{1}{2} m_p;jm}\right)^2  
E E^{\prime} \G^{\Delta=\frac{1}{2}}_{s,s'},
\ee

\be
\lb \Psi_{m',m'}^{E', j'} | e^{-\frac{\ell}{2}} D^{\frac{1}{2}}_{m_p m_p} |\chi_{m,m}^{E, j}\rb = \frac{1}{2j'+1}\left(C^{j'm'}_{\frac{1}{2} m_p;jm}\right)^2  \big(\Delta(\Delta+1)-\frac{3}{4}\big) \G^{\Delta+1}_{s,s'}\rvert_{\D=\frac{1}{2}} = 0,
\ee

\be
\lb H_{m',m'}^{E', j'} | e^{-\frac{\ell}{2}} D^{\frac{1}{2}}_{m_p m_p}|H_{m,m}^{E, j}\rb =  
\frac{j+j^{\prime}+\frac{3}{2}}{j+j^{\prime}+\frac{1}{2}} \frac{1}{2j'+1}\left(C^{j'm'}_{\frac{1}{2} m_p;jm}\right)^2 
(\Theta(j<j')(E')^2+\Theta(j>j')E^2)\G^{\Delta=\frac{1}{2}}_{s,s'},
\ee

\be
\begin{aligned}
    \lb L_{m',m'}^{E', j'} | e^{-\frac{\ell}{2}} D^{\frac{1}{2}}_{m_p m_p}|L_{m,m}^{E, j}\rb = 
    \frac{j+j^{\prime}+\frac{1}{2}}{j+j^{\prime}+\frac{3}{2}} \frac{1}{2j'+1}\left(C^{j'm'}_{\frac{1}{2} m_p;jm}\right)^2 
(\Theta(j>j')(E')^2+\Theta(j<j')E^2)
    \G^{\Delta=\frac{1}{2}}_{s,s'},
\end{aligned}
\ee

\be
\begin{aligned}
    \lb H_{m',m'}^{E', j'} | e^{-\frac{\ell}{2}} D^{\frac{1}{2}}_{m_p m_p}|L_{m,m}^{E, j}\rb =  \frac{\delta_{(j+\frac{1}{2})j^{\prime}}}{(2j'+1)(2j+1)} \frac{1}{2j'+1}\left(C^{j'm'}_{\frac{1}{2} m_p;jm}\right)^2 \big((2j+1)E-(2j'+1)E^{\prime}\big)^2 \G^{\Delta=\frac{1}{2}}_{s,s'},
\end{aligned}
\ee

\be
\lb \Psi_{m',m'}^{E', j'} | e^{-\frac{\ell}{2}} D^{\frac{1}{2}}_{m_p m_p} |L_{m,m}^{E, j}\rb = \lb \chi_{m',m'}^{E', j'} | e^{-\frac{\ell}{2}} D^{\frac{1}{2}}_{m_p m_p} | L_{m,m}^{E, j}\rb 
 =\delta_{(j+\frac{1}{2})j^{\prime}} \frac{j+j'+\frac{1}{2}}{j+j'+\frac{3}{2}} \frac{1}{2j'+1}\left(C^{j'm'}_{\frac{1}{2} m_p;jm}\right)^2   E \G^{\Delta=1}_{s,s'},
\ee

\be
\lb H_{m',m'}^{E', j'} | e^{-\frac{\ell}{2}} D^{\frac{1}{2}}_{m_p m_p} |\chi_{m,m}^{E, j}\rb = \lb H_{m',m'}^{E', j'} | e^{-\frac{\ell}{2}} D^{\frac{1}{2}}_{m_p m_p} |\Psi_{m,m}^{E, j}\rb 
 =\delta_{(j+\frac{1}{2})j^{\prime}} \frac{j+j'+\frac{3}{2}}{j+j'+\frac{1}{2}} \frac{1}{2j'+1}\left(C^{j'm'}_{\frac{1}{2} m_p;jm}\right)^2 E^{\prime} \G^{\Delta=1}_{s,s'},
\ee

\be
 \lb \mathrm{BPS}|e^{-\frac{\ell}{2}} D^{\frac{1}{2}}_{m_p m_p} | H^{E,\frac{1}{2}}_{m,m} \rangle =  \delta_{m_p,-m} \frac{1}{4}E^{2} \pi^2 \sech^2(\pi \sqrt{E-E_0(\mathbf{J}=1/2)}),
\ee

\paragraph{Identities.}
We will also make use of the following identities:
\be
\int dg (D^{j'}_{m',m'}(g))^*  D^{\ell}_{m_p,m_p}(g)  D^{j}_{m,m}(g) =\frac{1}{2j'+1}\left(C^{j' m'}_{j m; \ell m_p} \right)^2,
\ee
\be
\sum_{m_i,m',m_p} \frac{1}{2j'+1}{\left(C^{j' m'}_{j_p m_p; j,m_i}\right)^2} = 1\,,
\ee
where the last identity holds assuming $j' \supset j \otimes j_p$. In the above, the Haar measure for $SU(2)$ is unit normalized $\int d g = 1$. The last identity is useful when summing over all quantum numbers in transition rates.

\subsection{Bilocal operators and selection rules}

Below, we will discuss the non-trivial selection rules that follow from $\mathcal N=4$ super-Poincar\'e symmetry and are not already implied by angular momentum selection rules. Specifically, the following transitions are forbidden 
\begin{equation}\label{eqn:selection_rules}
\begin{aligned}
   & \langle H^{j^\prime}|e^{-\frac{\ell}{2}} D_{m_p m_p}^{\frac{1}{2}}|\chi^j\rangle =   \langle H^{j^\prime}|e^{-\frac{\ell}{2}} D_{m_p m_p}^{\frac{1}{2}}|\Psi^j\rangle = 0 \quad \text{if } j>j^{\prime}, \\
    & \langle L^{j^\prime}|e^{-\frac{\ell}{2}} D_{m_p m_p}^{\frac{1}{2}}|\chi^j\rangle =   \langle L^{j^\prime}|e^{-\frac{\ell}{2}} D_{m_p m_p}^{\frac{1}{2}}|\Psi^j\rangle = 0 \quad \text{if } j<j^{\prime}, \\
    & \langle \chi|e^{-\frac{\ell}{2}} D_{m_p m_p}^{\frac{1}{2}} | \Psi \rangle = 0\,,
\end{aligned}
\end{equation}
where all of the states in the above table can have arbitrary values of $j_z$. The reason for these selection rules is that the operator $e^{-\frac{\ell}{2}}D^{\frac{1}{2}}_{m_p m_p}$, whose origin is from the dimensional reduction of a higher dimensional fermionic field, itself falls into a representation of $\mathcal{N}=4$ super-Poincar\'e. In particular, half of the supercharges commute with it
\begin{equation} \label{eqn:selection_rules_commutators}
\begin{aligned}
       [e^{-\frac{\ell}{2}}D^{\frac{1}{2}}_{\frac{1}{2}\frac{1}{2}},\bar{Q}_{l}^1]=  [e^{-\frac{\ell}{2}}D^{\frac{1}{2}}_{\frac{1}{2}\frac{1}{2}},{Q}_{l}^2]= [e^{-\frac{\ell}{2}}D^{\frac{1}{2}}_{\frac{1}{2}\frac{1}{2}},Q_{r}^1]=  [e^{-\frac{\ell}{2}}D^{\frac{1}{2}}_{\frac{1}{2}\frac{1}{2}},\bar{Q}_{r}^2] = 0 \,,\\
\end{aligned}
\end{equation}
and conjuating the relations gives commutators of $Q,\bar{Q}$ with $e^{-\frac{\ell}{2}}D^{\frac{1}{2}}_{-\frac{1}{2}-\frac{1}{2}}$.
The selection rule \eqref{eqn:selection_rules} can be understood from this special feature of the operator.
We give one example proof of such a relation. 
First, notice that because of the Wigner-Eckart theorem  
\begin{equation}
    \langle H^{j^\prime}_{m_H m_H}|e^{-\frac{\ell}{2}} D_{m_p m_p}^{\frac{1}{2}}|\Phi^j_{m_\Psi m_\Psi}\rangle \propto \left(C^{j'm_H}_{\frac{1}{2}m_p;j m_\Psi}\right)^2  \times (\text{reduced matrix element independent of }m)\,,
\end{equation}
so that one can just focus on the highest $j_z$ state for $|H^{j'}\rangle$ and $|\Psi^j\rangle$.
If for the highest $j_z$ state the matrix element is 0, then it is true for every $m$.
Then if one insists on having $j'<j$, then $m_p$ has to be $-\frac{1}{2}$. Thus, one needs to check 
\begin{equation}
\begin{aligned}
    \Theta(j'>j)\langle H^{j'}_{j'j'}|e^{-\frac{\ell}{2}}D^{\frac{1}{2}}_{-\frac{1}{2}-\frac{1}{2}} | \Psi^j_{jj}\rangle  &= \frac{\delta_{j'(j+\frac{1}{2})}}{E} \langle H^{j'}|e^{-\frac{\ell}{2}}D^{\frac{1}{2}}_{-\frac{1}{2}-\frac{1}{2}} (\bar{Q}_l^1 Q_l^1+ {Q}_l^1 \bar{Q}_l^1)| \Psi^j\rangle \\
     & = \frac{\delta_{j'(j+\frac{1}{2})}}{E} \langle H^{j'}|e^{-\frac{\ell}{2}}D^{\frac{1}{2}}_{-\frac{1}{2}-\frac{1}{2}} ({Q}_l^1 \bar{Q}_l^1)| \Psi^j\rangle \\
     & = \frac{\delta_{j'(j+\frac{1}{2})}}{E} \langle H^{j'}|{Q}_l^1 e^{-\frac{\ell}{2}}D^{\frac{1}{2}}_{-\frac{1}{2}-\frac{1}{2}}  \bar{Q}_l^1| \Psi^j\rangle = 0\,,
\end{aligned}
\end{equation}
where we have used $Q_{l}^1|\Psi\rangle = \bar{Q}_{l}^1|H\rangle = 0$, and the commutator $[e^{-\frac{\ell}{2}}D^{\frac{1}{2}}_{-\frac{1}{2}-\frac{1}{2}},{Q}_{l}^1]=0$.
Note that $j$ has to be greater than $j'$ in this calculation; if $j<j'$, then one could again go to highest $j_z$ state and use $e^{-\frac{\ell}{2}}D^{\frac{1}{2}}_{\frac{1}{2}\frac{1}{2}}$ operator, which does not commute with $Q_l^1$, so that the argument above does not hold. In fact, the $j<j'$ matrix element is indeed non-zero. Thus, this example is a specific case of selection rules arising from the Clebsch-Gordan coefficients of $\mathcal{N}=4$ super-Poincar\'e.

\bibliographystyle{utphys}
\bibliography{evaporation}

\end{document}